\documentclass[pdflatex,sn-mathphys-num]{sn-jnl}

\usepackage{graphicx}%
\usepackage{multirow}%
\usepackage{amsmath,amssymb,amsfonts}%
\usepackage{amsthm}%
\usepackage{mathrsfs}%
\usepackage[title]{appendix}%
\usepackage{xcolor}%
\usepackage{textcomp}%
\usepackage{manyfoot}%
\usepackage{booktabs}%
\usepackage{algorithm}%
\usepackage{algorithmicx}%
\usepackage{algpseudocode}%
\usepackage{listings}%

\usepackage{amsmath,amssymb,amsfonts,bm}
\usepackage{amsthm}
\usepackage{mathrsfs}
\usepackage{empheq}
\usepackage{cancel}

\usepackage{graphicx}
\usepackage{subcaption}   
\usepackage{caption}  

\usepackage{booktabs}
\usepackage{multirow}
\usepackage{tabularx}
\usepackage{longtable}
\usepackage{adjustbox}
\usepackage{makecell}
\usepackage{rotating}
\usepackage{hyperref}

\theoremstyle{thmstyleone}%
\theoremstyle{thmstyletwo}%

\theoremstyle{thmstylethree}%

\begin{document}

\title[Rethinking RSSI for WiFi Sensing]{Rethinking RSSI for WiFi Sensing}

\author*[1,2]{\fnm{Zhongqin} \sur{Wang}}
\email{zhongqin.wang@uts.edu.au}

\author*[1,2]{\fnm{J. Andrew} \sur{Zhang}}
\email{andrew.zhang@uts.edu.au}

\author[1,2]{\fnm{Kai} \sur{Wu}}
\email{kai.wu@uts.edu.au}

\author[2]{\fnm{Y. Jay} \sur{Guo}}
\email{jay.guo@uts.edu.au}

\affil[1]{%
\orgdiv{School of Electrical and Data Engineering},
\orgname{University of Technology Sydney},
\orgaddress{\city{Sydney}, \state{NSW}, \postcode{2007}, \country{Australia}}}

\affil[2]{%
\orgdiv{Global Big Data Technologies Centre},
\orgname{University of Technology Sydney},
\orgaddress{\city{Sydney}, \state{NSW}, \postcode{2007}, \country{Australia}}}


\abstract{The Received Signal Strength Indicator (RSSI) is ubiquitously available on commodity WiFi devices but is commonly regarded as too coarse for fine-grained sensing. This paper revisits its sensing potential and presents \textit{WiRSSI}, a bistatic WiFi sensing framework that enables RSSI-only passive human tracking and motion sensing. WiRSSI employs a transmitter and a receiver equipped with a three-antenna array (1Tx-3Rx), and is readily extensible to Multiple-Input Multiple-Output (MIMO) deployments. We first show how Channel State Information (CSI) power implicitly preserves phase-related motion modulation and how this relationship carries over to RSSI, indicating that RSSI can retain exploitable Doppler, Angle-of-Arrival (AoA), and delay cues. WiRSSI extracts Doppler-AoA features via a lightweight 2D Fast Fourier Transform (FFT) pipeline and infers bistatic delay from amplitude-only information in the absence of subcarrier-level phase. The estimated AoA and delay are then mapped to Cartesian coordinates and denoised to recover motion trajectories. Experiments in practical environments show that WiRSSI achieves median XY localization errors of 0.905~m, 0.784~m, and 0.785~m for elliptical, linear, and rectangular trajectories, respectively, compared with 0.574~m, 0.599~m, and 0.514~m from a representative CSI-based method. We further demonstrate RSSI-only gesture recognition on the Widar3.0 dataset, where WiRSSI features provide meaningful discriminative performance. These results suggest that, despite lacking subcarrier-level information compared with CSI, RSSI can support practical WiFi sensing as a complementary and hardware-friendly option when CSI is restricted, unreliable, or privacy-sensitive.}

\maketitle

\section{Introduction}
Integrated Sensing and Communication (ISAC) has emerged as a key paradigm for next-generation communication systems, aiming to integrate environmental sensing and data communication into a unified framework \cite{zhang2021overview, liu2023integrated}. By enabling wireless signals to simultaneously convey information and perceive the surrounding environment, ISAC opens new opportunities in diverse applications such as environment sensing \cite{feng2022lte, wu2025isac, wang2025water, masood2025efficient, xu2025smartphone}, human-computer interaction \cite{pegoraro2024jump, miao2025wi, wang2025towards}, and healthcare monitoring \cite{feng2021lte, chen2023cross}. These applications demand low-cost, device-free, and easily deployable sensing solutions that can operate within existing communication infrastructure. At present, WiFi is particularly appealing for realizing ISAC, as the IEEE 802.11bf amendment \cite{du2024overview} introduces Channel State Information (CSI)-based sensing functionality into WiFi systems. {However, CSI access on commodity platforms is often limited by firmware/driver support and vendor policies. Beyond availability, in human-centric ISAC applications, fine-grained CSI may raise privacy and compliance concerns because it can encode detailed motion signatures. As a result, some deployments intentionally limit or avoid exposing fine-grained channel measurements, which can hinder the large-scale deployment of WiFi sensing in practice.}

{The Received Signal Strength Indicator (RSSI) is ubiquitously exposed on commodity WiFi devices and can serve as a complementary, hardware-friendly sensing modality, offering a low-cost baseline capability and a practical fallback option when CSI is restricted, unreliable, or privacy-sensitive. However, RSSI has long been regarded as inadequate for fine-grained sensing. This view mainly stems from its coarse representation of the wireless channel: RSSI reports only bandwidth-aggregated received power, typically quantized at low resolution (e.g., 8-10 bits, $\sim$1~dB per level), thereby discarding multipath diversity and limiting sensitivity to small-scale fading and micro-motion. In addition, RSSI is influenced by hardware-dependent factors such as automatic gain control, antenna polarization mismatch, and RF front-end variations, leading to inconsistent power scaling across time and devices. As a result, RSSI has been widely considered incapable of capturing meaningful environmental or human-motion dynamics.} 

Over the past decades, RSSI-based WiFi sensing has been extensively studied \cite{yang2013rssi, dubey2021enhanced}. Early work primarily relied on classical path-loss models \cite{rappaport2002wireless, srinivasa2009path} to estimate the distance between an active transmitter and receiver. While such models can capture large-scale attenuation, they face significant challenges for passive target localization. Low-resolution RSSI reflects only bandwidth-aggregated received power from all propagation paths, making it difficult to isolate reflections associated with the target of interest. Some studies explore RSSI fingerprinting \cite{yiu2017wireless, chatzimichail2019rssi, wang2025learning}, which builds a database of RSSI patterns at known locations and localizes by matching new measurements to stored fingerprints. Although fingerprinting can achieve meter-level accuracy in a fixed environment, it requires site-specific calibration and extensive data collection, and its performance often degrades under environmental changes or when the target moves in unseen regions. These limitations motivate a more principled investigation of how RSSI encodes spatial, temporal, and geometric information beyond empirical mapping and data-driven matching.

{
In recent years, WiFi sensing has increasingly shifted toward CSI because it provides fine-grained amplitude and phase measurements across subcarriers. However, bistatic CSI sensing is highly sensitive to random phase distortions, including clock asynchrony between the transmitter and receiver and hardware-induced phase discontinuities (e.g., $\pi$-radian jumps caused by phase-locked loop (PLL) behaviour across receive chains), which degrade sensing accuracy. A range of methods have been proposed to mitigate these effects \cite{wu2024sensing}, including inter-antenna correlation- or ratio-based processing \cite{qian2018widar2, feng2021lte, chen2023development, wang2023single, 10737138}, regression-based compensation \cite{Navid2019}, and reference-signal construction \cite{meneghello2022sharp, 11079818, wang2025towards, wang2025bistaticpassivetrackingcsi}. These advances have significantly improved the robustness of CSI-based sensing on commodity devices. In contrast, RSSI is a quantized, bandwidth-aggregated power measurement and does not provide subcarrier-level information. Nevertheless, RSSI is derived from the same received signal energy that underlies CSI, and the interaction between dominant static paths and target-induced dynamic paths can imprint motion-dependent modulation in the power domain. Despite this physical basis, RSSI-centric sensing has received far less attention than CSI-based approaches. \textit{Is RSSI truly as unsuitable for fine-grained sensing as commonly believed?}}

In this work, we revisit the physical-layer modeling of RSSI and show that, with appropriate CSI-inspired signal processing, RSSI can serve as an effective signal source for passive WiFi sensing. We present \textit{WiRSSI}, a bistatic WiFi sensing framework that enables passive target tracking using only RSSI measurements. WiRSSI employs a 1Tx-3Rx setup, where the receiver is equipped with a three-antenna array, and the transmitter and receiver are spatially separated. It achieves joint AoA and delay estimation from power-domain signals and supports robust, continuous target tracking. The framework is readily extensible to Multiple-Input Multiple-Output (MIMO) deployments. {Beyond tracking, we show that WiRSSI features provide effective motion signatures for RSSI-only sensing tasks.} The main contributions of this work are summarized as follows:

\textit{1)} Unlike conventional path-loss RSSI models, we derive an RSSI signal-power model from aggregated CSI power, showing that RSSI, as a subcarrier-integrated power representation, implicitly encodes phase-related motion information and captures the relationship among Doppler, AoA, and delay.

\textit{2)} We design a low-complexity feature extraction method based on a 2D Fast Fourier Transform (FFT) to jointly estimate Doppler and AoA features from multi-antenna RSSI measurements. In the absence of subcarrier-level information, amplitude-only features are further exploited to estimate target delay. Based on these features, WiRSSI enables target localization and continuous tracking in dynamic scenarios.

\textit{3)} We implement a bistatic WiFi system using Intel 5300 NICs \cite{halperin2011tool}, which provide per-antenna RSSI readings alongside CSI measurements, enabling synchronized data collection and fair, side-by-side performance comparison under identical conditions. Experiments are conducted on elliptical, linear, and rectangular trajectories, achieving median XY errors of 0.905~m, 0.784~m, and 0.785~m, respectively. For comparison, a representative CSI-based sensing framework~\cite{wang2025towards} achieves median XY errors of 0.574~m, 0.599~m, and 0.514~m. {We further evaluate WiRSSI features for RSSI-only gesture recognition on the Widar3.0 dataset \cite{zhang2021widar3}, showing that RSSI-only features retain discriminative motion cues beyond tracking.} Despite this accuracy gap, the results demonstrate the potential of low-resolution RSSI for low-cost and widely deployable ISAC applications.

\section{CSI and RSSI Signal Models}
This section examines the physical-layer relationship between CSI and RSSI, and shows how RSSI can be approximately related to aggregated CSI power across subcarriers.

\subsection{Bistatic CSI Model}
In bistatic systems, the transmitter and receiver are spatially separated and operate without a shared clock. As a result, the measured CSI at the receiver suffers from random phase distortions on every CSI sample, including a Timing Offset (TO) $\tau_k^{\text{TO}}$ and a Carrier Frequency Offset (CFO) $\phi_k^{\text{CFO}}$. In addition, each receiving antenna experiences a hardware-induced Phase Offset (PO) $\phi_j^{\text{PO}}$, initialized by the hardware in the receiving chain, such as PLL,  local oscillator, and RF cables, which may vary across devices, receiving antennas, and power cycles. Moreover, the received signal undergoes Automatic Gain Control (AGC) adjustment, represented by a time-varying amplitude factor $\alpha_k$. The AGC dynamically scales the baseband signal amplitude based on the measured received power to prevent Analog-to-Digital Converter (ADC) saturation and to maintain an appropriate signal level for demodulation. Consequently, the measured CSI is subject to multiple impairments, including random phase shifts from clock asynchrony, hardware-dependent per-antenna phase offsets, and AGC-induced amplitude scaling, all of which jointly distort the observed channel response.

Within a short-time coherent processing interval (CPI) \cite{zhang2021overview} (e.g., 0.1 seconds interval), the AGC-induced time-varying amplitude factor $\alpha_k$ is typically assumed to be constant. And let $\mathit{CSI}_{i,j,k}$ denote the measured CSI at the $i$-th Rx antenna, the $j$-th subcarrier, and the $k$-th time, which is modeled as:
\begin{equation}
\mathit{CSI}_{i,j,k} =
\alpha e^{-\bm{J} \left( 2\pi f_j \tau_k^{\text{TO}} + \phi_k^{\text{CFO}} + \phi_i^{\text{PO}} \right)}\left( H_{i,j}^{S} + H_{i,j,k}^{X} \right),
\label{eq:csi_model}
\end{equation}
where
\begin{equation}
\left\{
\begin{aligned}
&H_{i,j}^{S} = \sum_{l_1}\rho_{i,j}^{S}[l_1]\,e^{-\bm{J}2\pi f_j \tau^{S}_i[l_1]},\\
&H_{i,j,k}^{X} = \sum_{l_2}\rho_{i,j,k}^{X}[l_2]\,e^{-\bm{J}2\pi \left( f_j\tau_k^{X}[l_2] + f_k^{D}[l_2](k-1)\Delta t + \frac{(i-1)}{2} \sin\theta_k^X[l_2] \right)}.
\end{aligned}
\right.
\label{eq:csi_static_dynamic}
\end{equation}
Here, $H_{i,j}^{S}$ denotes the channel frequency response (CFR) of static paths, including Line-of-Sight (LOS), Non-Line-of-Sight (NLOS), and reflections from stationary objects such as walls, floors, and furniture. Each static path is characterized by an attenuation $\rho_{i,j}^{S}$ and delay $\tau^{S}_i$. In addition, $H_{i,j,k}^{X}$ represents the CFR of dynamic paths induced by moving objects, each described by attenuation $\rho_{i,j,k}^{X}$, delay $\tau^{X}$, Doppler frequency shift $f^{D}$, and AoA $\theta_k^{X}$. We assume a uniform linear array where the antenna spacing is set to half of the wavelength. $\Delta t$ denotes the CSI sample interval, which determines the temporal sampling resolution for Doppler estimation. The static and dynamic CFRs jointly encode the multipath delay, Doppler, and AoA across subcarriers, time, and antennas, forming the foundation for extracting delay-Doppler-AoA sensing features.

\subsection{CSI Power Modelling}
\label{subsec:csi_power_model}
According to Eq.~\eqref{eq:csi_model}, the CSI can be multiplied by its complex conjugate to obtain the CSI power, which removes random phase components while implicitly retaining variations induced by underlying phase evolution,
\begin{equation}
\begin{aligned}
& \text{P}_{i,j,k}=CSI_{i,j,k} \overline{CSI}_{i,j,k} \\
& = \alpha^2 \left( H^S_{i,j} + H^X_{i,j,k} \right) 
\left( \overline{H}^S_{i,j} + \overline{H}^X_{i,j,k} \right) \\
&= \alpha^2 \left[ \left| H^S_{i,j} \right|^2 + \left| H^X_{i,j,k} \right|^2 + 2 \left| H^S_{i,j} \overline{H}^X_{i,j,k} \right| \cos \left( \angle H^S_{i,j} \overline{H}^X_{i,j,k} \right)  \right],
\end{aligned}
\label{equation3}
\end{equation}
where $\left| \cdot \right|$ denotes the magnitude, and the TO, CFO, and PO are all eliminated. We then rely on Eq. \eqref{eq:csi_static_dynamic} to obtain
\begin{equation}
\begin{aligned}
& \left| H_{i,j}^S H_{i,j,k}^X \right| \cos \left( \angle H_{i,j}^S H_{i,j,k}^X \right) = \\
& \sum_{l_1} \sum_{l_2} \rho_{i,j}^S[l_1] \rho_{i,j,k}^X[l_2] \cos \big( 
    \varphi^{\text{Doppler}}_{k}[l_2]+ \varphi^{\text{Delay}}[l_1,l_2] +\varphi^{\text{AoA}}_{i,j}[l_1,l_2] \big),
\end{aligned}
\label{equation4}
\end{equation}
where 
\begin{equation}
\left\{ 
\begin{aligned}
&\varphi^{\text{Doppler}}_{k}[l_2] = 2 \pi f^D[l_2] \left(k - 1\right) \Delta t, \\
&\varphi^{\text{Delay}}_j{[l_1,l_2]} = 2 \pi f_j \left( \tau^X[l_2] - \tau^S[l_1] \right), \\
&\varphi^{\text{AoA}}_{i,j}[l_1,l_2] = \pi \left(i - 1\right) \left( \sin \theta^X[l_2] - \sin \theta^S[l_1] \right).
\end{aligned}
\right.
\label{equation5}
\end{equation}

From the above equations, it can be observed that the static component $H^S_{i,j}$ contains the direct propagation path between the transmitter and receiver, and its power ${\left| H^S_{i,j} \right|^2}$ typically dominates over other reflections. Furthermore, the mixed term ${\left| H^S_{i,j} \overline{H}^X_{i,j,k} \right|}$, which captures the interaction between the static and dynamic components, often remains stronger than the purely dynamic reflection term ${\left| H^X_{i,j,k} \right|^2}$. Therefore, the latter can be typically treated as noise in subsequent processing.

\begin{figure}[t]
    \centering
    \begin{subfigure}[t]{0.4\textwidth}
        \centering
        \includegraphics[width=\textwidth]{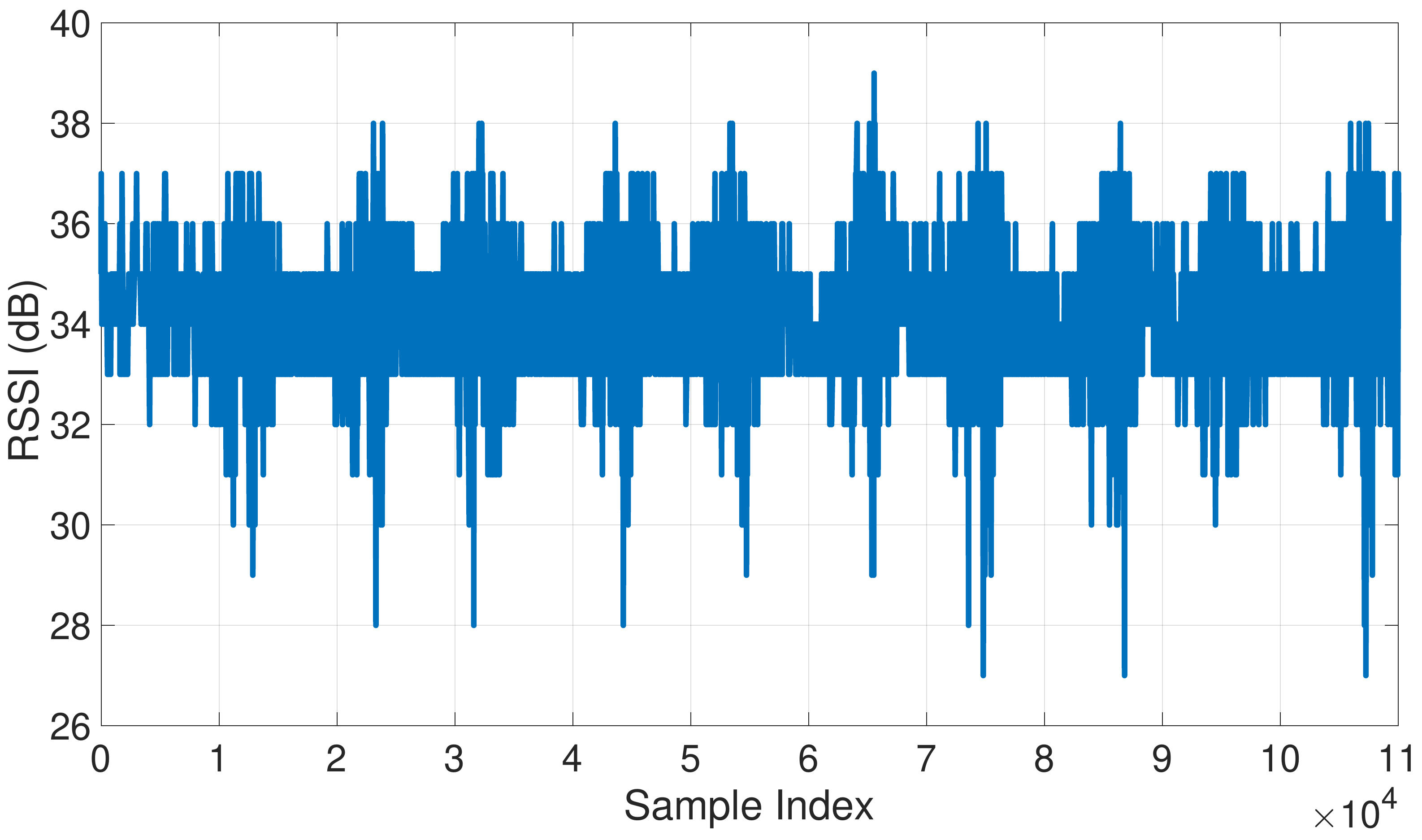}
        \subcaption{RSSI}
        \label{fig:rssi_timeseries}
    \end{subfigure}
    ~
    \begin{subfigure}[t]{0.4\textwidth}
        \centering
        \includegraphics[width=\textwidth]{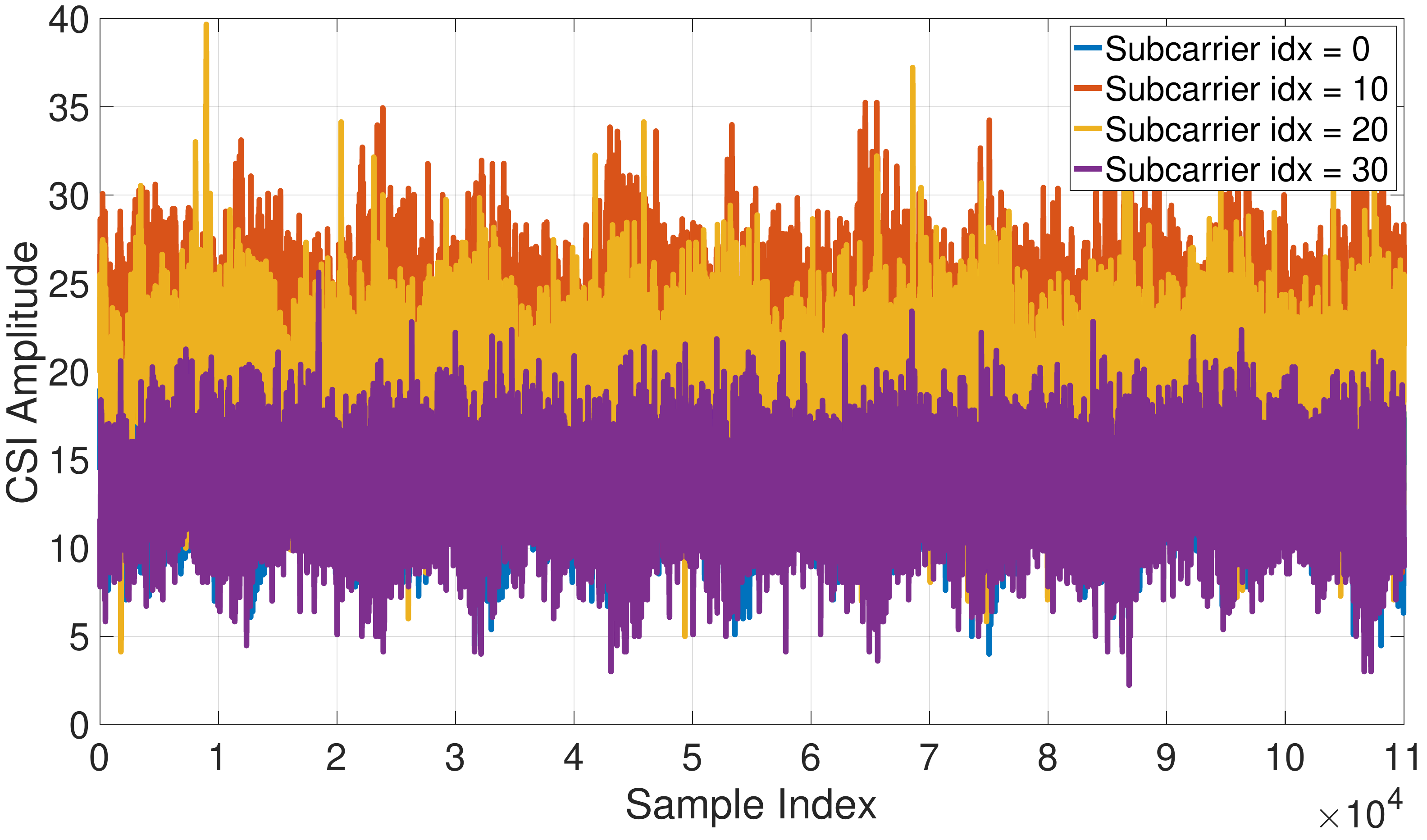}
        \subcaption{Raw CSI amplitude}
        \label{fig:csi_timeseries}
    \end{subfigure}
    \caption{Comparison of RSSI and CSI during human motion.}
    \label{fig:timeseries_comparison}
    \vspace{-1.5em}
\end{figure}

\subsection{RSSI Derivation from CSI Power}
We consider an approximate mapping from CSI power to RSSI. Let $\eta$ denote an effective power-scaling factor that accounts for transmit power, receiver gains, and other RF-related losses. Under the common assumption that such factors vary slowly compared with channel fluctuations, $\eta$ is treated as constant. Accordingly, the linear-scale RSSI (in milliwatts) $\mathcal{R}_{i,k}$ measured at the $i$-th antenna and the $k$-th time can be approximated from CSI power as
\begin{equation}
\mathcal{R}_{i,k} \approx \frac{1}{L}\sum_{j=1}^{L} \eta \left|\text{P}_{i,j,k}\right|,
\label{eq:RSSI_from_CSI_perantenna}
\end{equation}
where $L$ is the number of subcarriers. It can be converted from the hardware-reported RSSI measurement (in dB) to a linear scale as $\mathcal{R} = 10^{\frac{\text{RSSI}_{\text{dB}}}{10}}$. Note that in our subsequent experiments, we use the Intel 5300 NIC, for which the reported $\text{RSSI}_{\text{dB}}$ values are logarithmic power indicators expressed in dB \cite{halperin2011tool}, rather than absolute received power in dBm as commonly used in practical WiFi devices. This distinction affects only the absolute power reference and does not alter the linear conversion, since both dB and dBm follow the same logarithmic relationship up to a constant offset.

In commercial WiFi hardware, RSSI is quantized into integer steps with a resolution of about 1 dB. It is obtained by averaging the received signal power over a short interval within each packet, typically during the Orthogonal Frequency-Division Multiplexing (OFDM) preamble, thereby yielding a packet-level and relatively stable power measurement, as opposed to the rapidly varying per-symbol amplitudes available in CSI. Due to this quantization, the effective amplitude resolution is limited to roughly 10-12 distinguishable levels per decade of received power. Moreover, RSSI reflects the total received power integrated across the entire communication bandwidth (e.g., 20 MHz), resulting in lower temporal and spectral resolution compared with subcarrier-resolved CSI. Fig.~\ref{fig:rssi_timeseries} and Fig.~\ref{fig:csi_timeseries} show the reported RSSI (in dB) and the raw CSI amplitudes across selected subcarriers, including AGC-induced variations, measured by the Intel 5300 NIC, when a human subject repeatedly moves along an elliptical trajectory. RSSI mainly captures large-scale power variations correlated with CSI measurements, whereas CSI amplitudes provide higher-resolution signal details essential for fine-grained sensing. As indicated by the signal model in Section \ref{subsec:csi_power_model}, sensing information is preserved in the CSI power, and thus RSSI computed from the CSI power retains low-resolution sensing information, which is further validated below.

\section{Passive Human Tracking and Sensing via RSSI}
\label{sec:rssi_tracking}
This section presents the overall processing pipeline, outlining how RSSI can be transformed into Doppler-AoA features, delay estimates, and finally continuous  trajectories.

\subsection{System Overview}
This work re-examines the physical-layer foundations of RSSI and shows that, when properly processed, RSSI can be effectively leveraged for sensing. Building upon these insights, we develop \textit{WiRSSI}, a passive tracking framework that relies solely on low-resolution RSSI measurements for passive human tracking. The system adopts a bistatic 1Tx-3Rx configuration, while remaining naturally scalable to MIMO deployments. Despite the coarse resolution of RSSI, the proposed pipeline enables reliable AoA and delay estimation, and ultimately robust target tracking. {WiRSSI focuses on single-target passive tracking using a multi-antenna receiver array. Extending the framework to robust multi-target scenarios would require explicit multi-peak association and additional modeling, and is therefore left for future work.} The proposed WiRSSI framework consists of four main components:

\textit{1) Static Clutter Removal}: Convert RSSI to linear scale, suppress the dominant static component, and extract motion-induced variations. 

\textit{2) Doppler-AoA Feature Extraction:} Apply sequential temporal and spatial FFTs to obtain a 2D Doppler-AoA spectrum and locate the dominant target-related peak.

\textit{3) Delay Estimation:} Estimate the human-reflection delay from the amplitude of the detected Doppler-AoA peak by using the amplitude ratio between the dynamic (target) component and the static reference component.

\textit{4) Tracking and Sensing:} Convert the estimated AoA and delay into Cartesian coordinates and apply window-based filtering to obtain smoothed target trajectories. {In addition, the Doppler-AoA features can be compacted into time-Doppler signatures for sensing tasks such as gesture recognition.}

\subsection{Static Clutter Removal}
Within a CPI, the RSSI measurements can be denoted as $\mathbf{R} \in \mathbb{R}^{N \times M}$, where $N$ and $M$ are the number of receiving antennas and RSSI samples, respectively. Each RSSI value is first converted to the linear power domain for subsequent processing. As discussed earlier, both the transmit power and the amplitude of each propagation path typically remain constant within the CPI. And it is reasonable to assume that the static component between the transmitter and receiver is dominant, as it corresponds to the LOS or strong NLOS propagation paths, whereas the dynamic component induced by moving targets typically propagates over longer and more attenuated paths. Under these assumptions, we compute the temporal mean of the RSSI at each antenna to suppress the static clutter, denoted by $\mathcal{R}_i^{S}$,
\begin{equation}
\mathcal{R}_i^{S} \approx \frac{1}{M}\sum_{k=1}^{M}\mathcal{R}_{i,k}.
\label{eq:rssi_static_mean}
\end{equation}
And the dynamic component is then extracted by subtracting the mean from the instantaneous RSSI:
\begin{equation}
\mathcal{R}_{i,k}^{X} \approx \mathcal{R}_{i,k} - \mathcal{R}_i^{S}.
\label{eq:rssi_dynamic}
\end{equation}  
This approach robustly suppresses static clutter concentrated at zero Doppler frequency with low computational complexity. Furthermore, the dynamic component is normalized by the average $\mathcal{R}_i^{S}$:
\begin{equation}
\Delta\mathcal{R}_{i,k} = 
\frac{\mathcal{R}_{i,k}^{X}}{\mathcal{R}_i^{S}}.
\label{eq:rssi_normalized_dynamic}
\end{equation}
This normalization mitigates unknown amplitude-related variations, including those caused by hardware imperfections and transmit power fluctuations, and facilitates subsequent delay estimation, which will be detailed in Section \ref{subsec:delay_estimation}.

\subsection{Doppler-AoA Estimation via 2D FFT}
The Doppler-AoA features are then obtained by applying 2D FFTs along the temporal and spatial dimensions, achieving low computational complexity and stable feature extraction. 

\subsubsection{Doppler FFT} First, a Doppler FFT is applied over the sampling time for each antenna:
\begin{equation}
\begin{aligned}
X_i(f^D) &= \mathcal{F}_{D} \{\Delta \mathcal{R}_{i,k}\} \\
&= \sum_{k=0}^{M-1} \Delta\mathcal{R}_{i,k} 
e^{-\bm{J}2\pi {f}^D (k-1)\Delta t},
\end{aligned}
\label{eq:rssi_doppler_fft}
\end{equation}
where the Doppler frequency range is set within $\pm f_{\text{max}}$, with 
$f_{\text{max}} = {v_{\text{max}}}/{\lambda}$ and $v_{\text{max}}$ denoting the maximum target Doppler velocity. A Doppler range of $\pm100$~Hz is used, corresponding to a maximum velocity of about $6$~m/s at a 5~GHz center frequency, which adequately covers indoor human motion.

\begin{figure}
    \centering
        \includegraphics[width=0.4\textwidth]{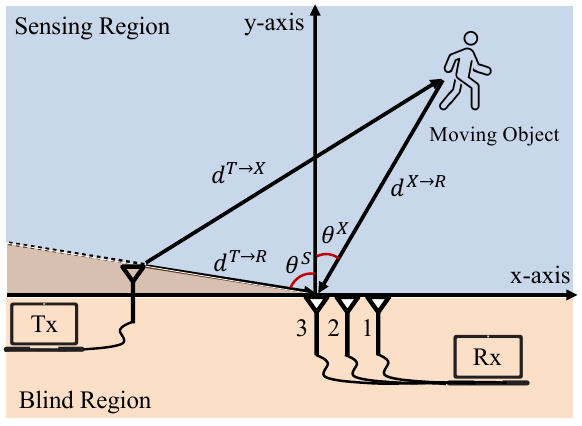}  
        \caption{{Bistatic geometry of the 1Tx-3Rx WiFi setup.}}
    \label{fig:coordinate_system}
    \vspace{-1.5em}
\end{figure}

\subsubsection{AoA FFT} An AoA FFT is applied over the antenna dimension at each Doppler bin to extract angle information:
\begin{equation}
\begin{aligned}
Y(f^D,\theta^X) &= \mathcal{F}_{AoA} \{X_i(f^D)e^{\bm{J}\pi (i-1)\sin\theta^S}\} \\
&= \sum_{i=0}^{N-1} \left[ X_i(f^D) e^{\bm{J}\pi (i-1)\sin\theta^S}\right] e^{-\bm{J}\pi (i-1)\sin\theta^X},
\end{aligned}
\label{eq:rssi_aoa_fft}
\end{equation}
where $\theta^{S}$ denotes the AoA of the transmitter with respect to the receiver's linear antenna array, which can be pre-determined from the known transceiver geometry. We define $\theta^{X}\in[-\pi/2,\pi/2]$ as the target AoA within the visible field-of-view of the linear array. 

{Since RSSI is real-valued, the Doppler FFT exhibits conjugate symmetry, and the magnitudes at $\pm f^{D}$ are identical. In our bistatic setup, we assume that the target moves within the designated sensing region shown in Fig.~\ref{fig:coordinate_system}. We restrict the search to one angular sector relative to the known transmitter direction $\theta^{S}$, and discard its mirrored sector. Under our AoA definition, this corresponds to keeping the sector $0 \le \sin\theta^{X}-\sin\theta^{S} \le 2$ and discarding $-2 \le \sin\theta^{X}-\sin\theta^{S} < 0$, which removes the Doppler-AoA mirror component induced by real-valued RSSI measurements. A blind configuration occurs when the transmitter direction provides no effective spatial reference, i.e., $\sin\theta^{S}=0$ (equivalently, $\theta^{S}$ is close to the array broadside direction under our angle convention). In this case, the transmitter-induced phase progression across antennas vanishes, and the joint spectrum becomes nearly symmetric, so the mirrored pair $(f^{D},\theta^{X})$ and $(-f^{D},-\theta^{X})$ cannot be reliably distinguished from RSSI magnitudes alone, as analyzed in Appendix~\ref{appendix:asymmetry}. Resolving the ambiguity would then require an additional prior constraint (e.g., restricting the allowed motion side or using extra geometry information). Therefore, in practice we place/configure the transmitter such that $\sin\theta^{S}\neq 0$, avoiding this blind configuration.}

\subsubsection{Doppler-AoA Spectrum}
The resulting 2D spectrum $|Y(f^D,\theta^X)|$ characterizes the power distribution over Doppler and AoA. Let $\mathbf{Y} \in \mathbb{C}^{L_{\text{Doppler}} \times L_{\text{AoA}}}$ denote the Doppler-AoA matrix, where $L_{\text{Doppler}}$ and $L_{\text{AoA}}$ are the numbers of Doppler and angular bins, respectively. The location of each peak corresponds to a moving target, capturing its radial motion through Doppler and its spatial direction through AoA. {Beyond tracking, this Doppler-AoA representation can also be compacted into motion signatures for sensing tasks (Section~\ref{subsec:object_sensing}).}

\subsection{Delay Estimation from Signal Amplitude}
\label{subsec:delay_estimation}
Unlike CSI, which enables delay estimation via subcarrier phase information, RSSI aggregates power over the entire bandwidth and thus lacks the phase information required for conventional delay extraction. To address this limitation, we instead explore the use of signal amplitude as an alternative cue for delay estimation. However, beyond its dependence on propagation distance, the amplitude of each propagation path is jointly influenced by multiple factors, including the target reflection coefficient, incident angle, surface material, and antenna polarization. These factors complicate the direct use of raw amplitude measurements for accurate sensing. 

\subsubsection{Amplitude Modeling of Static and Dynamic Paths} 
We first briefly explain the rationale for exploiting the path amplitude to infer the propagation delay. From a propagation perspective, the received signal amplitude inherently depends on the propagation distance due to path-loss and attenuation effects. Since the propagation delay is directly proportional to the path length, the amplitude provides an indirect yet informative cue for delay estimation. According to the amplitude-delay dependence derived in Appendix~\ref{appendix:LOS} and Appendix~\ref{appendix:NLOS}, the amplitudes of the dominant static Tx-Rx path and the target-induced dynamic path can be expressed as
\begin{equation}
\left\{
\begin{aligned}
\rho^{S} &= \frac{\Gamma^{S}}{\tau^{T \rightarrow R}},\\
\rho^{X} &= \frac{\tilde{\Gamma}^{X}}{\tau^{T \rightarrow X} + \tau^{X \rightarrow R}},
\end{aligned}
\right.
\label{eq:rho_proportionality}
\end{equation}
where $\tau^{T \rightarrow R}$ denotes the propagation delay of the static
Tx-Rx path, and $\tau^{T \rightarrow X}$ and $\tau^{X \rightarrow R}$ are the propagation delays from the transmitter to the target and from the target to the receiver, respectively. The coefficients $\Gamma^{S}$ and $\tilde{\Gamma}^{X}$ capture path-dependent amplitude factors, including reflection characteristics, antenna gains, polarization effects, and other hardware-related terms. 

In particular, Appendix~\ref{appendix:NLOS} also shows that the dynamic-path amplitude can be decomposed into a slowly varying geometric term and a reflection-related term. Specifically, we write
\begin{equation}
\tilde{\Gamma}^{X}
\triangleq
\frac{\tau^{T \rightarrow X} + \tau^{X \rightarrow R}}
{\tau^{T \rightarrow X} \, \tau^{X \rightarrow R}}
\, \Gamma^{X},
\end{equation}
where ${\Gamma}^{X}$ denotes a delay-independent amplitude factor associated with the target's scattering characteristics. Since these characteristics, together with the underlying hardware effects, vary on a much slower time scale than the propagation delays induced by target motion, ${\Gamma}^{X}$ can be treated as quasi-static. Moreover, for a fixed transceiver setup and a target whose motion is confined to a finite region relative to the Tx-Rx separation, the delay ratio $(\tau^{T \rightarrow X}_k + \tau^{X \rightarrow R}_k)/(\tau^{T \rightarrow X}_k \tau^{X \rightarrow R}_k)$ varies much more slowly over time than the total bistatic delay $\tau^{T \rightarrow X}_k + \tau^{X \rightarrow R}_k$. As a result, the combined coefficient $\tilde{\Gamma}^{X}_k$ can be approximated as a constant in typical indoor tracking scenarios.

\subsubsection{Delay Estimation} Recall from Eq.~\eqref{eq:rssi_aoa_fft} that, under the single-target assumption, the dominant peak in the extracted 2D Doppler-AoA spectrum corresponds to the moving target. Accordingly, the amplitude at the Doppler-AoA bin $(f^{D}, \theta^{X})$ reflects the contribution of the target-induced dynamic path. In contrast, the static component is assumed to be dominated by the LoS Tx-Rx path. Using the amplitude models in Eq. \eqref{eq:rho_proportionality}, the ratio between the  static and dynamic path amplitudes can be expressed as
\begin{equation}
\begin{aligned}
|Y(f^{D}, \theta^{X})|
&= \frac{\rho^{X}}{\rho^{S}} = \frac{\tilde{\Gamma}^{X}}{\Gamma^{S}}
\frac{\tau^{T \rightarrow R}} {\tau^{X}},
\end{aligned}
\label{eq:amp_ratio_full}
\end{equation}
where $\tau^{X} \triangleq \tau^{T \rightarrow X} + \tau^{X \rightarrow R}$ denotes the total bistatic propagation delay. Under the above approximation, the reflection-coefficient ratio $\gamma \triangleq \tilde{\Gamma}^{X}/\Gamma^{S}$ can be treated as a constant and obtained through prior calibration, as detailed below. This ratio effectively absorbs and cancels hardware-dependent amplitude scaling effects, thereby improving robustness across different devices and deployment configurations. In addition, the static-path delay $\tau^{T \rightarrow R}$ can be measured in advance. Consequently, the bistatic delay $\tau^{X}$ associated with the target-induced reflection path can be directly inferred.

{In indoor multipath environments, the instantaneous RSSI amplitude can exhibit Rayleigh/Rician fading and is not strictly monotonic with distance. Eq. \ref{eq:amp_ratio_full} is therefore not applied to raw RSSI. Instead, it is applied to the target-related component after joint Doppler-AoA processing, which acts as a temporal-spatial filter to suppress strong static clutter and off-bin multipath contributions. The focused peak amplitude shows an approximately monotonic trend with the effective bistatic path length, while residual fading-induced fluctuations are treated as noise and mitigated by the following CPI-level aggregation and temporal smoothing.}

\subsubsection{Prior Calibration of Reflection Coefficient Ratio} The reflection-coefficient ratio $\gamma$ is obtained through a one-time calibration under the same Tx-Rx deployment. During calibration, the bistatic delay of the target-induced path, $\tau^{X} = \tau^{T \rightarrow X} + \tau^{X \rightarrow R}$, is measured using an auxiliary ranging modality that provides accurate position information, such as a co-located mmWave radar. The static-path delay $\tau^{T \rightarrow R}$ is measured in advance. By collecting measurements at multiple known target locations within the sensing area, $\gamma$ is estimated at each location, and the final calibration value is obtained by averaging these estimates. During subsequent localization and tracking, the pre-estimated ratio $\gamma$ is treated as a known parameter and used to infer the bistatic delay associated with the human-reflection path.

{\textit{Remark:} The calibration of $\gamma$ is only required to obtain an absolute range scale. Without calibration, WiRSSI still provides relative range variations over time (up to an unknown scale factor) together with AoA trends, which is sufficient for many ISAC applications that prioritize motion tracking and change detection over absolute bistatic ranging.}

\subsection{Object Tracking and Sensing}
\subsubsection{Object Tracking}
As shown in Fig.~\ref{fig:coordinate_system}, the estimated polar coordinates of each detection, including the delay and the AoA, are first converted into Cartesian coordinates for target localization. Let $d^{T \rightarrow X}$ and $d^{X \rightarrow R}$ denote the propagation distances from the transmitter to the target and from the target to the receiver, respectively. These distances are obtained from the corresponding delays $\tau^{T \rightarrow X}$ and $\tau^{X \rightarrow R}$ by multiplying with the speed of light. 

Given the estimated AoA $\theta^{X}$ of the target-induced reflection and the known Tx-Rx separation, the target-receiver distance $d^{X \rightarrow R}$ can be computed according to the law of cosines as
\begin{equation}
d^{X \rightarrow R}
= \frac{(d^{X})^{2} - (d^{S})^{2}}
{2[d^{X} - d^{S}\cos(\theta^{X} - \theta^{S})]}.
\label{eq:coordinate_cosine}
\end{equation}
Then, the target coordinate can be represented in the receiver's local Cartesian coordinate system as:
\begin{equation}
\left\{
\begin{aligned}
x &= d^{X \rightarrow R}\sin\theta^{X},\\
y &= d^{X \rightarrow R}\cos\theta^{X}.
\end{aligned}
\right.
\label{eq:coordinate_transform}
\end{equation}

Before performing continuous tracking, the per-frame position estimates may contain outliers. To mitigate these effects, a Hampel filter is first applied to identify and remove outliers in both the X- and Y- coordinates. After that, a Savitzky-Golay (SG) filter is used to smooth the remaining position sequence, producing a locally smoothed trajectory while preserving the underlying motion trend. Unlike some CSI-based systems \cite{wu2025isac, 10737138} that employ an Extended Kalman Filter (EKF) to jointly refine delay, AoA, and Doppler\footnote{{Our CSI-based single-target tracking and sensing demo: \url{https://youtu.be/ldF2vq5x0P4}.}},  {WiRSSI adopts a window-based smoothing strategy to improve robustness under practical interference and RSSI variability. This design is motivated by the fact that RSSI-based Doppler estimates are typically noisier (see Section \ref{sec:results}), which can destabilize EKF updates. In dynamic environments, occasional movers (e.g., fans or other people) may cause intermittent feature jumps or short-term target loss. To suppress these transient disturbances, we apply Hampel outlier removal followed by SG smoothing, yielding robust and temporally consistent single-target tracking. However, persistent and high-amplitude motion interference is fundamentally a multi-target problem. Robust multi-target tracking is challenging even for CSI-based sensing, as it requires reliable feature extraction and data association\footnote{{Our CSI-based multi-target tracking demo: \url{https://youtu.be/DVib9wOY48k}.}}; the challenge is greater for RSSI due to its lower resolution. We therefore leave robust RSSI-based multi-target tracking for future work.}

{
\subsubsection{Object Sensing}
\label{subsec:object_sensing}
Beyond tracking, the Doppler-AoA spectrum extracted by WiRSSI also provides motion signatures that can be used for sensing tasks such as gesture recognition. Specifically, for each CPI, WiRSSI produces a complex Doppler-AoA response $Y(f^D,\theta)$ as in Eq. \eqref{eq:rssi_aoa_fft}. We first convert it to a nonnegative energy map:
\begin{equation}
S(f^D,\theta) = \left|Y(f^D,\theta)\right|^2 .
\label{eq:sensing_energy}
\end{equation}

To obtain a compact representation that is less sensitive to viewpoint changes, we compress the AoA dimension and form a Doppler profile for each CPI:
\begin{equation}
p(f^D) = \sum_{\theta} S(f^D,\theta).
\label{eq:aoa_collapse_sum}
\end{equation}
Let $t$ index the CPI sequence. Stacking the Doppler profiles over time yields a time-Doppler feature map:
\begin{equation}
\mathbf{P}(t,f^D) = p_t(f^D),
\label{eq:time_doppler_map}
\end{equation}
which captures motion-dependent Doppler evolution patterns. The resulting $\mathbf{P}$ can be used as an input feature for learning-based sensing modules, enabling RSSI-only sensing without requiring coordinate reconstruction.
}

\begin{figure}
    \centering
    \begin{subfigure}[t]{0.329\textwidth}
        \centering
        \includegraphics[width=\textwidth]{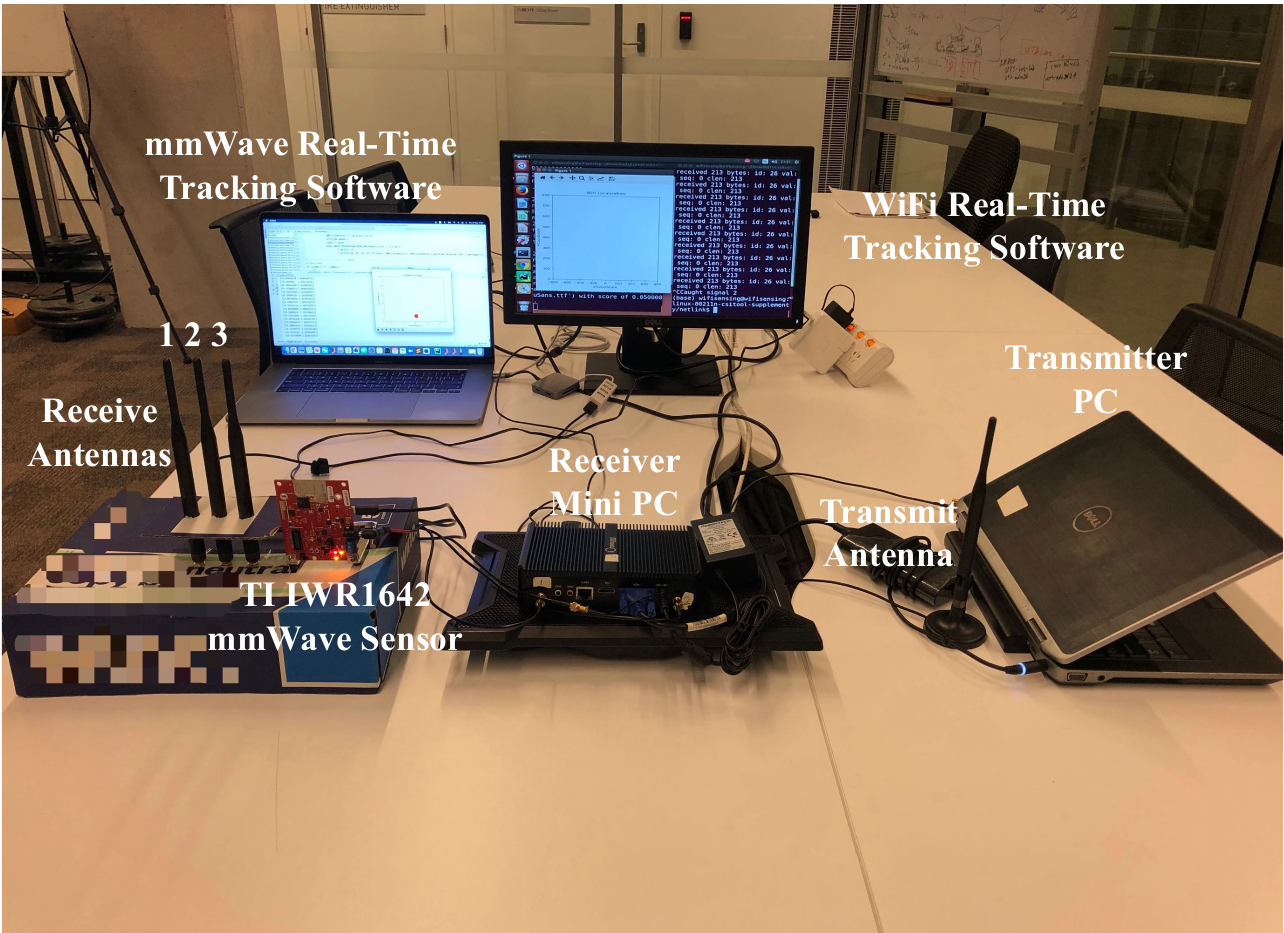}
        \subcaption{WiFi devices}
    \end{subfigure}
    \begin{subfigure}[t]{0.329\textwidth}
        \centering
        \includegraphics[width=0.98\textwidth]{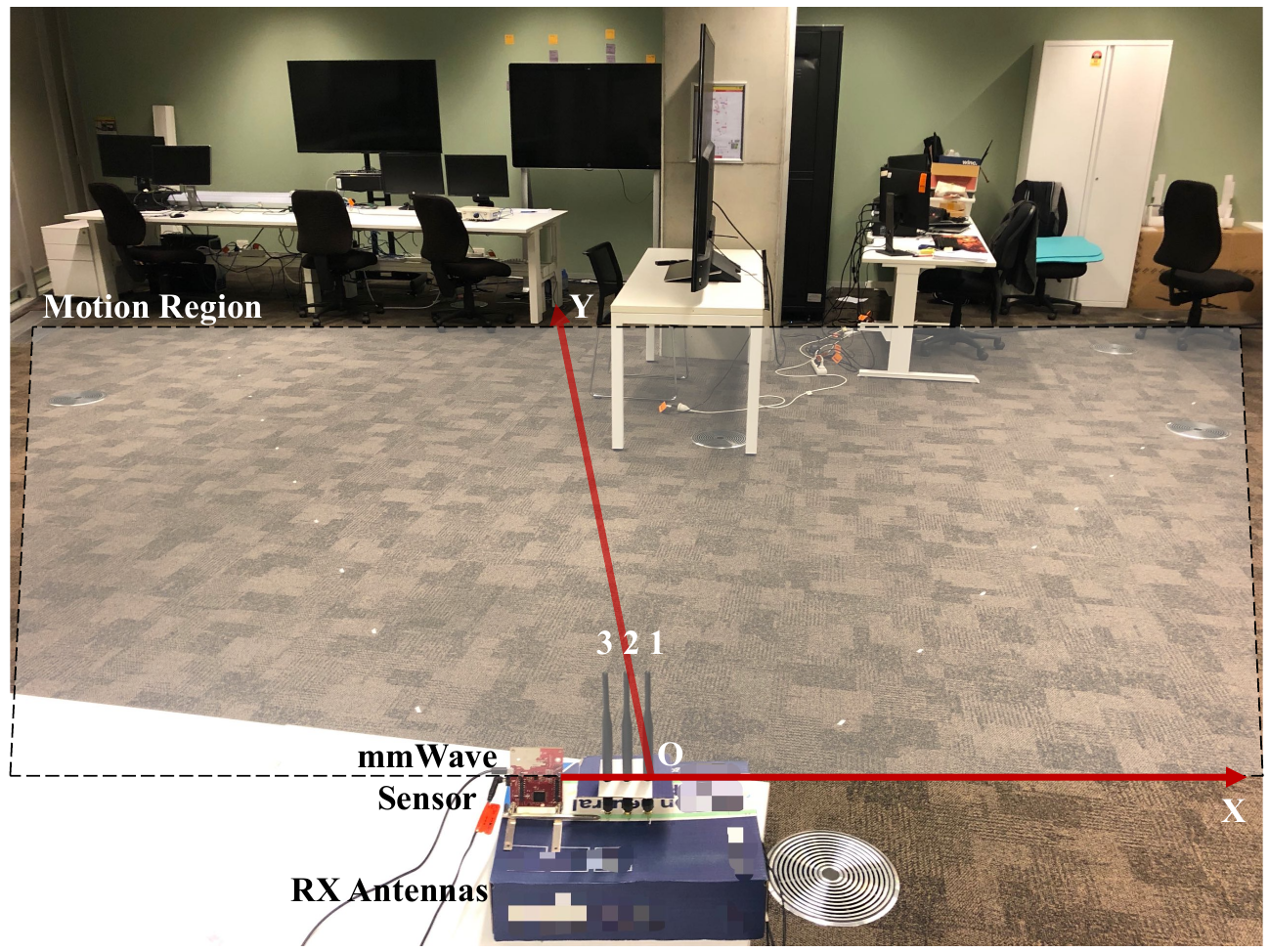}
        \subcaption{Receiving antenna array}
    \end{subfigure}
    \caption{Experimental setup of the WiFi sensing system.}
    \label{fig:experiment_setup}
    \vspace{-1.5em}
\end{figure}

\begin{figure*}
    \centering
    \begin{subfigure}[t]{0.329\textwidth}
        \centering
        \includegraphics[width=0.83\textwidth]{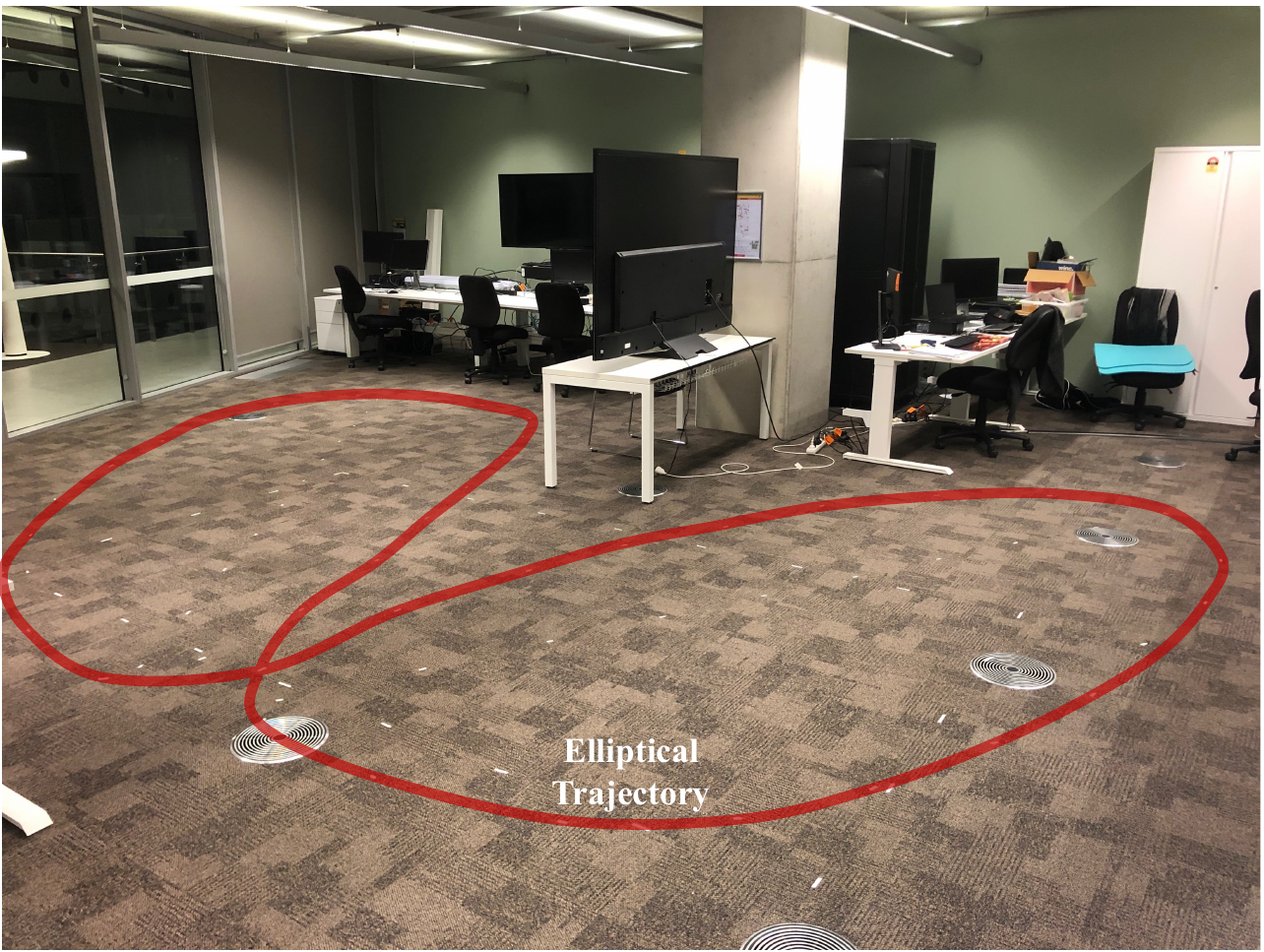}
        \subcaption{Elliptical trajectory}
        \label{fig:traj_ellipse_3}
    \end{subfigure}
    \hfill
    \begin{subfigure}[t]{0.329\textwidth}
        \centering
        \includegraphics[width=0.83\textwidth]{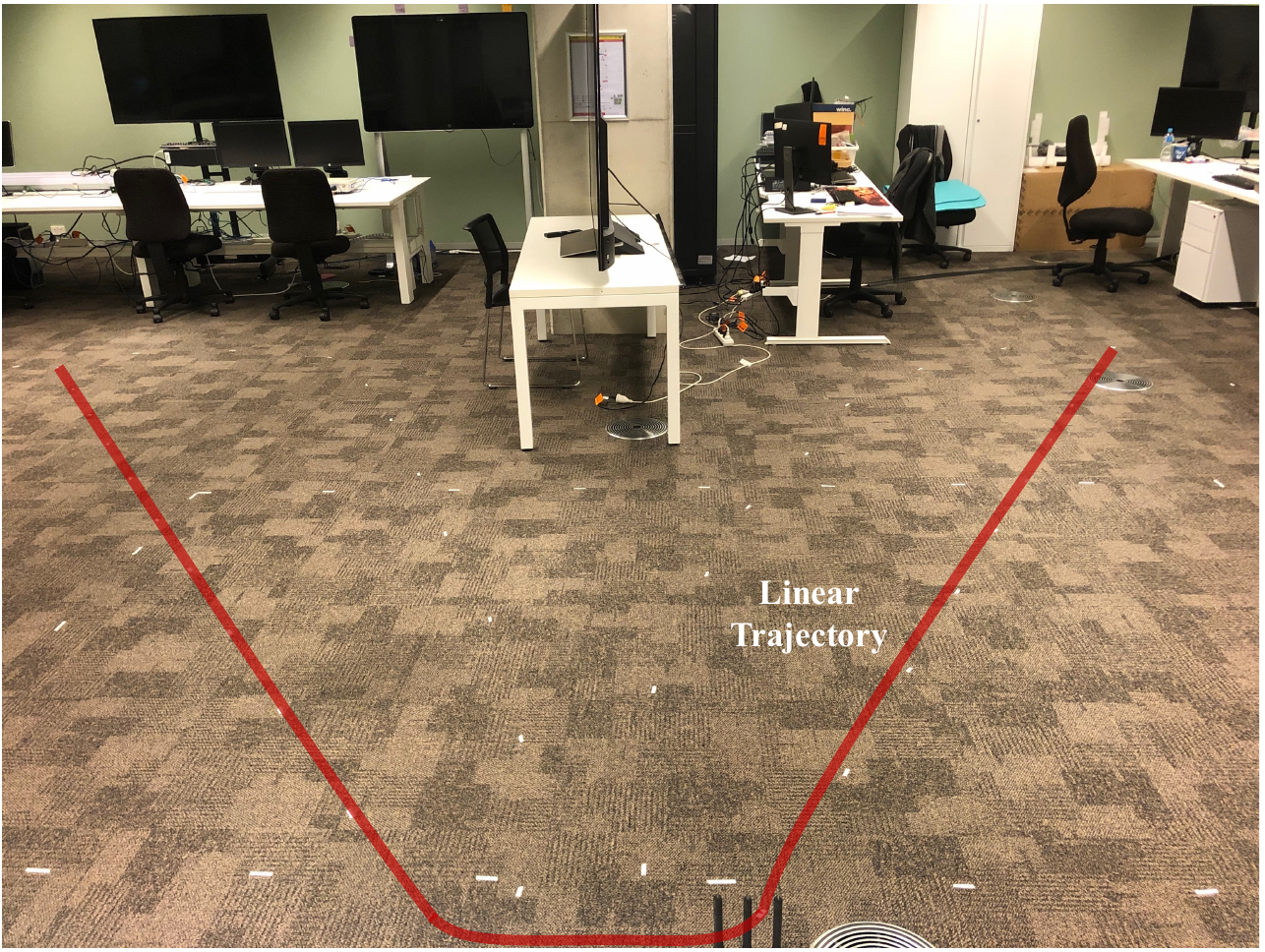}
        \subcaption{Linear trajectory}
        \label{fig:traj_linear_3}
    \end{subfigure}
    \hfill
    \begin{subfigure}[t]{0.329\textwidth}
        \centering
        \includegraphics[width=\textwidth]{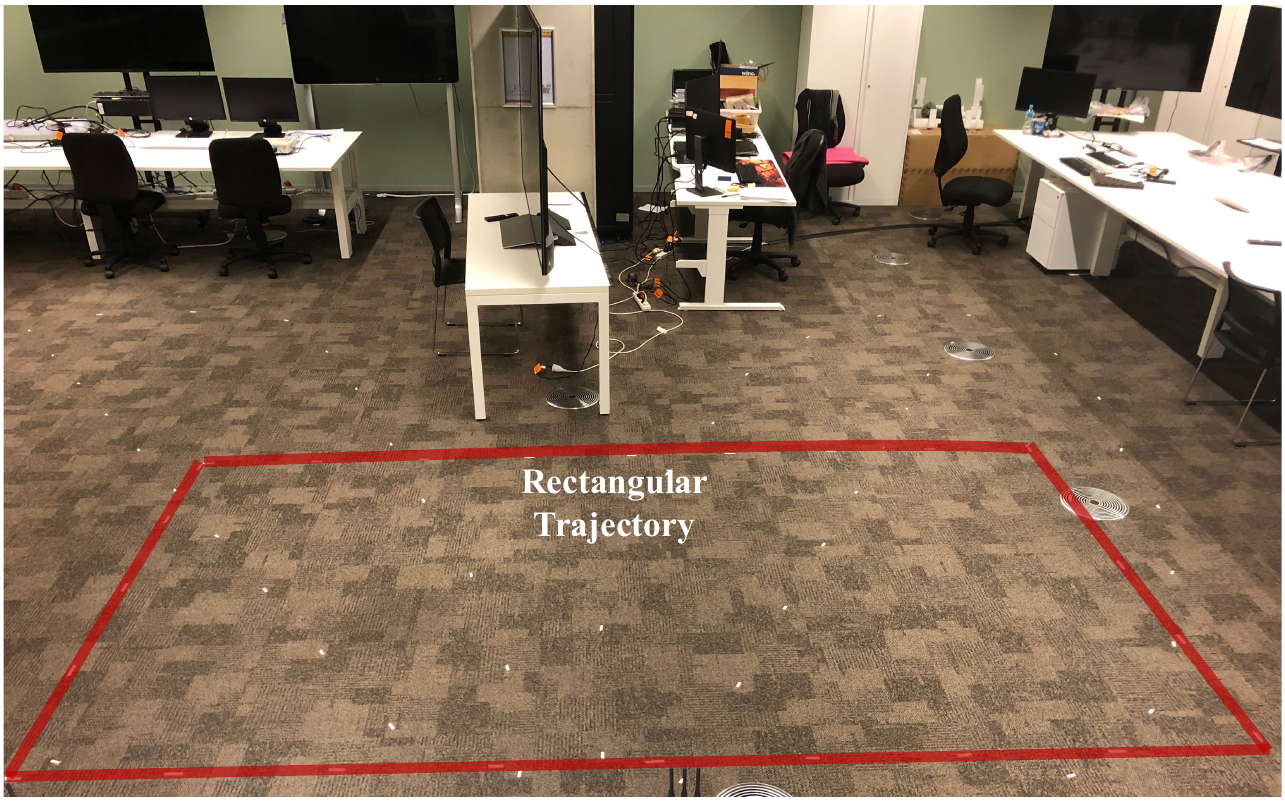}
        \subcaption{Rectangular trajectory}
        \label{fig:traj_rectangle_3}
    \end{subfigure}
    \caption{Human motion trajectories: (a) elliptical, (b) linear, and (c) rectangular.}
    \label{fig:traj_summary_3}
    \vspace{-1.5em}
\end{figure*}

\section{Implementation}

\subsection{Tracking Setup}
\subsubsection{Dataset}
As shown in Fig.~\ref{fig:experiment_setup}, the WiFi tracking dataset is collected using two PCs equipped with Intel 5300 NICs in a bistatic 1Tx-3Rx setup, where the receiver is equipped with a three-antenna array. The system operates at a center frequency of 5.32~GHz, with the transmitter and receiver separated by approximately 2.3~m. The Linux 802.11n CSI Tool \cite{halperin2011tool} is configured in monitor mode with a sampling rate of 1~kHz. {Both CSI and per-antenna RSSI are recorded by the NIC driver; WiRSSI uses only RSSI measurements, and CSI is used only to implement CSI-based baselines for fair comparison.} The dataset contains three representative motion trajectories: (1) elliptical, (2) linear, and (3) rectangular, as illustrated in Fig.~\ref{fig:traj_ellipse_3}, Fig.~\ref{fig:traj_linear_3}, and Fig.~\ref{fig:traj_rectangle_3}, respectively. A human subject moves at an average walking speed of approximately 1~m/s within the monitored area during data collection.

\subsubsection{Default Configuration}
Each CPI contains 128 RSSI samples per Rx antenna. A step size of 32 samples (approximately 32~ms between consecutive CPIs) is used, resulting in an overlap of 96 samples between adjacent CPIs. The AoA domain is discretized into 64 bins spanning $[-90^{\circ},\,90^{\circ}]$, corresponding to an angular resolution of approximately $2.8^{\circ}$. For Doppler estimation, 128 Doppler bins are used over the range of $[-100,\,100]$~Hz, yielding a Doppler resolution of approximately $1.56$~Hz. The FFT-based processing ensures that the average processing time per CPI remains shorter than the step-size interval, enabling real-time operation. For trajectory refinement, outlier suppression is performed using a 1D Hampel filter with a window size of 7 and a threshold of 1 standard deviation. The denoised sequence is then smoothed using a SG filter with a window length of 101 and a polynomial order of 2.

\subsubsection{Baselines}
For performance comparison, we consider two representative CSI-based methods: SRCC \cite{wang2025towards} and CASR \cite{feng2021lte, wu2024sensing}. SRCC estimates delay, AoA, and Doppler by constructing a per-antenna reference signal to suppress random phase distortions, followed by delay-domain beamforming and Doppler-AoA 2D FFT for parameter extraction. CASR removes random phase offsets by forming ratios between CSI measurements from two antennas, and we apply a Doppler-delay 2D FFT to estimate Doppler and propagation delay (range). Due to residual hardware-induced phase inconsistencies across antennas, accurate AoA estimation is challenging for CASR; therefore, we report only Doppler and range results for CASR. Ground-truth trajectories with centimeter-level accuracy are obtained using a TI IWR1642 mmWave radar \cite{TI_SWRU521C_IWR1642BOOST}.

\subsection{{Sensing Setup}}
\subsubsection{{Dataset}}
{To evaluate RSSI-only sensing beyond tracking, we conduct gesture recognition on the Widar3.0 dataset (Dataset~1) \cite{zhang2021widar3} and compare WiRSSI features with representative CSI-based features. Widar3.0 was collected at 5~GHz using multiple WiFi receivers, each equipped with three antennas. It contains six gestures, including Clap (25,039 samples), Draw-O (26,543 samples), Draw-Zigzag (34,046 samples), Push\&Pull (25,799 samples), Slide (29,248 samples), and Sweep (25,049 samples), performed by 16 subjects across 5 environments and 5 orientations (165,724 samples in total).}

\subsubsection{{Model and Training Configuration}}
{All classifiers are implemented using MobileViT-XXS \cite{mehta2022mobilevit} for fair comparison. We choose MobileViT-XXS as a lightweight, deployment-oriented backbone with low inference cost suitable for resource-constrained edge devices. All models are trained for 512 epochs with a batch size of 128 and an initial learning rate of 0.001 using the Adam optimizer. A step scheduler halves the learning rate every 196 epochs. We report Accuracy (Acc.), Macro Precision (Prec.), Macro Recall (Rec.), and Macro F1-score (F1).}

\begin{figure*}
    \centering
    \begin{minipage}[t]{0.329\linewidth}
    \vspace{0pt}
        \centering
        \begin{subfigure}{\textwidth}
            \centering
            \includegraphics[width=0.9\textwidth]{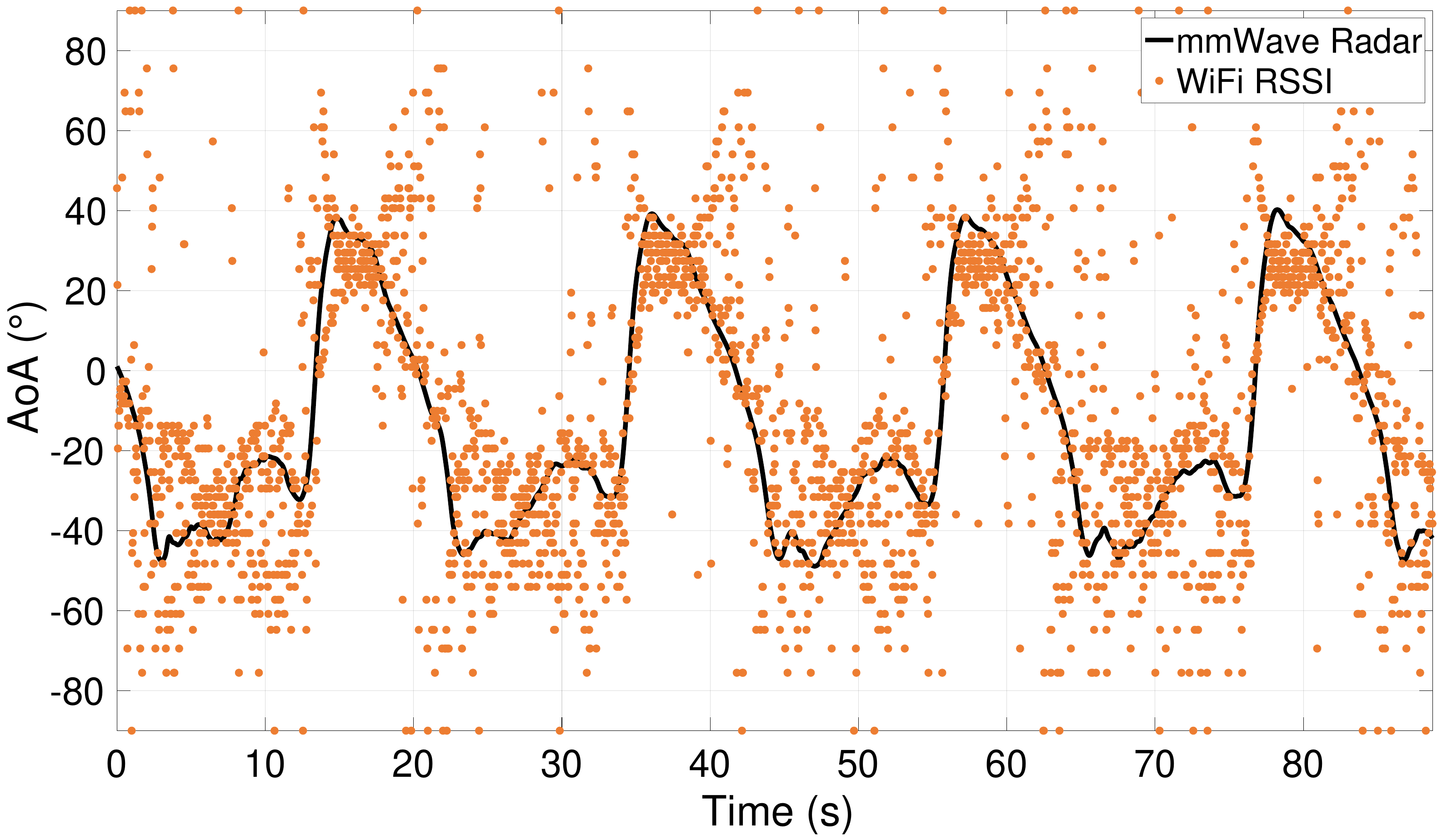}
            \subcaption{Raw AoA (RSSI)}
            \label{Fig_Ellipse_Block_a}
            \vspace{0.3em}
        \end{subfigure}\\
        \begin{subfigure}{\textwidth}
            \centering
            \includegraphics[width=0.9\textwidth]{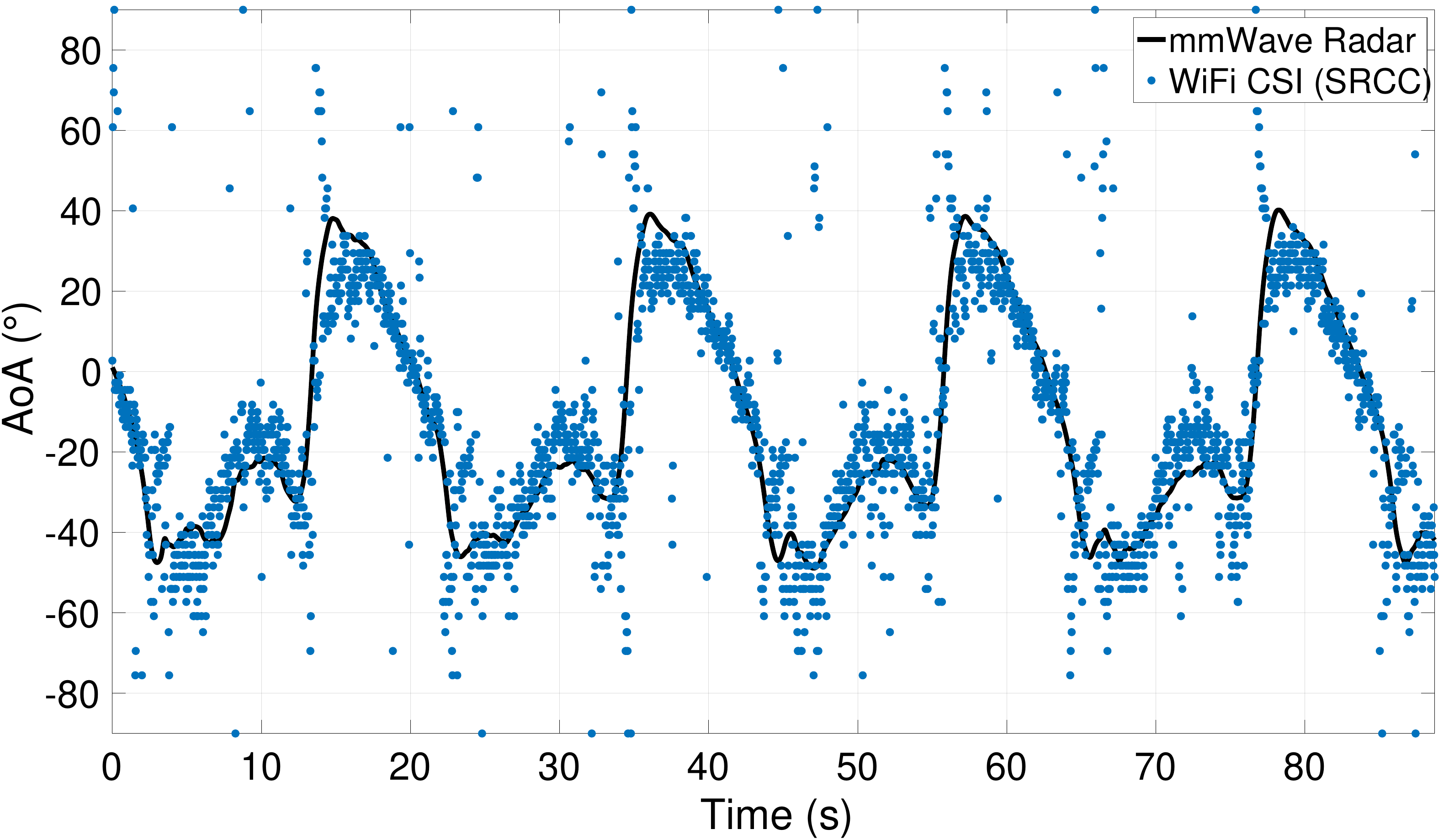}
            \subcaption{Raw AoA (CSI)}
            \label{Fig_Ellipse_Block_b}
            \vspace{0.3em}
        \end{subfigure}\\
        \begin{subfigure}{\textwidth}
            \centering
            \includegraphics[width=0.9\textwidth]{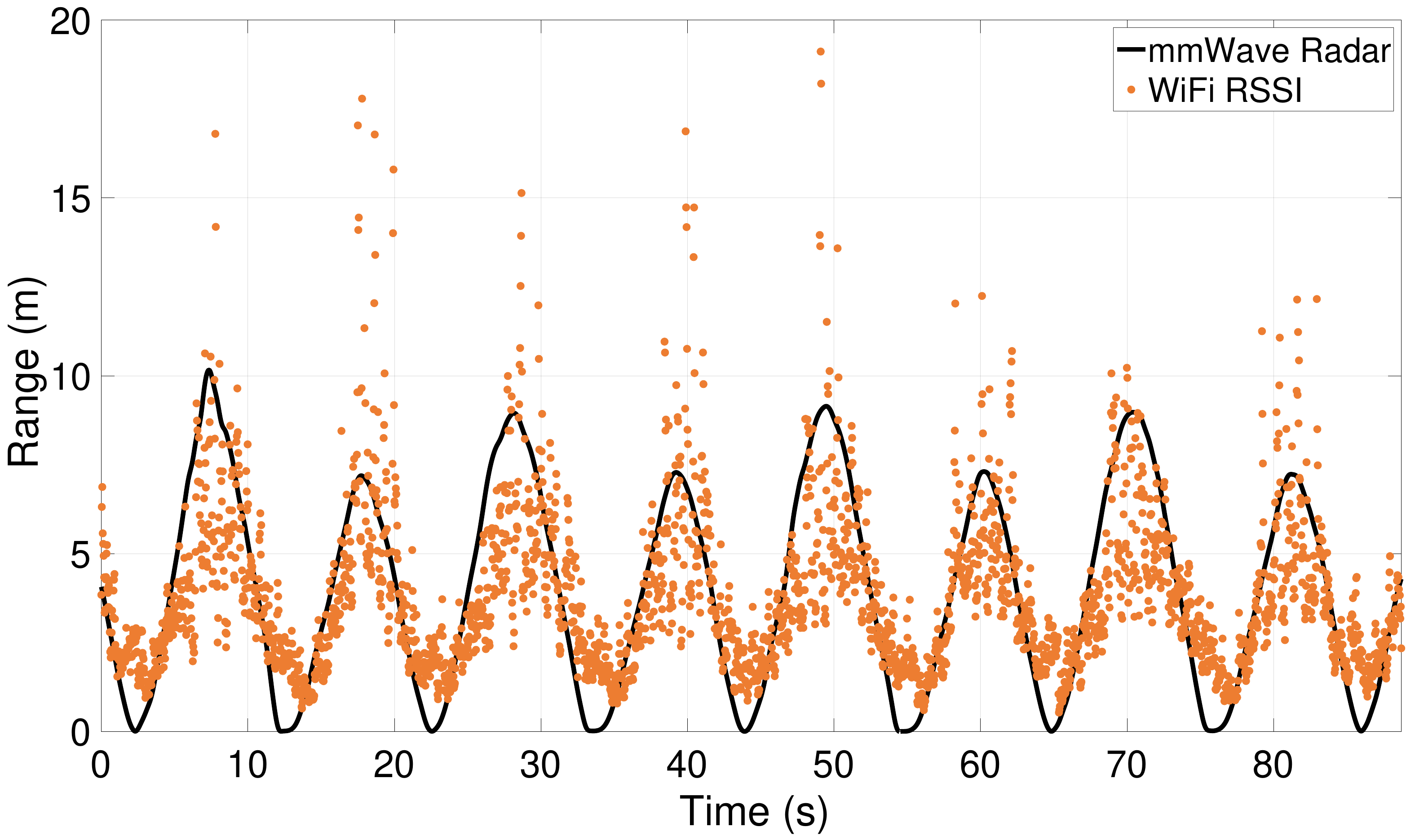}
            \subcaption{Raw Range (RSSI)}
            \label{Fig_Ellipse_Block_c}
            \vspace{0.3em}
        \end{subfigure}\\
        \begin{subfigure}{\textwidth}
            \centering
            \includegraphics[width=0.9\textwidth]{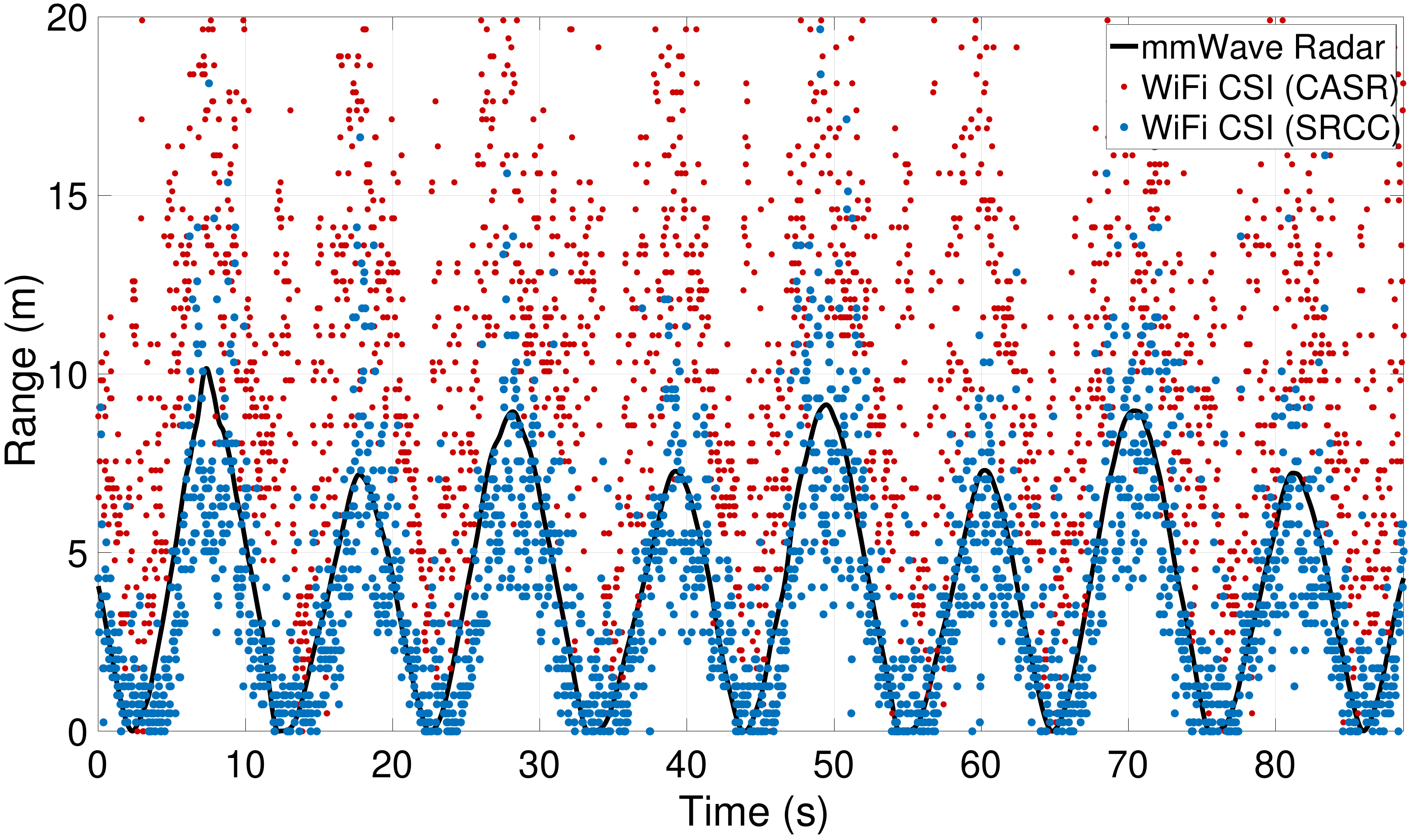}
            \subcaption{Raw Range (CSI)}
            \label{Fig_Ellipse_Block_d}
            \vspace{0.3em}
        \end{subfigure}\\
        \begin{subfigure}{\textwidth}
            \centering
            \includegraphics[width=0.9\textwidth]{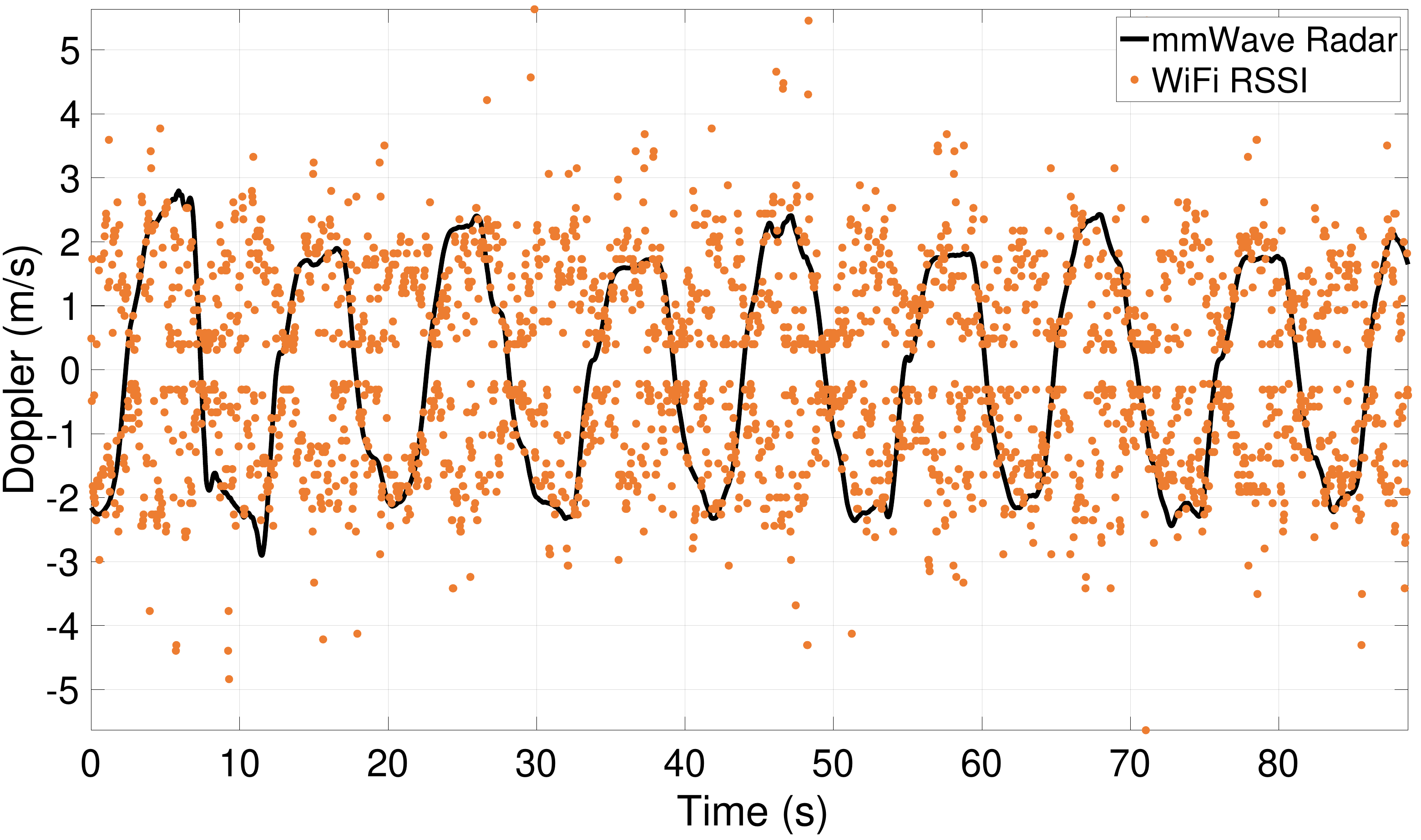}
            \subcaption{Raw Doppler (RSSI)}
            \label{Fig_Ellipse_Block_e}
            \vspace{0.3em}
        \end{subfigure}\\
        \begin{subfigure}{\textwidth}
            \centering
            \includegraphics[width=0.9\textwidth]{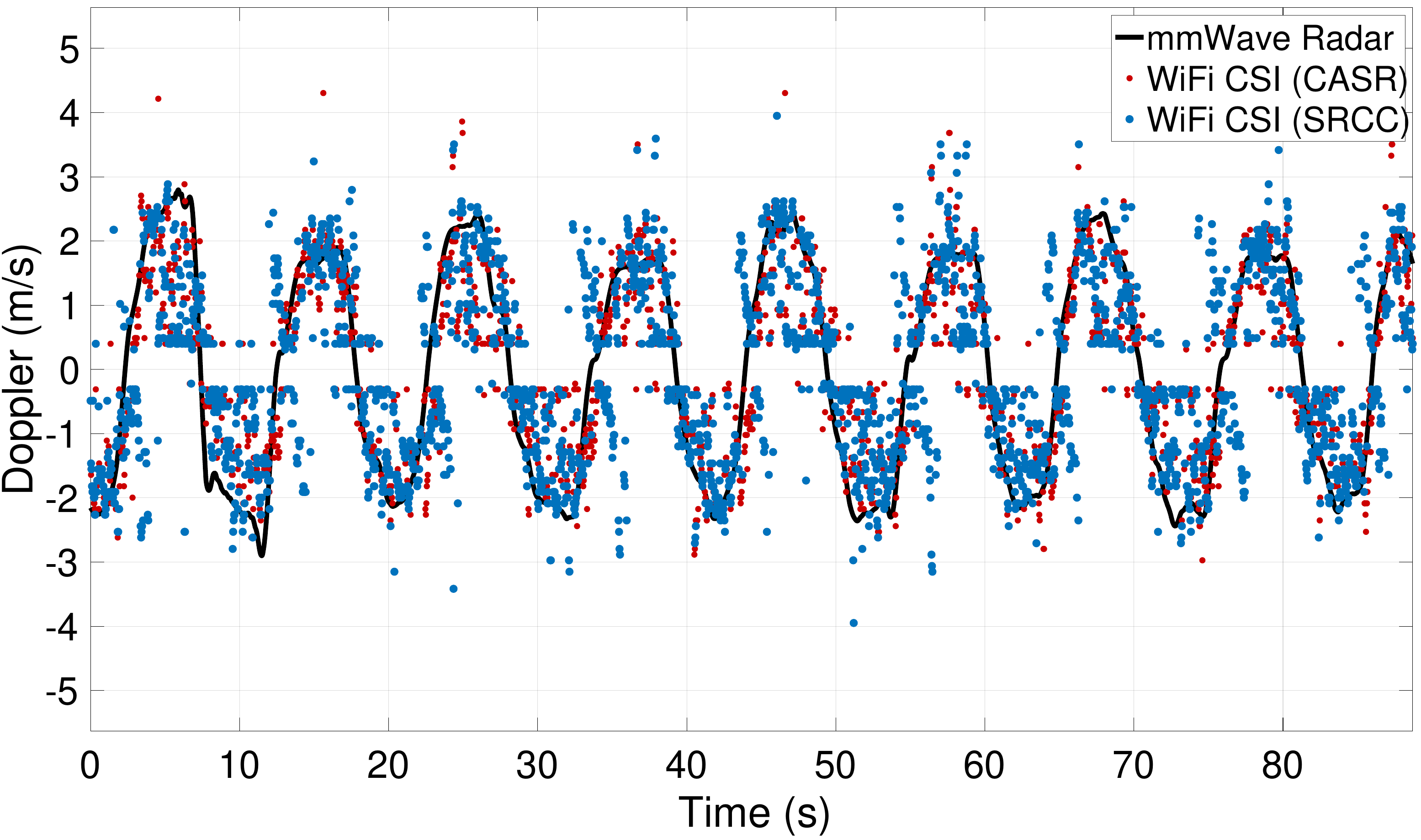}
            \subcaption{Raw Doppler (CSI)}
            \label{Fig_Ellipse_Block_f}
        \end{subfigure}
        \caption{Elliptical trajectory}
        \label{Fig_Ellipse_Block}
    \end{minipage}
    \begin{minipage}[t]{0.329\linewidth}
    \vspace{0pt}
        \centering
        \begin{subfigure}{\textwidth}
            \centering
            \includegraphics[width=0.9\textwidth]{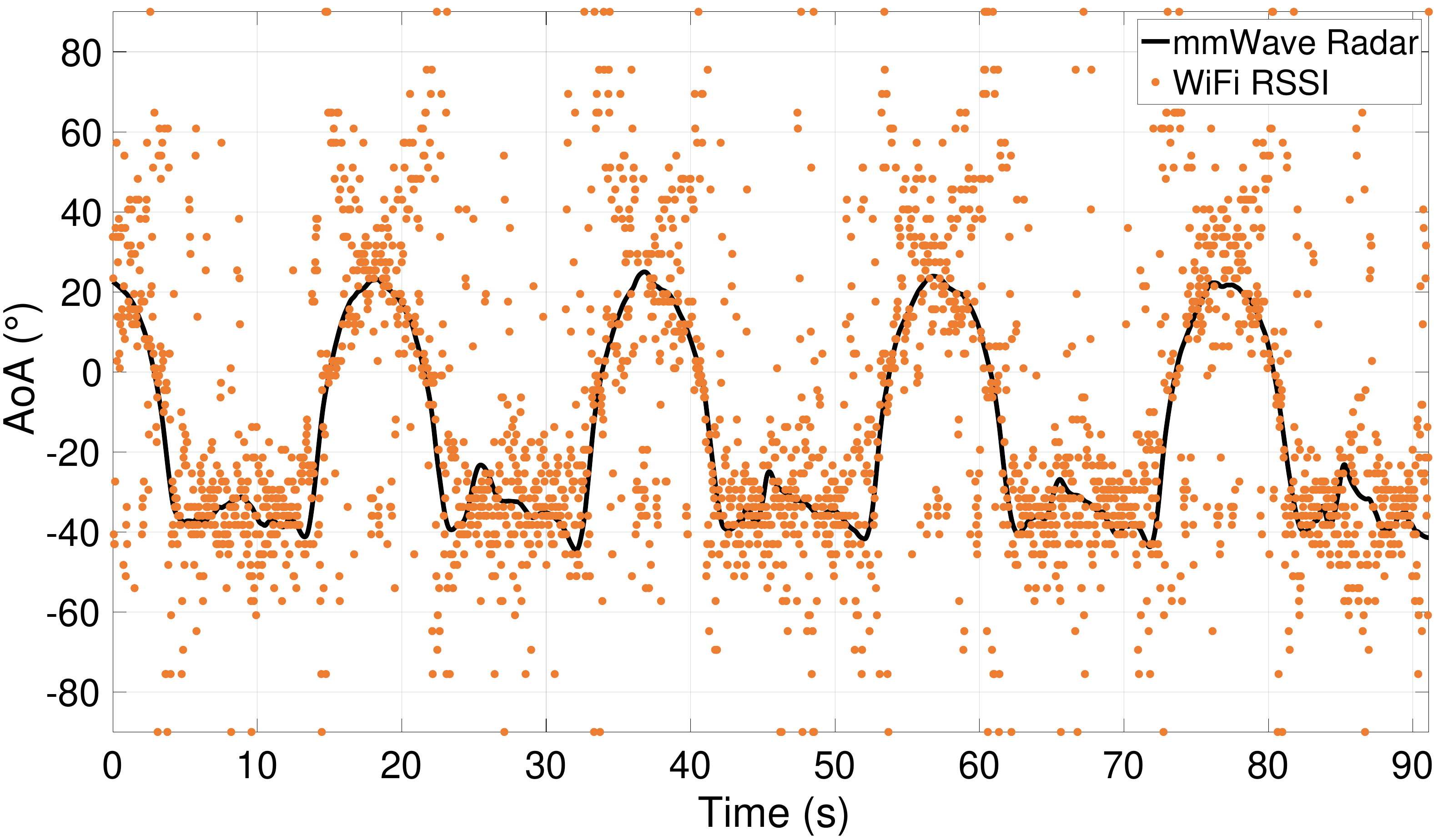}
            \subcaption{Raw AoA (RSSI)}
            \label{Fig_Linear_Block_a}
            \vspace{0.3em}
        \end{subfigure}\\
        \begin{subfigure}{\textwidth}
            \centering
            \includegraphics[width=0.9\textwidth]{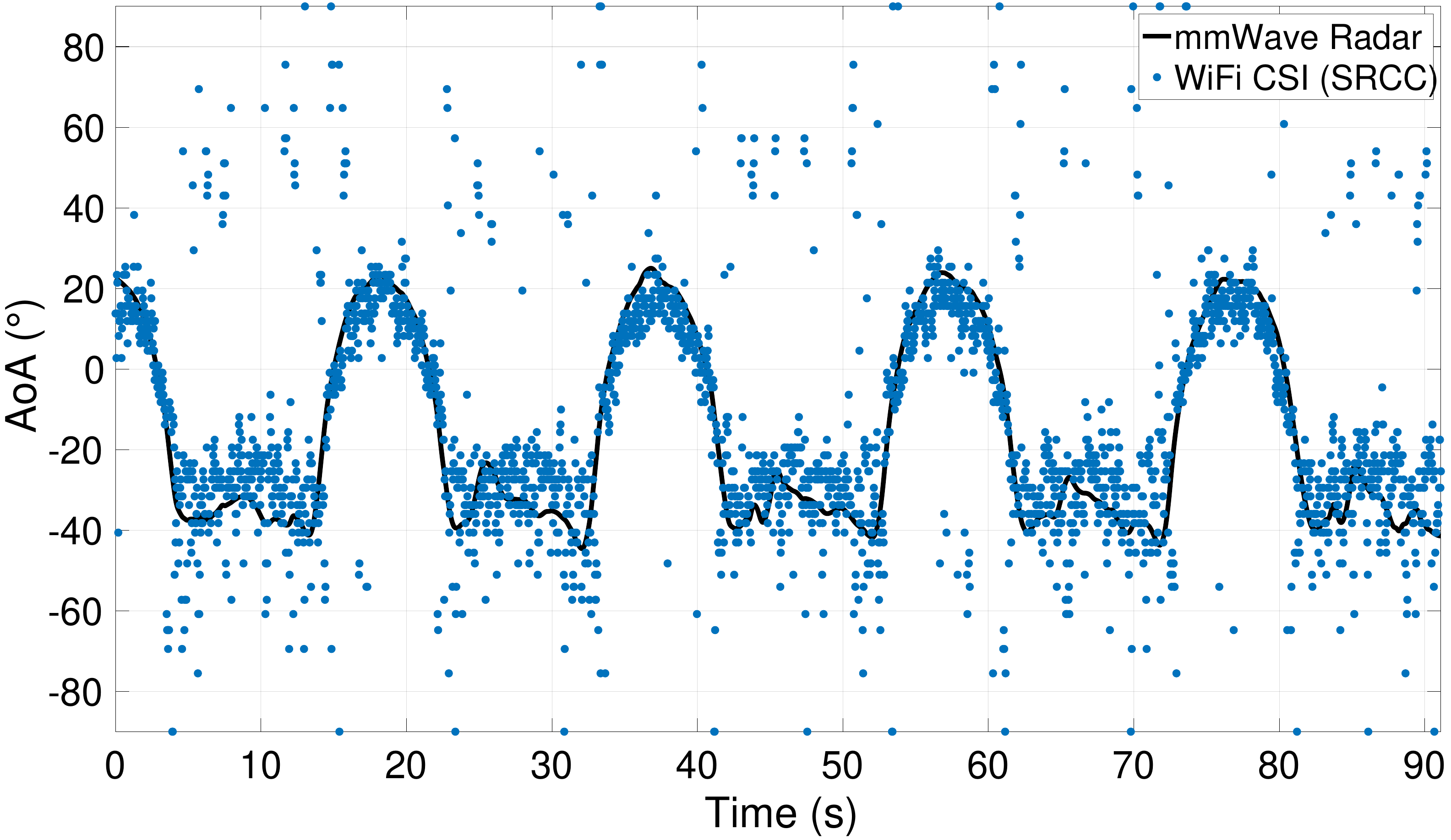}
            \subcaption{Raw AoA (CSI)}
            \label{Fig_Linear_Block_b}
            \vspace{0.3em}
        \end{subfigure}\\
        \begin{subfigure}{\textwidth}
            \centering
            \includegraphics[width=0.9\textwidth]{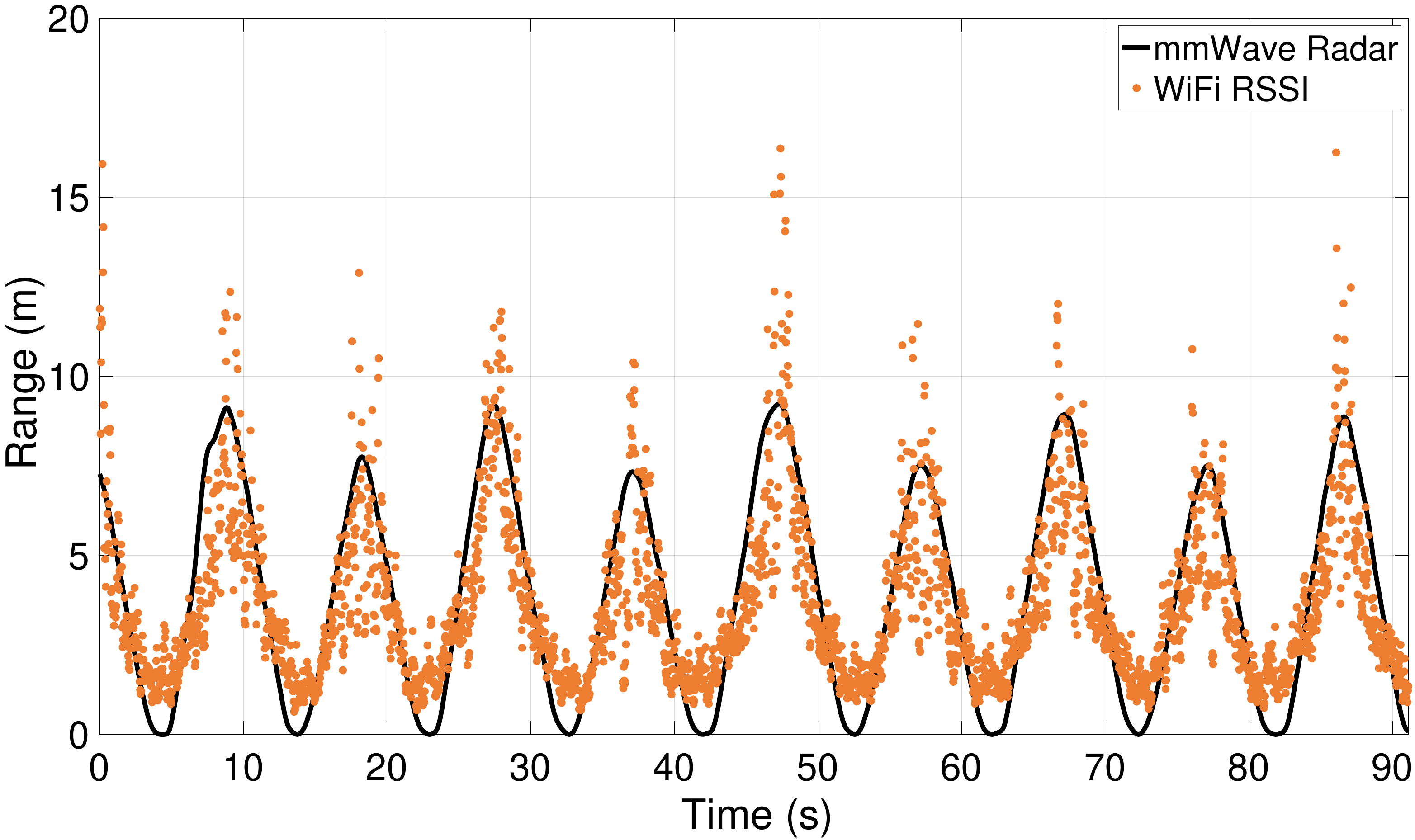}
            \subcaption{Raw Range (RSSI)}
            \label{Fig_Linear_Block_c}
            \vspace{0.3em}
        \end{subfigure}\\
        \begin{subfigure}{\textwidth}
            \centering
            \includegraphics[width=0.9\textwidth]{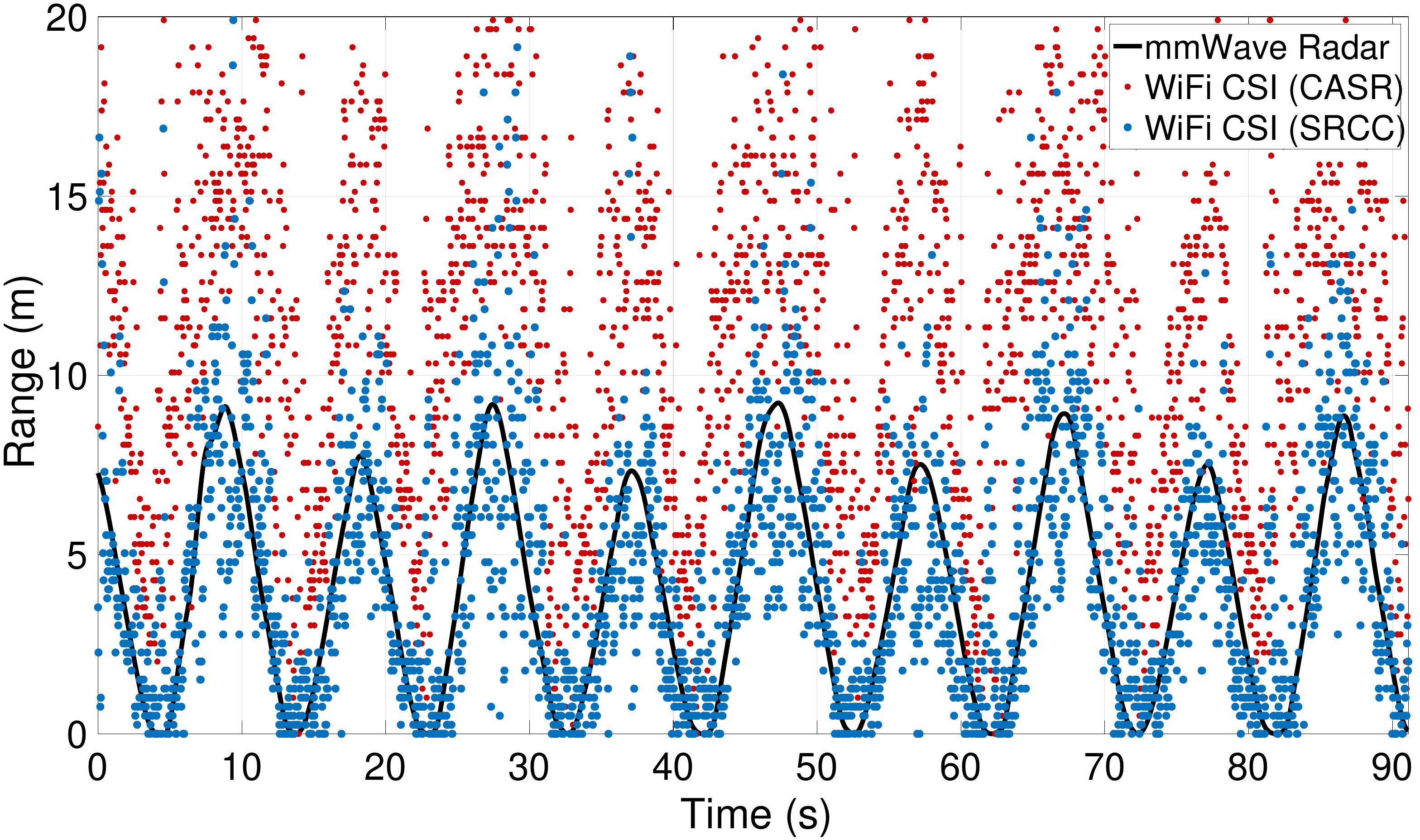}
            \subcaption{Raw Range (CSI)}
            \label{Fig_Linear_Block_d}
            \vspace{0.3em}
        \end{subfigure}\\
        \begin{subfigure}{\textwidth}
            \centering
            \includegraphics[width=0.9\textwidth]{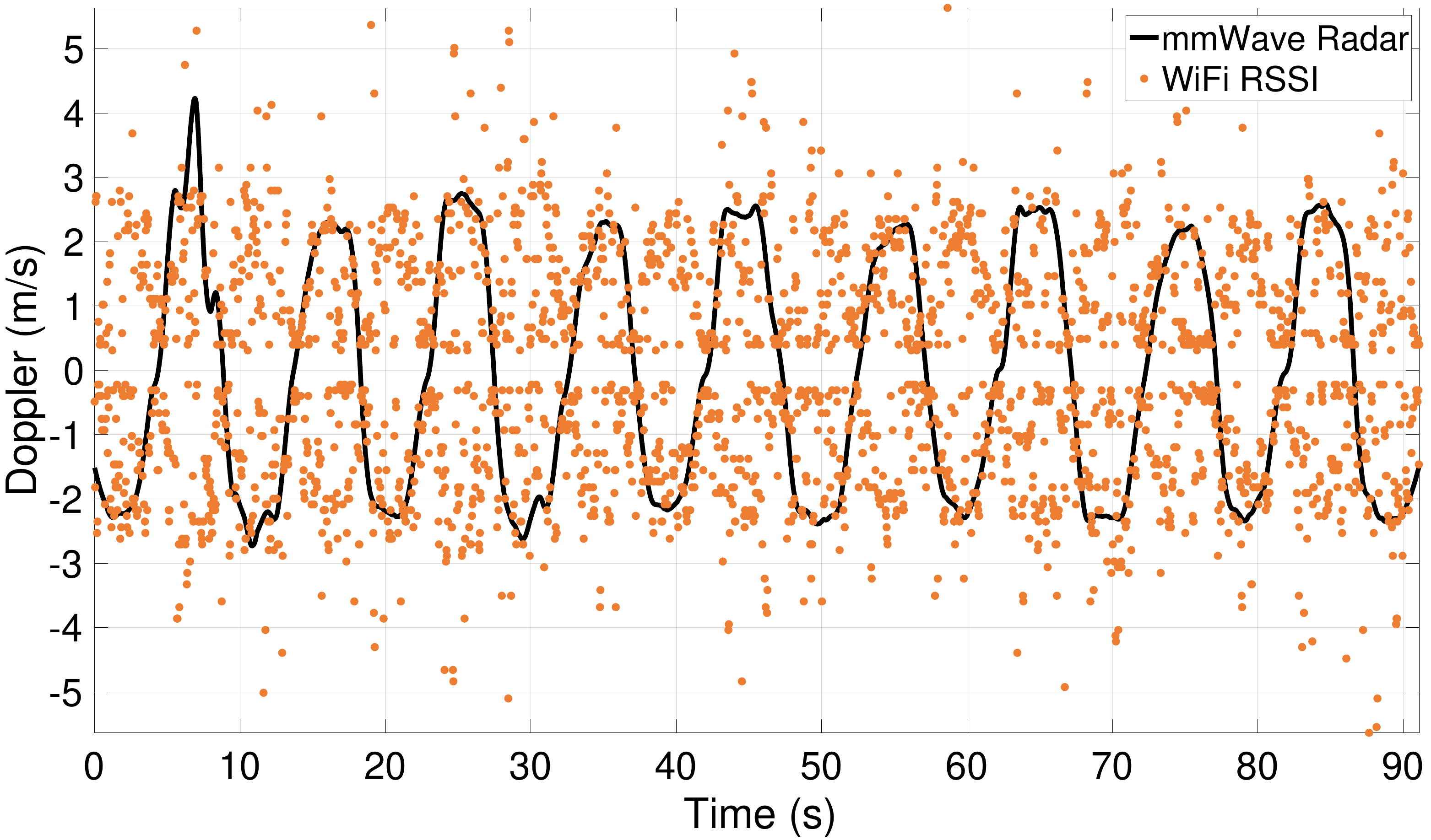}
            \subcaption{Raw Doppler (RSSI)}
            \label{Fig_Linear_Block_e}
            \vspace{0.3em}
        \end{subfigure}\\
        \begin{subfigure}{\textwidth}
            \centering
            \includegraphics[width=0.9\textwidth]{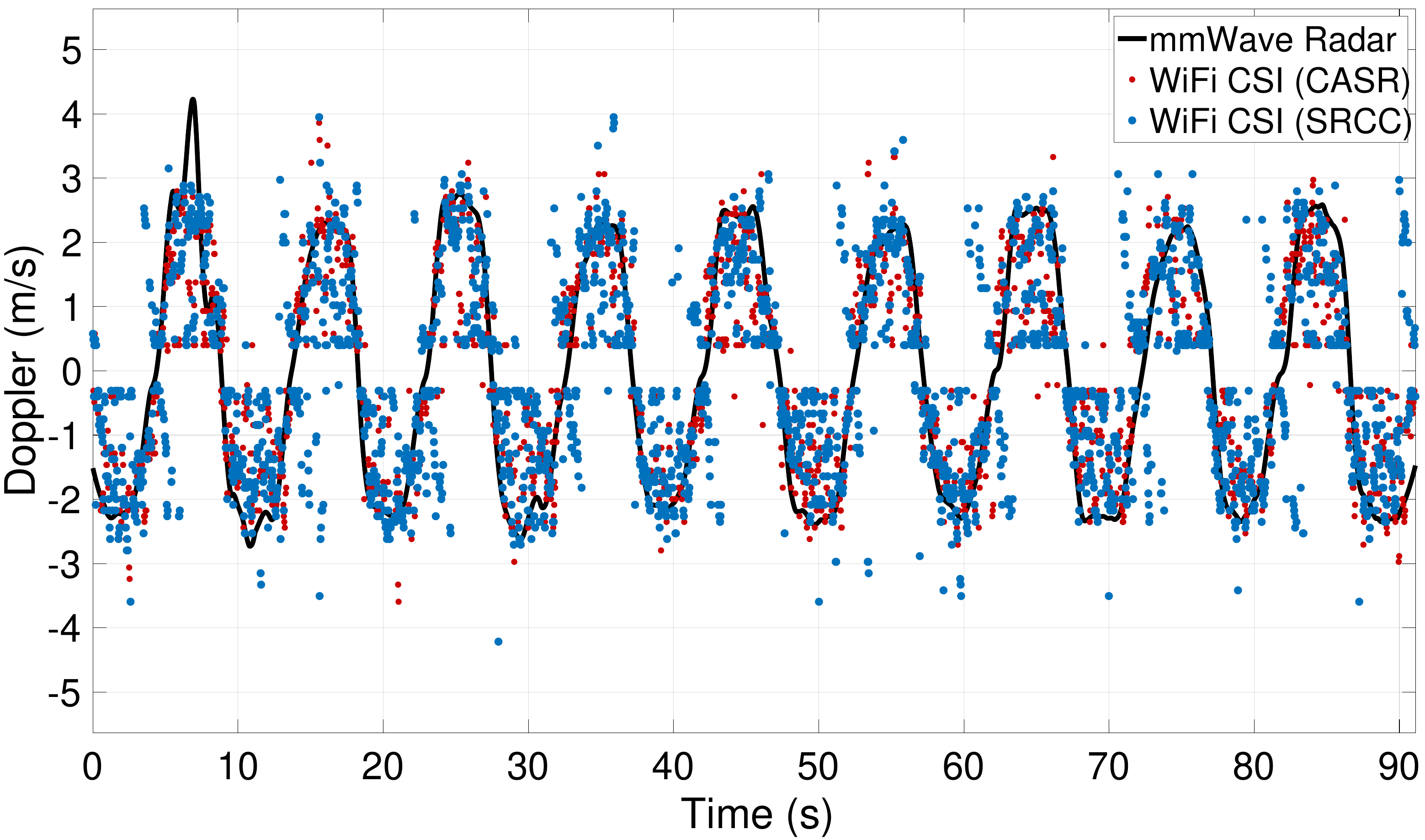}
            \subcaption{Raw Doppler (CSI)}
            \label{Fig_Linear_Block_f}
        \end{subfigure}
        \caption{Linear trajectory}
        \label{Fig_Linear_Block}
    \end{minipage}
    \begin{minipage}[t]{0.329\linewidth}
    \vspace{0pt}
        \centering
        \begin{subfigure}{\textwidth}
            \centering
            \includegraphics[width=0.9\textwidth]{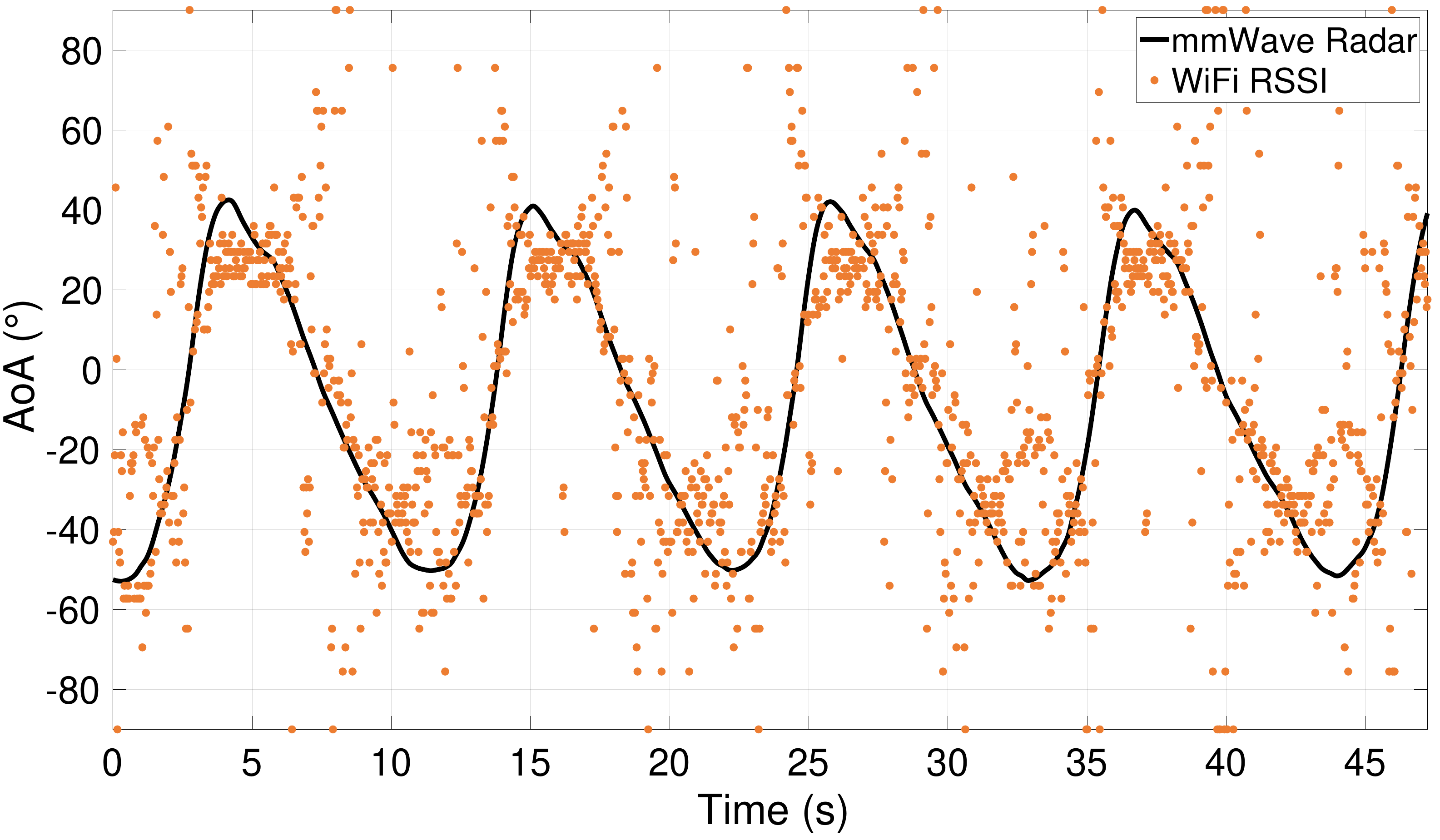}
            \subcaption{Raw AoA (RSSI)}
            \label{Fig_Rectangle_Block_a}
            \vspace{0.3em}
        \end{subfigure}\\
        \begin{subfigure}{\textwidth}
            \centering
            \includegraphics[width=0.9\textwidth]{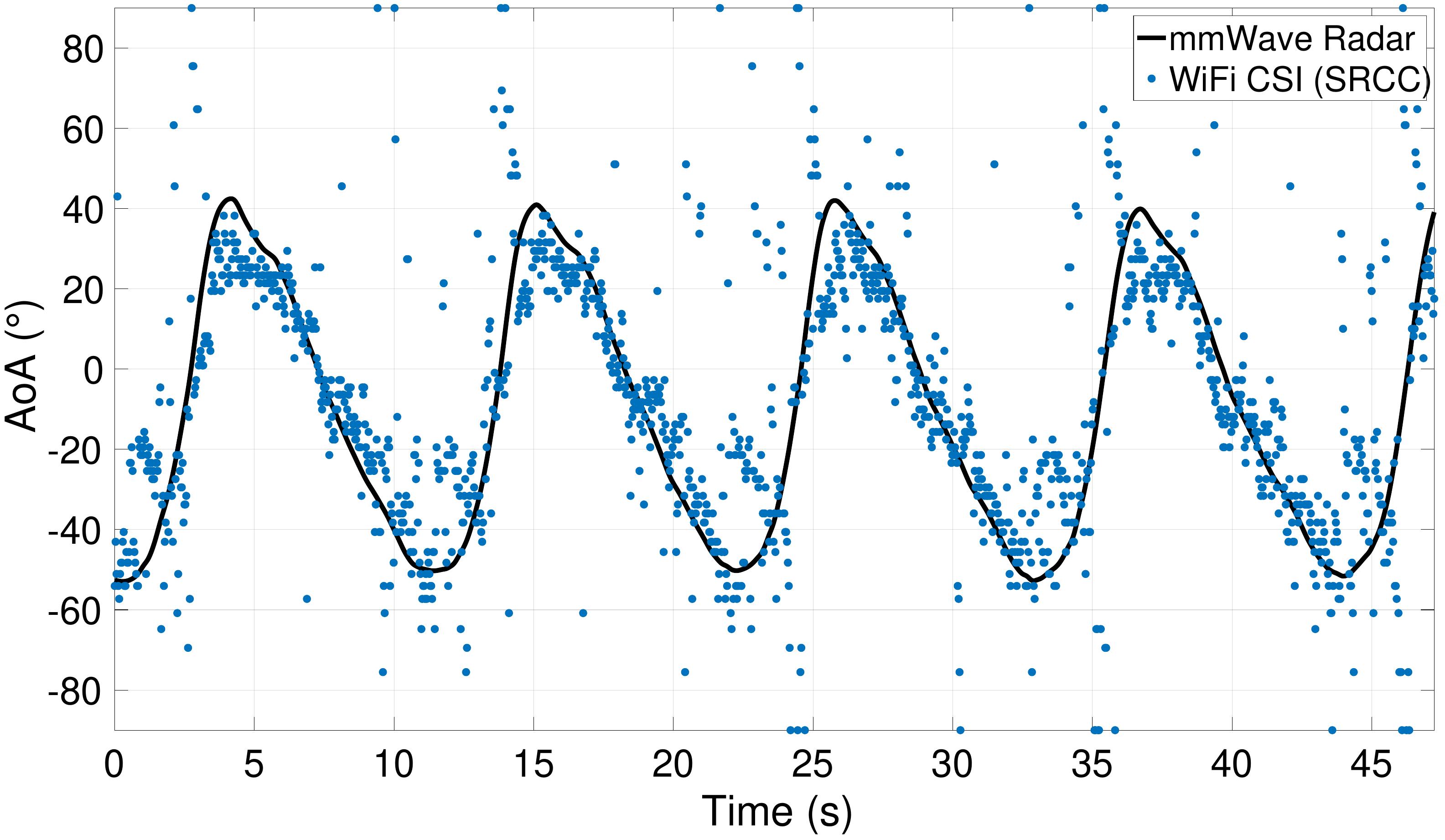}
            \subcaption{Raw AoA (CSI)}
            \label{Fig_Rectangle_Block_b}
            \vspace{0.3em}
        \end{subfigure}\\
        \begin{subfigure}{\textwidth}
            \centering
            \includegraphics[width=0.9\textwidth]{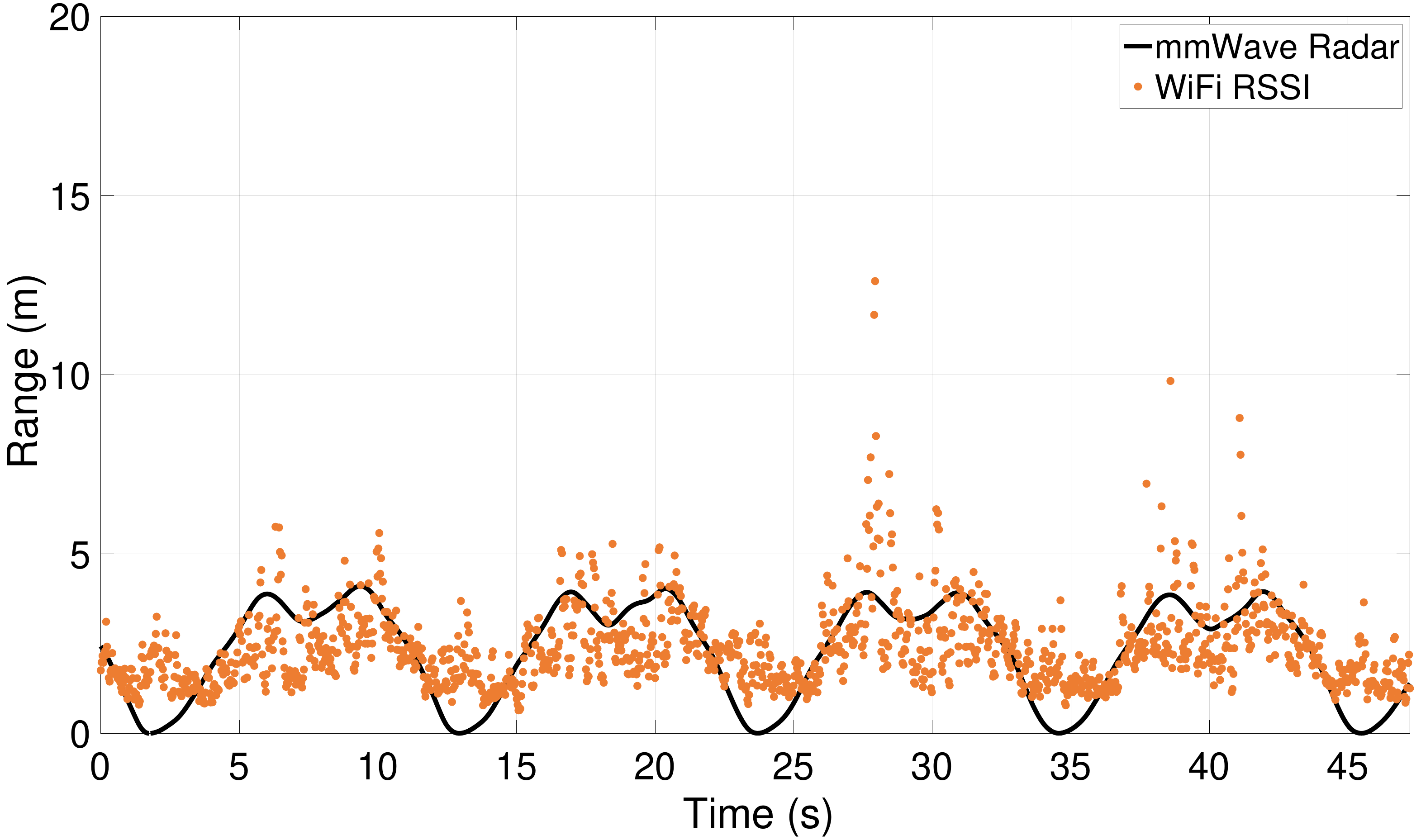}
            \subcaption{Raw Range (RSSI)}
            \label{Fig_Rectangle_Block_c}
            \vspace{0.3em}
        \end{subfigure}\\
        \begin{subfigure}{\textwidth}
            \centering
            \includegraphics[width=0.9\textwidth]{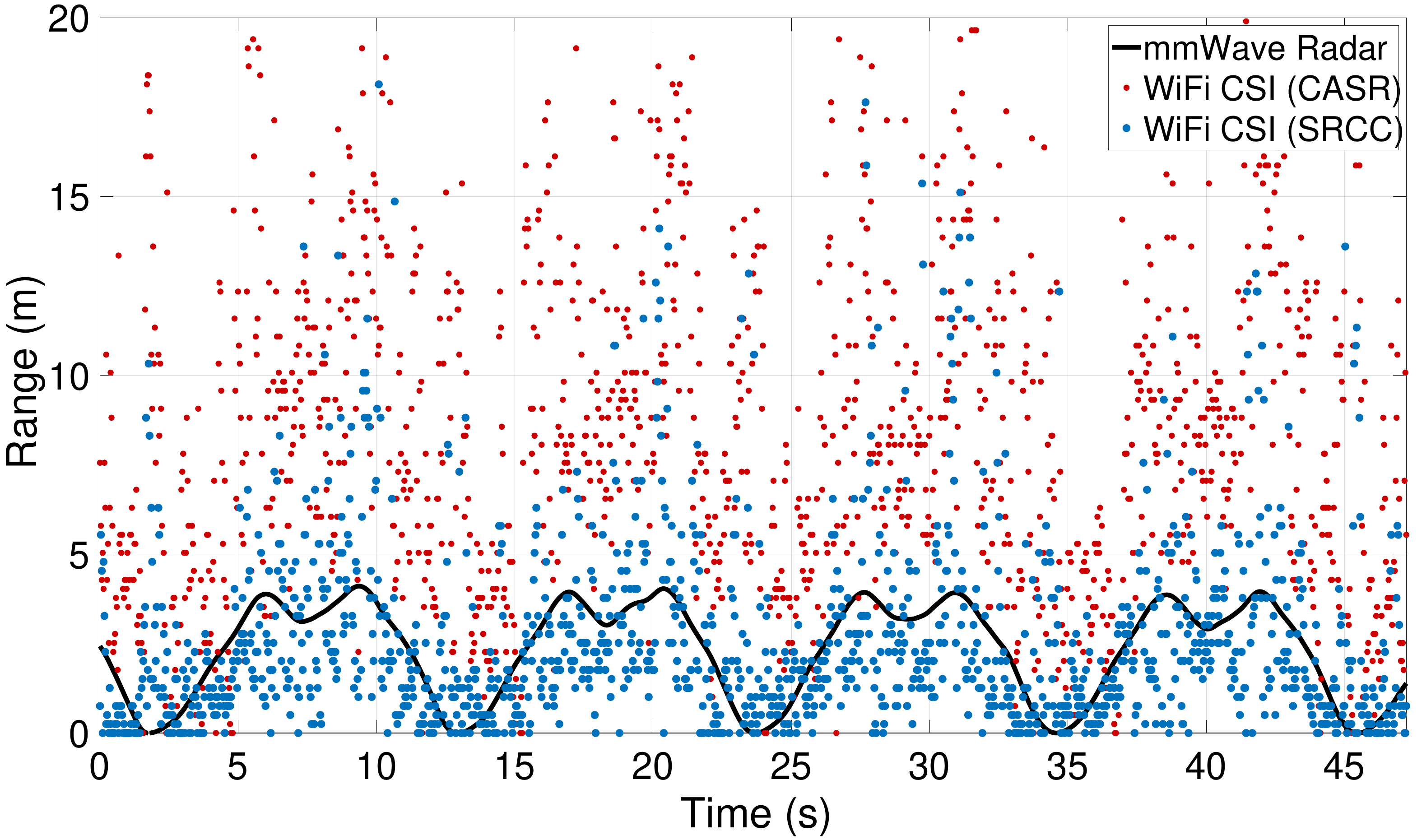}
            \subcaption{Raw Range (CSI)}
            \label{Fig_Rectangle_Block_d}
            \vspace{0.3em}
        \end{subfigure}\\
        \begin{subfigure}{\textwidth}
            \centering
            \includegraphics[width=0.9\textwidth]{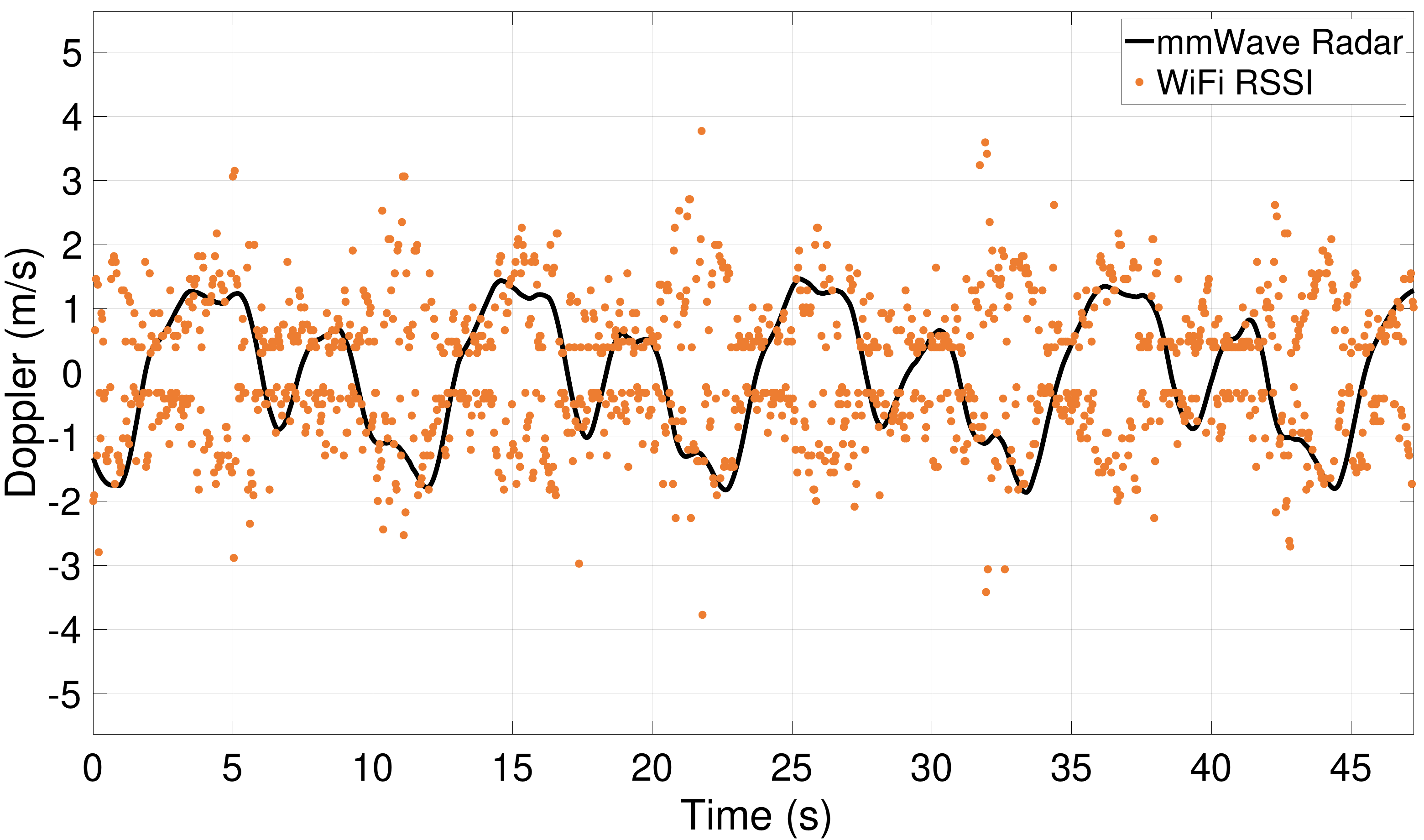}
            \subcaption{Raw Doppler (RSSI)}
            \label{Fig_Rectangle_Block_e}
        \end{subfigure}\\
        \begin{subfigure}{\textwidth}
            \centering
            \includegraphics[width=0.91\textwidth]{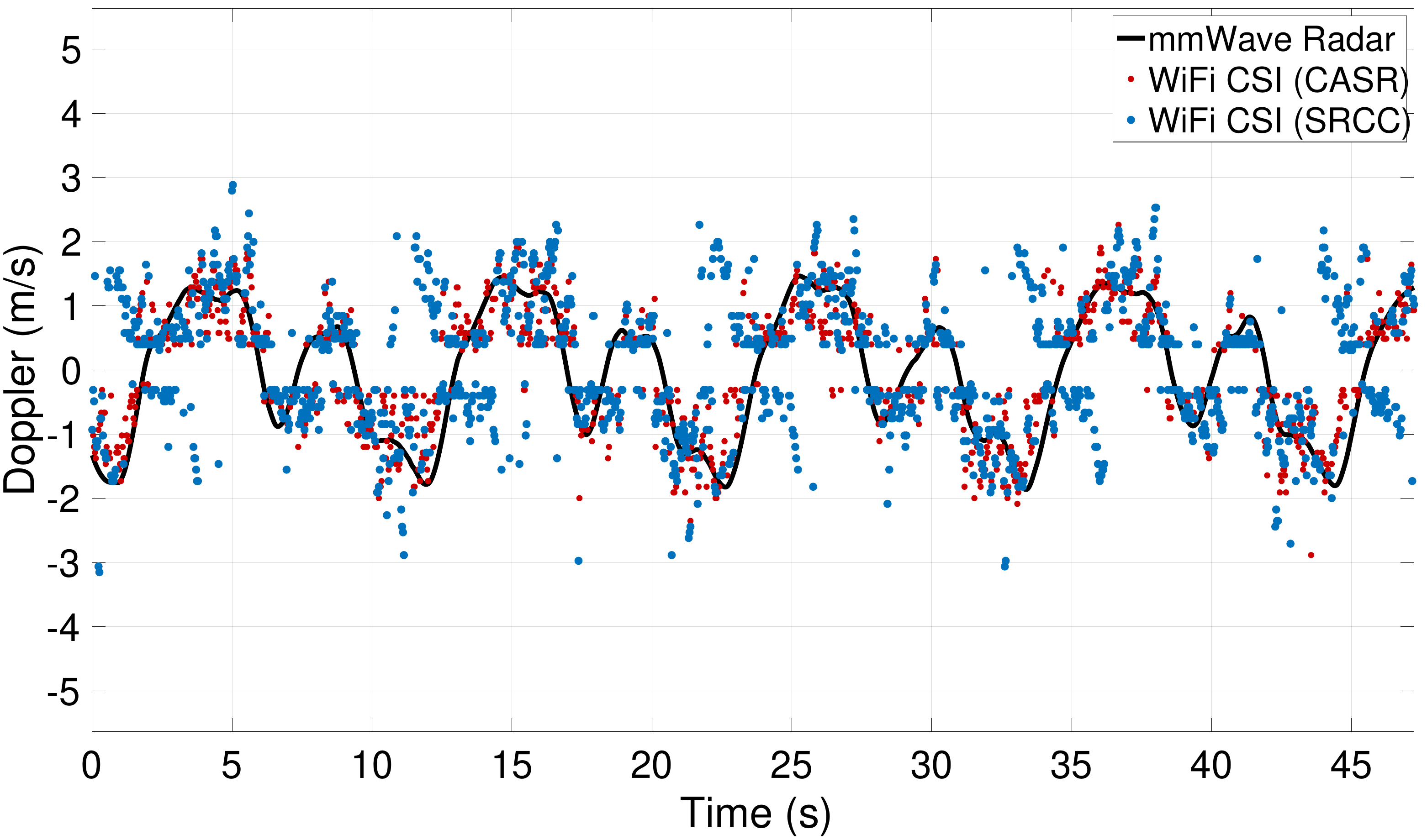}
            \subcaption{Raw Doppler (CSI)}
            \label{Fig_Rectangle_Block_f}
        \end{subfigure}
        \caption{Rectangular trajectory}
        \label{Fig_Rectangle_Block}
    \end{minipage}
    \vspace{-1em}
\end{figure*}

\section{Results}
\label{sec:results}
This section presents the experimental results of WiRSSI, including comparisons with CSI-based baselines, tracking accuracy and robustness studies, and {an additional gesture recognition evaluation for RSSI-only sensing beyond tracking}.

\subsection{Comparison with CSI-based Tracking Method}
We first compare the performance of the proposed WiRSSI against the CSI-based baselines in terms of Doppler, AoA, and delay estimation. All CSI-derived features are processed using the same configuration for fair comparison.

\subsubsection{AoA Estimation}
Fig.~\ref{Fig_Ellipse_Block_a}, Fig.~\ref{Fig_Linear_Block_a}, and Fig.~\ref{Fig_Rectangle_Block_a} compare RSSI-based AoA estimates with the ground-truth mmWave radar measurements, while Fig.~\ref{Fig_Ellipse_Block_b}, Fig.~\ref{Fig_Linear_Block_b}, and Fig.~\ref{Fig_Rectangle_Block_b} compare CSI-based AoA estimates with the same ground truth. Due to the coarse resolution of RSSI, the RSSI-derived AoA traces exhibit larger fluctuations and more outliers compared with the CSI-based results. Nevertheless, the dominant angular trend extracted from RSSI still closely follows the ground-truth trajectory across all motion patterns, indicating that the proposed Doppler-AoA processing can recover reliable directional information even under low-resolution measurements. It is also observed that RSSI relies solely on the Tx-Rx geometric AoA to suppress Doppler-AoA mirror ambiguities, whereas CSI leverages both spatial- and spectral-domain phase information to jointly constrain the angle estimation and remove mirrored components. Overall, while CSI provides cleaner and more stable AoA estimates, RSSI still retains sufficient angular consistency to support accurate trajectory tracking.

\subsubsection{Delay Estimation}
Fig.~\ref{Fig_Ellipse_Block_c}, Fig.~\ref{Fig_Linear_Block_c}, and Fig.~\ref{Fig_Rectangle_Block_c} compare RSSI-based range variation estimates with the mmWave radar ground truth, while Fig.~\ref{Fig_Ellipse_Block_d}, Fig.~\ref{Fig_Linear_Block_d}, and Fig.~\ref{Fig_Rectangle_Block_d} compare the corresponding CSI-based results. Here, the range represents the distance difference between the Tx-target-Rx reflection path and the direct Tx-Rx path. Because RSSI integrates power over the entire channel bandwidth, it loses the subcarrier-dependent phase diversity typically required for delay estimation in CSI-based sensing. Surprisingly, the amplitude-based delay estimation derived from RSSI achieves results that are highly consistent with the ground truth, which is often overlooked in prior work. This capability arises from the power-delay relationship exploited in Section \ref{sec:rssi_tracking}, although it requires pre-measuring the reflection coefficient ratio. In contrast, CSI naturally preserves subcarrier-level phase information and can recover propagation delay directly without prior calibration. In the CSI-based implementation, the range is searched from 0 to 32 m and uniformly discretized into 128 bins, yielding a range resolution of 0.25 m. However, in the CASR method, the nonlinear transformation applied across subcarriers introduces noticeable delay distortion, as indicated by the red points. By contrast, the SRCC method (shown in blue) relies on linear processing and preserves the intrinsic subcarrier structure, thereby achieving more accurate delay estimation. The results show that despite lacking phase information, the RSSI-based amplitude method is still able to capture the overall range evolution.

\subsubsection{Doppler Estimation}
Fig.~\ref{Fig_Ellipse_Block_e}, Fig.~\ref{Fig_Linear_Block_e}, and Fig.~\ref{Fig_Rectangle_Block_e} compare RSSI-based Doppler trace estimates with the bistatic Doppler derived from the mmWave radar ground truth, while Fig.~\ref{Fig_Ellipse_Block_f}, Fig.~\ref{Fig_Linear_Block_f}, and Fig.~\ref{Fig_Rectangle_Block_f} compare the corresponding CSI-based results. It is worth noting that human motion produces multiple Doppler components due to different body parts (e.g., legs, arms, and torso), and the dominant component may vary over time. The mmWave-derived bistatic Doppler therefore represents an overall motion trend, and slight discrepancies between WiFi and mmWave observations are expected. The Doppler extracted from RSSI exhibits irregular temporal patterns and is strongly affected by coarse amplitude quantization and limited resolution. Consequently, the RSSI-based Doppler estimates are noisy and fail to clearly reveal the underlying periodic motion. In contrast, CSI-based Doppler estimation yields significantly cleaner and more stable traces that closely follow the mmWave-derived Doppler across different motion trajectories. Both SRCC and CASR produce comparable Doppler estimation performance, highlighting the advantage of preserving fine-grained CSI information for Doppler sensing. {Nevertheless, even with noisier RSSI-based Doppler, the resulting time-Doppler patterns still provide useful motion signatures for sensing tasks such as gesture recognition, as evaluated in Section~\ref{subsec:gesture_recognition}.}

\begin{figure*}
    \centering
    \begin{subfigure}[b]{0.32\textwidth}
        \centering
        \includegraphics[width=\textwidth]{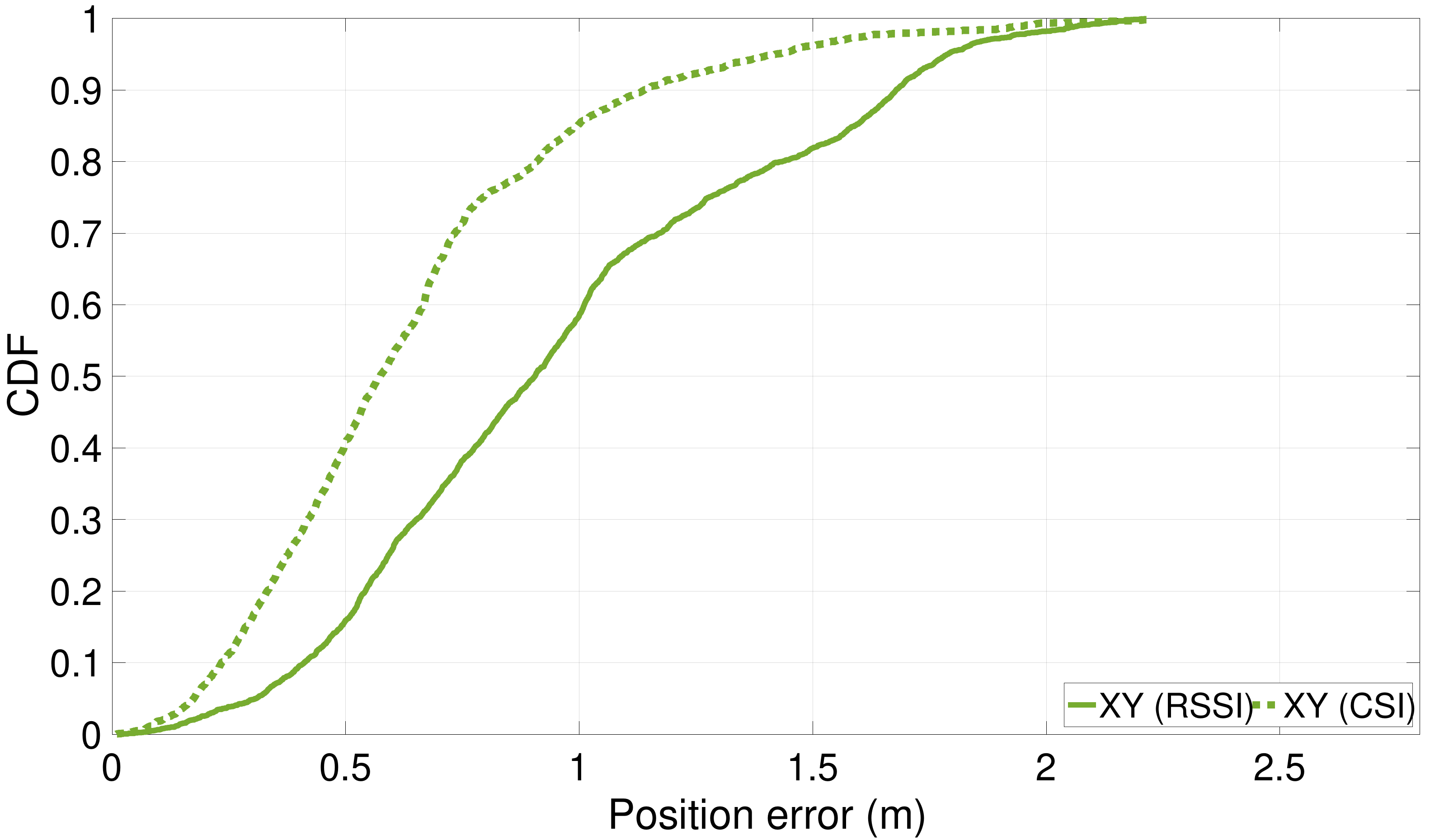}
        \caption{Ellipse}
    \end{subfigure}
    \begin{subfigure}[b]{0.32\textwidth}
        \centering
        \includegraphics[width=\textwidth]{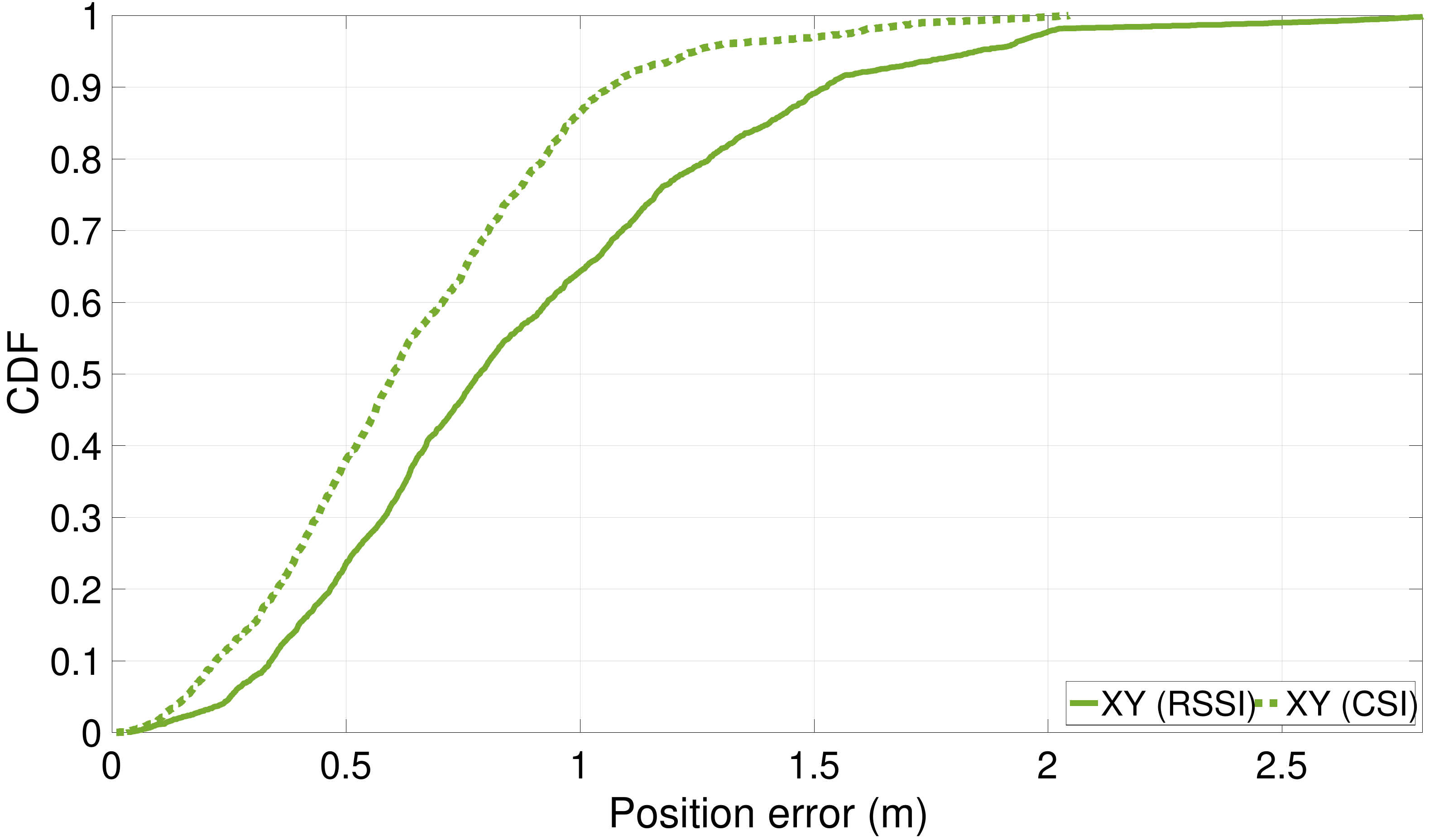}
        \caption{Linear}
    \end{subfigure}
    \begin{subfigure}[b]{0.32\textwidth}
        \centering
        \includegraphics[width=\textwidth]{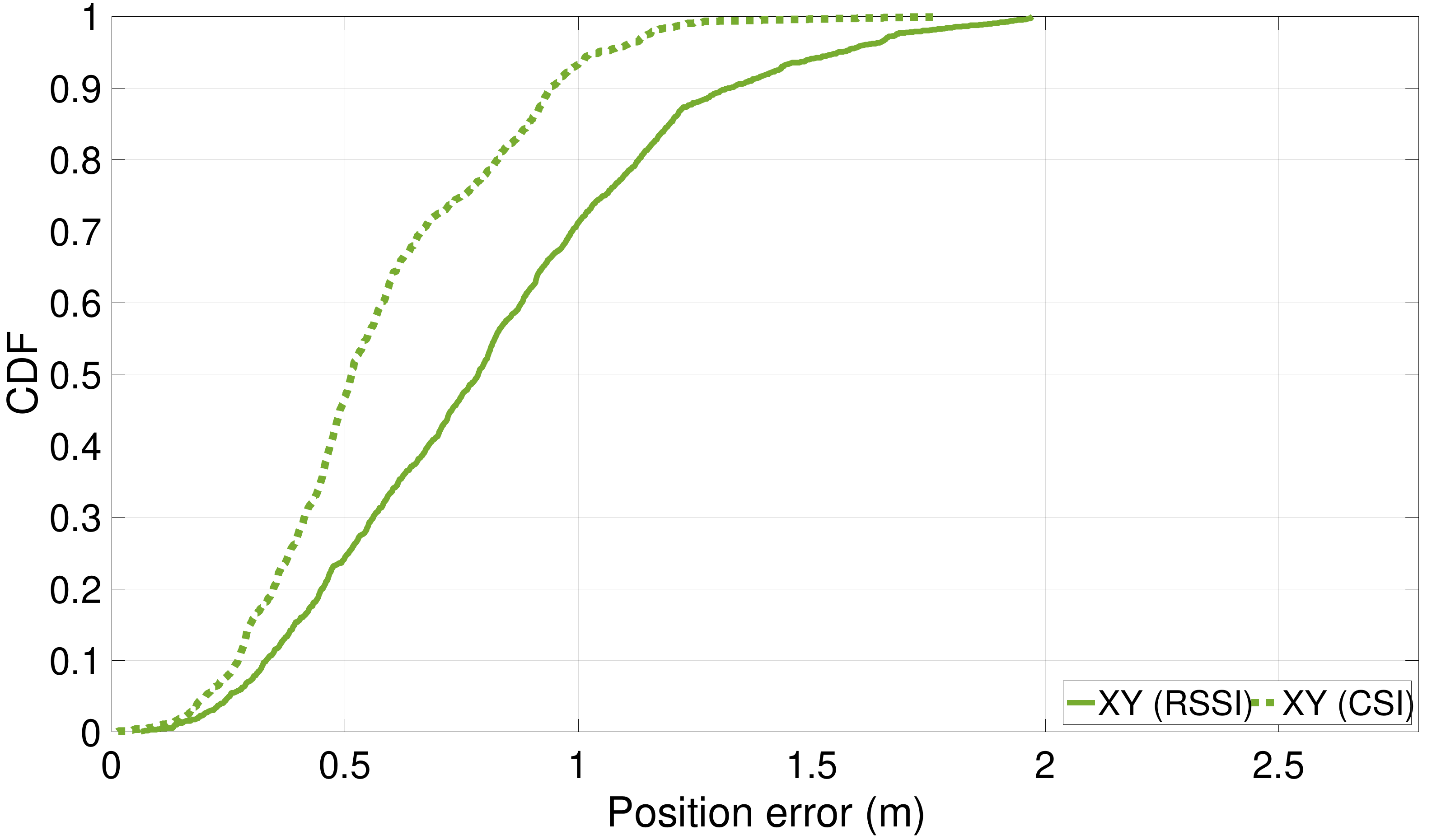}
        \caption{Rectangle}
    \end{subfigure}
    \caption{CDFs of trajectory errors.}
    \label{fig:CDF_traj_combined}
    \vspace{-1.5em}
\end{figure*}

\begin{table}[t]
\centering
\caption{{Median (CDF=0.5) and 90\% (CDF=0.9) position errors of smoothed RSSI and CSI results under different trajectories.}}
\label{tab:CDF_errors}
\renewcommand{\arraystretch}{1.2}
\setlength{\tabcolsep}{10pt}
\begin{tabular}{l|ccc|ccc}
\toprule
\multirow{2}{*}{Trajectory} & \multicolumn{3}{c|}{RSSI Error (m)} & \multicolumn{3}{c}{CSI Error (m)} \\ 
\cmidrule(lr){2-4} \cmidrule(lr){5-7}
 & X & Y & XY & X & Y & XY \\ 
\midrule
\textbf{CDF = 0.5} \\
Ellipse   & 0.492 & 0.607 & \textbf{0.905} & 0.386 & 0.318 & \textbf{0.574} \\
Linear    & 0.312 & 0.667 & \textbf{0.784} & 0.378 & 0.342 & \textbf{0.599} \\
Rectangle & 0.321 & 0.627 & \textbf{0.785} & 0.407 & 0.218 & \textbf{0.514} \\
\midrule
\textbf{CDF = 0.9} \\
Ellipse   & 1.113 & 1.313 & \textbf{1.678} & 0.887 & 0.784 & \textbf{1.139} \\
Linear    & 1.017 & 1.201 & \textbf{1.527} & 0.886 & 0.780 & \textbf{1.062} \\
Rectangle & 0.769 & 1.092 & \textbf{1.321} & 0.857 & 0.664 & \textbf{0.940} \\
\bottomrule
\end{tabular}
\vspace{-1.5em}
\end{table}

\begin{figure*}
    \centering
    \begin{minipage}[t]{0.329\linewidth}
    \vspace{0pt} 
        \centering
        \begin{subfigure}{\textwidth}
            \centering
            \includegraphics[width=0.95\textwidth]{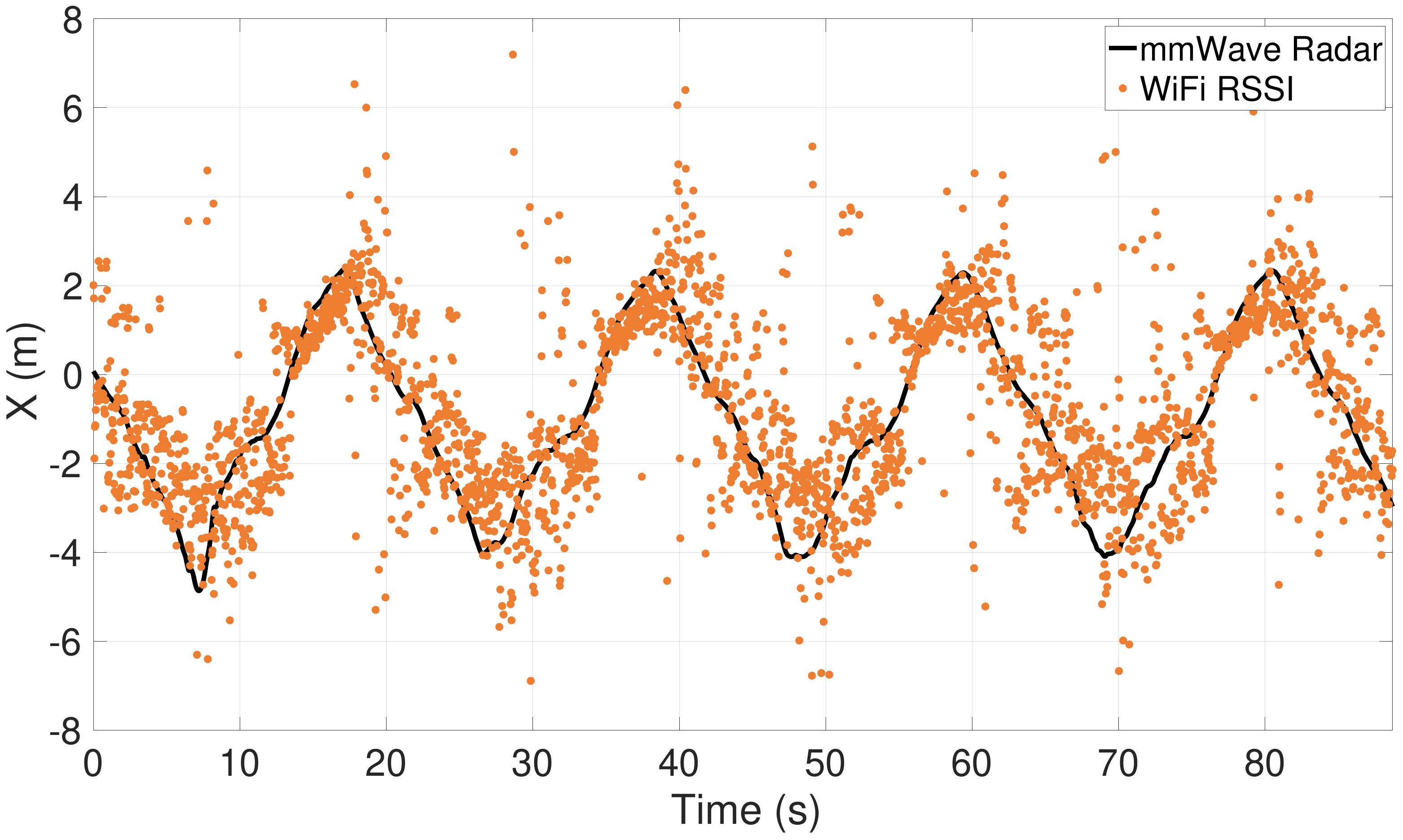}
            \vspace{-.5em}
            \subcaption{Raw x-axis coordinate (RSSI)}
        \end{subfigure}\\
        \begin{subfigure}{\textwidth}
            \centering
            \includegraphics[width=0.95\textwidth]{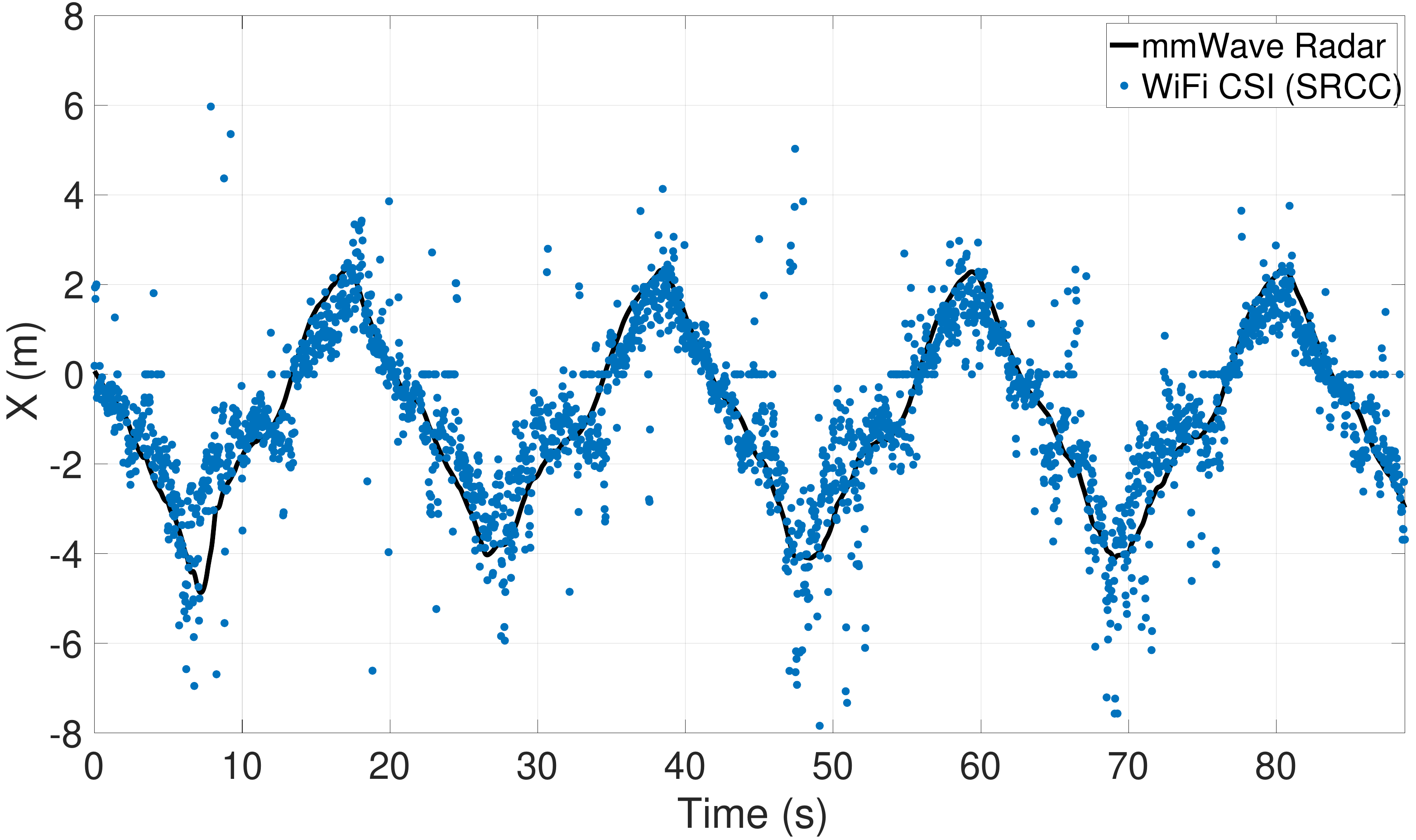}
            \vspace{-.5em}
            \subcaption{Raw x-axis coordinate (CSI)}
        \end{subfigure}\\
        \begin{subfigure}{\textwidth}
            \centering
            \includegraphics[width=0.95\textwidth]{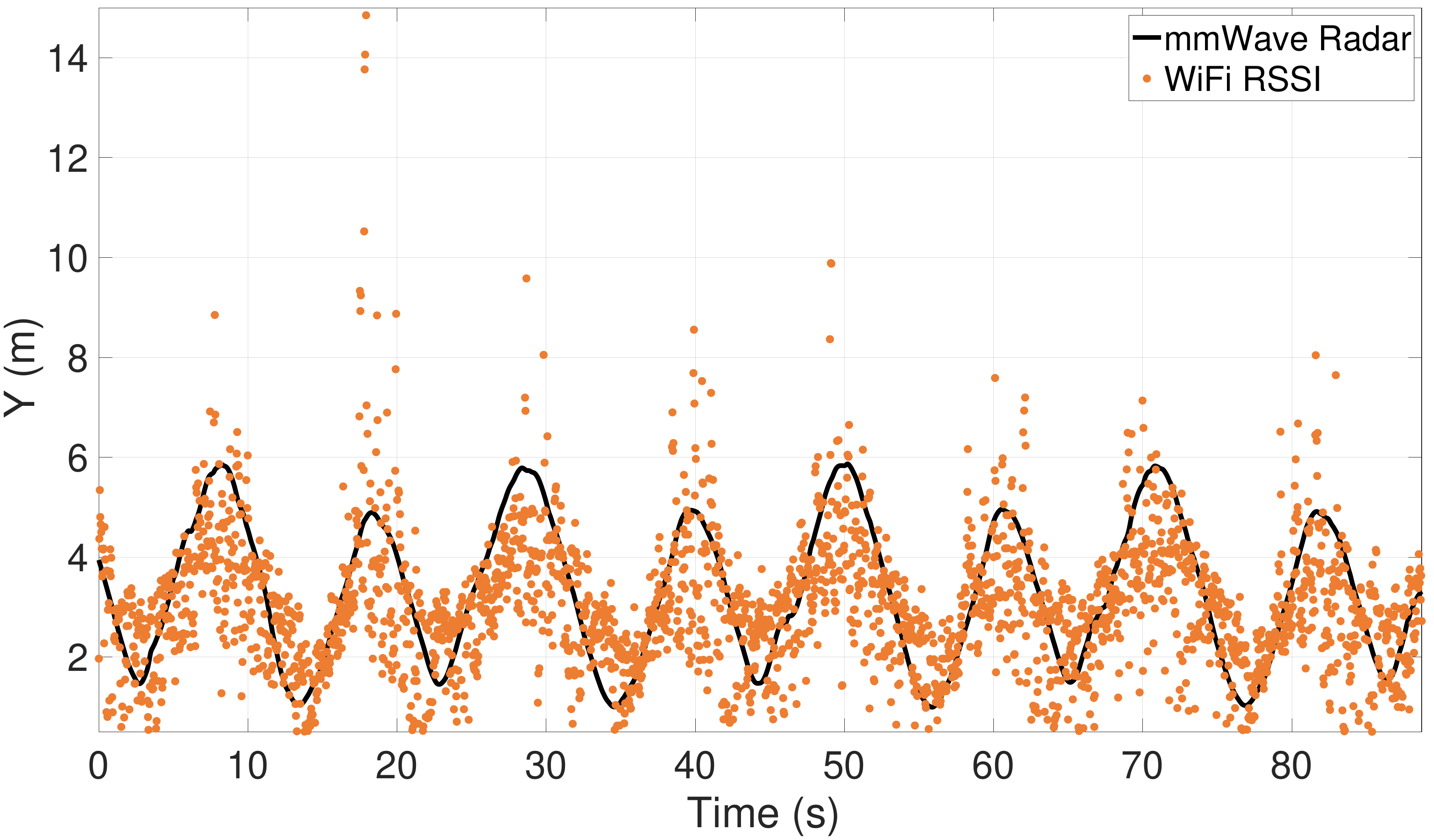}
            \vspace{-.5em}
            \subcaption{Raw y-axis coordinate (RSSI)}
        \end{subfigure}\\
        \begin{subfigure}{\textwidth}
            \centering
            \includegraphics[width=0.95\textwidth]{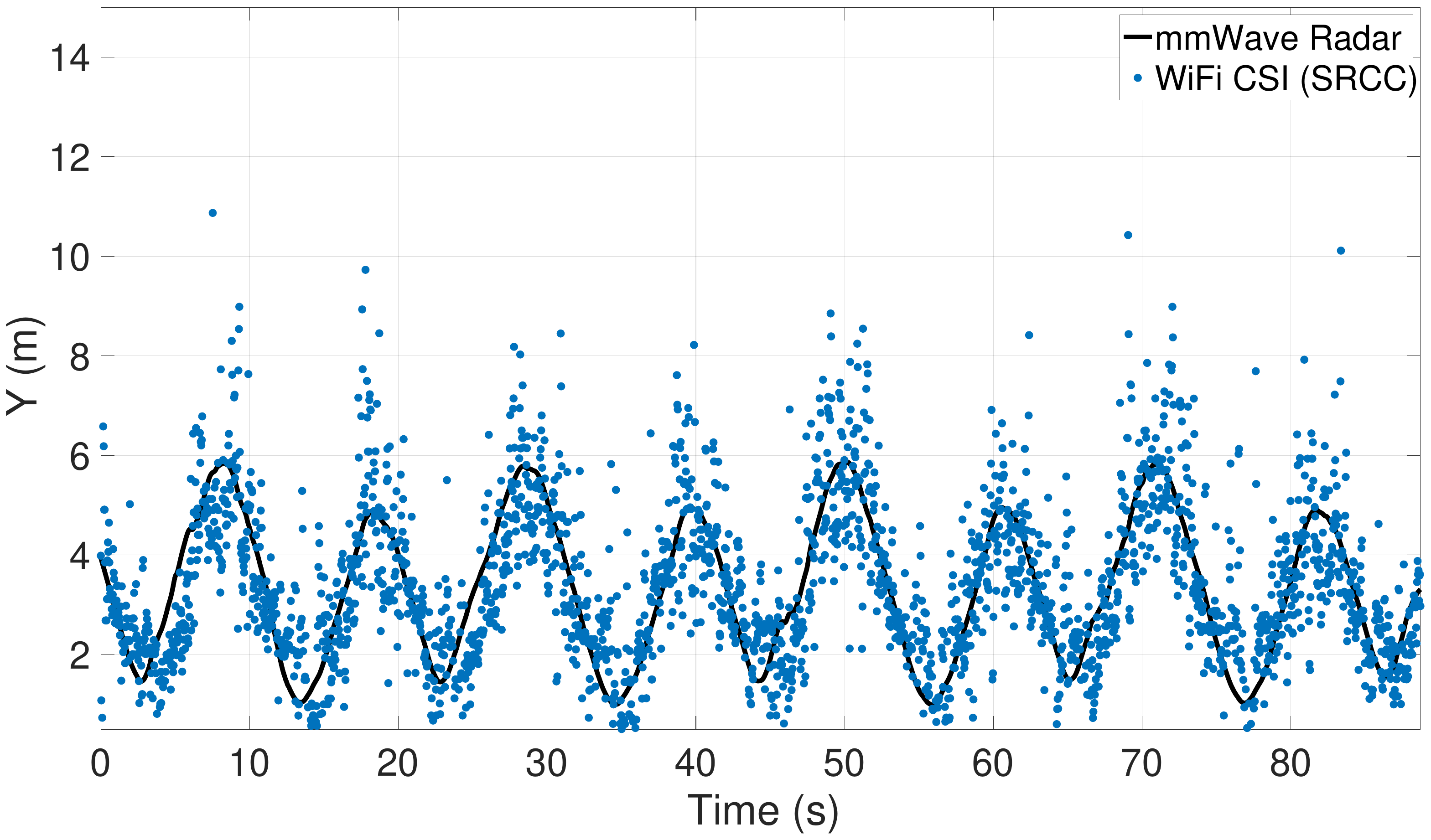}
            \vspace{-.5em}
            \subcaption{Raw y-axis coordinate (CSI)}
        \end{subfigure}\\
        \begin{subfigure}{\textwidth}
            \centering
            \includegraphics[width=0.955\textwidth]{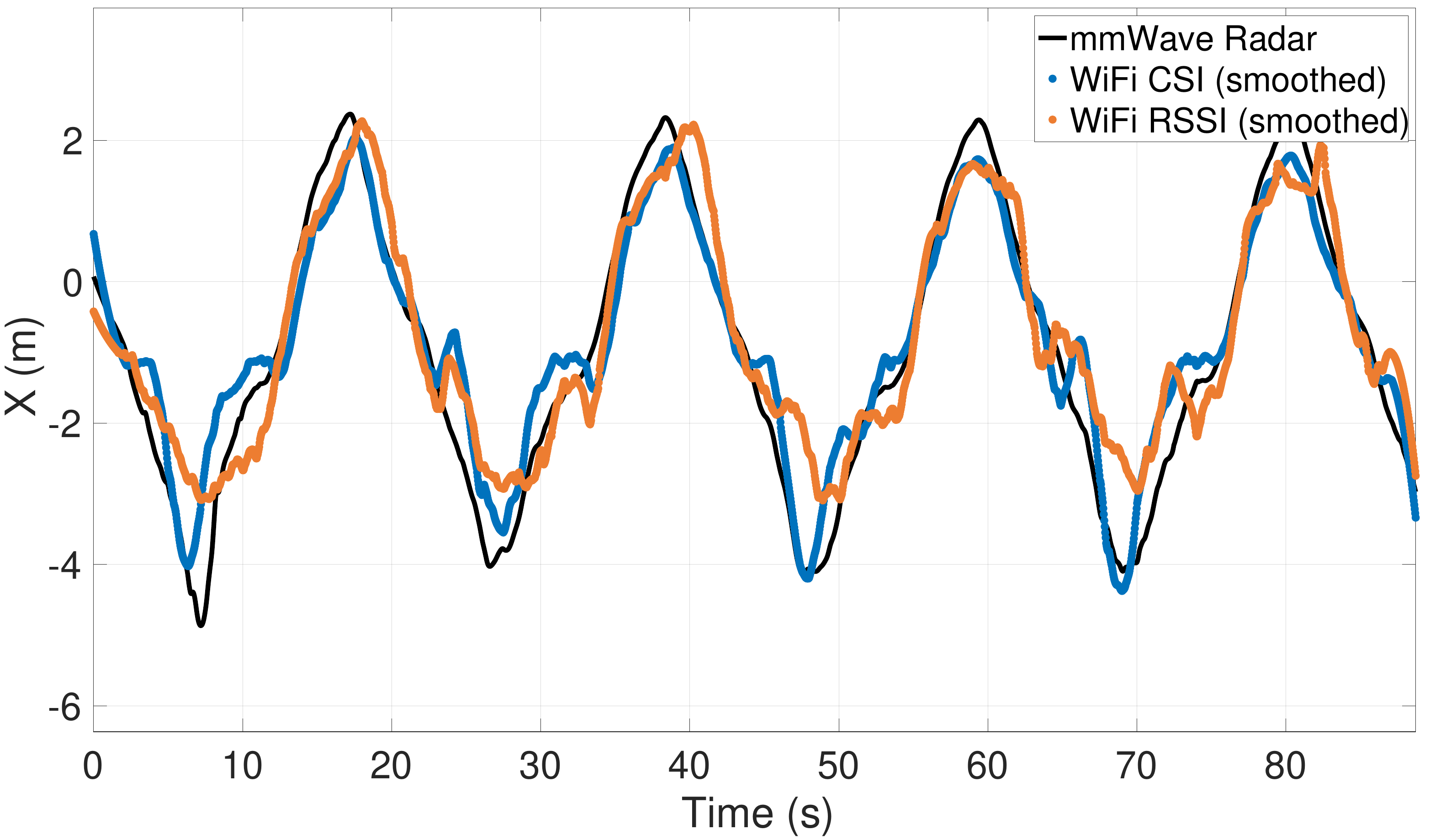}
            \vspace{-.5em}
            \subcaption{Smoothed x-axis coordinate}
        \end{subfigure}\\
        \begin{subfigure}{\textwidth}
            \centering
            \includegraphics[width=0.95\textwidth]{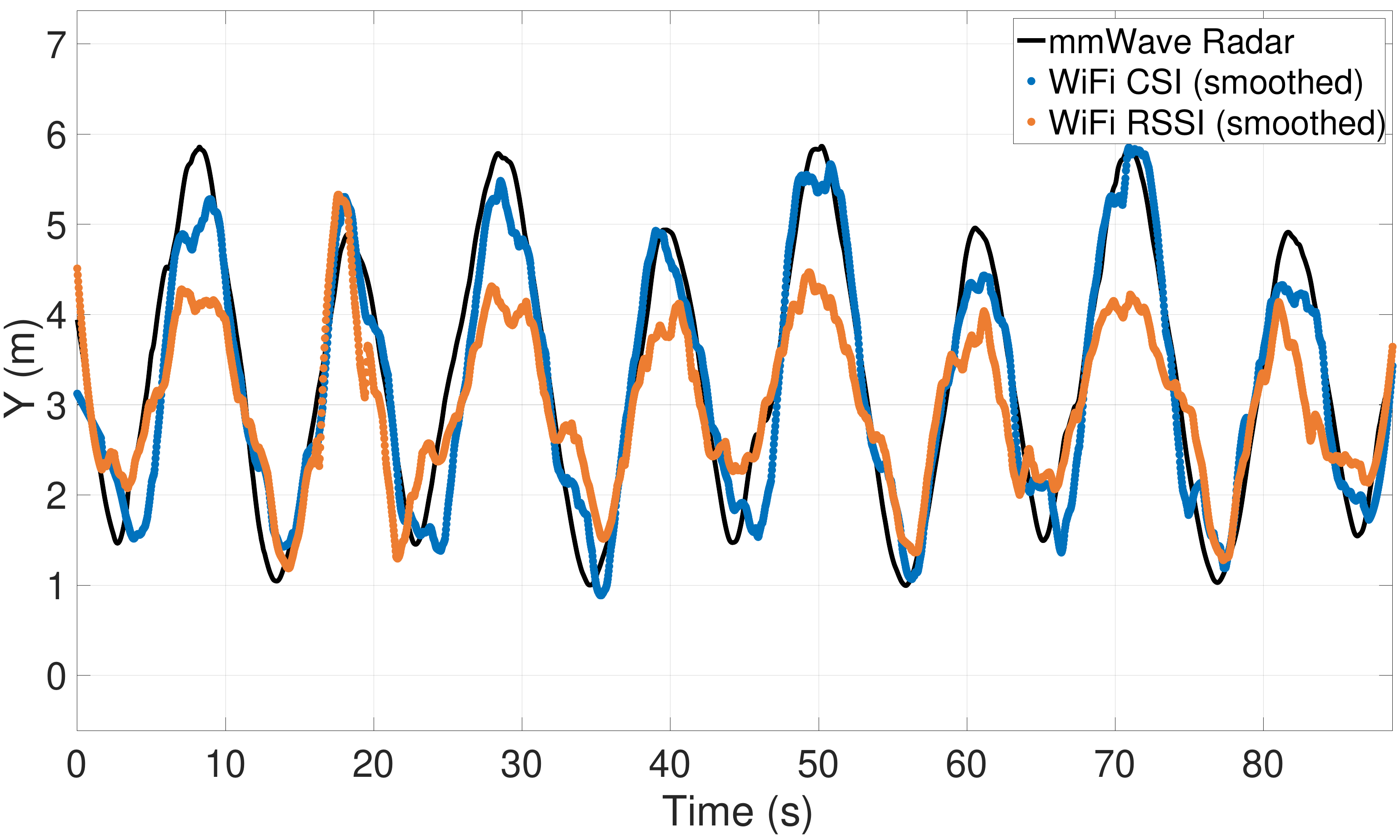}
            \vspace{-.5em}
            \subcaption{Smoothed y-axis coordinate}
        \end{subfigure}
        \caption{Elliptical trajectory.}
        \label{fig:raw_ellipse}
    \end{minipage}
    \begin{minipage}[t]{0.329\linewidth}
    \vspace{0pt} 
        \centering
        \begin{subfigure}{\textwidth}
            \centering
            \includegraphics[width=0.95\textwidth]{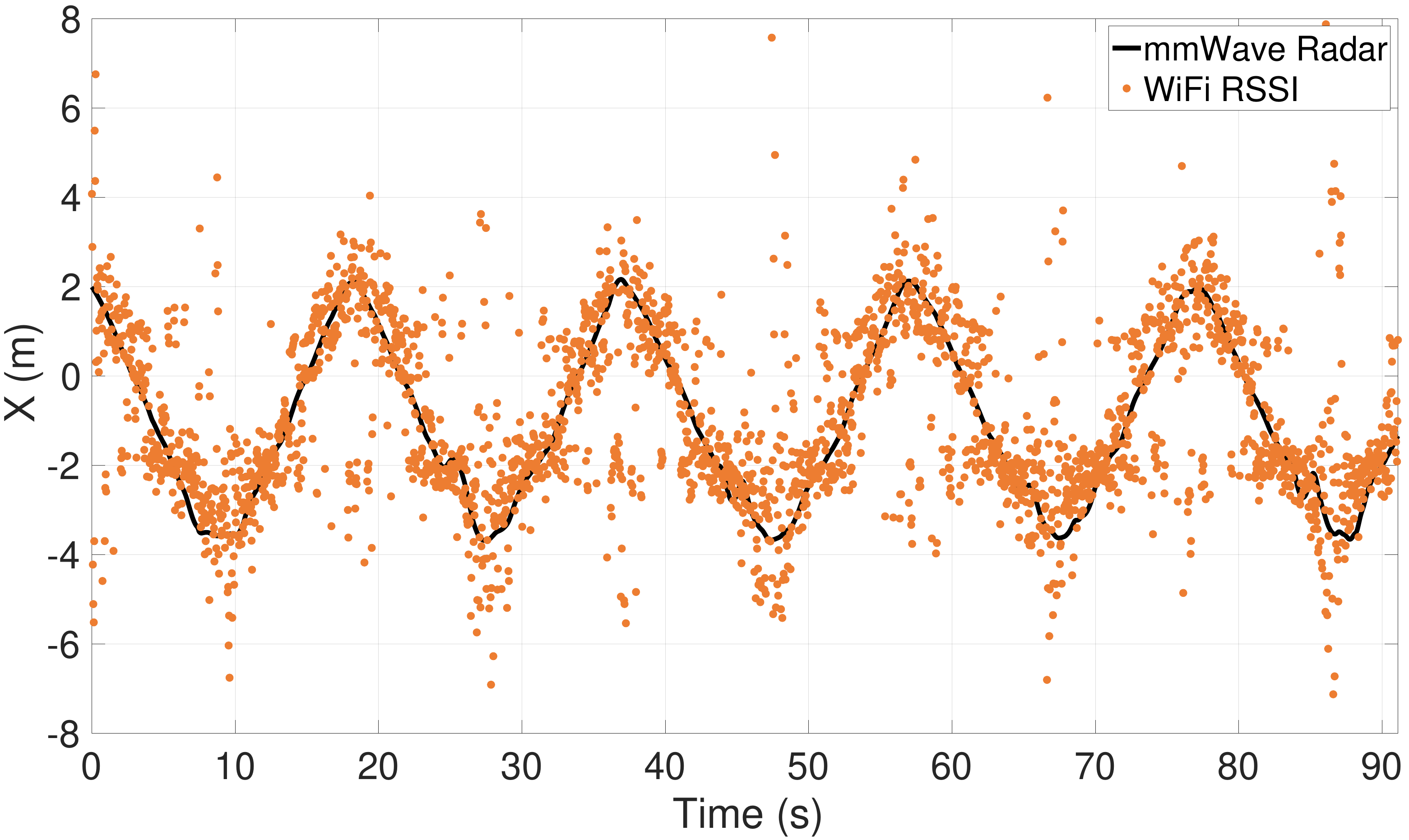}
            \vspace{-.5em}
            \subcaption{Raw x-axis coordinate (RSSI)}
        \end{subfigure}\\
        \begin{subfigure}{\textwidth}
            \centering
            \includegraphics[width=0.95\textwidth]{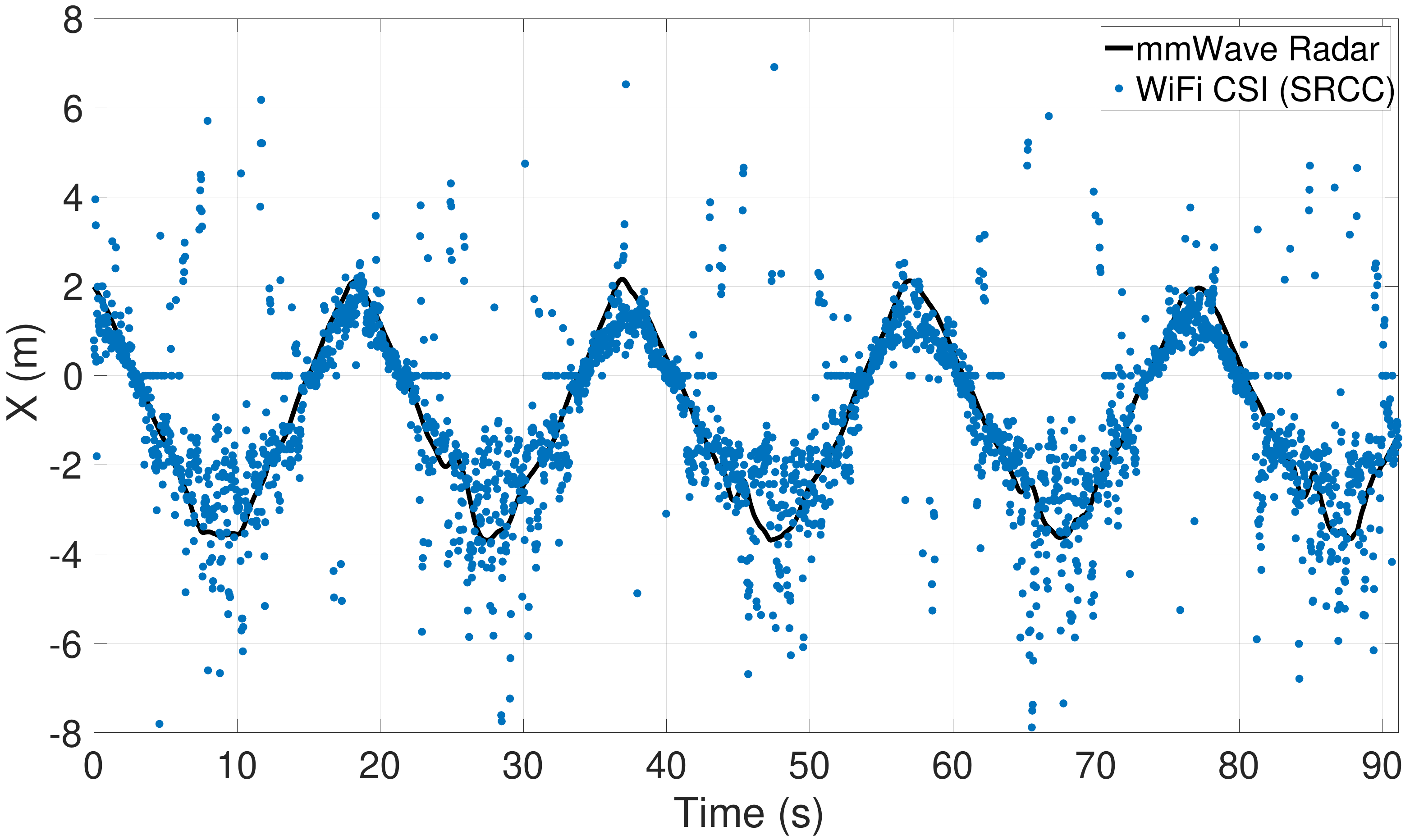}
            \vspace{-.5em}
            \subcaption{Raw x-axis coordinate (CSI)}
        \end{subfigure}\\
        \begin{subfigure}{\textwidth}
            \centering
            \includegraphics[width=0.95\textwidth]{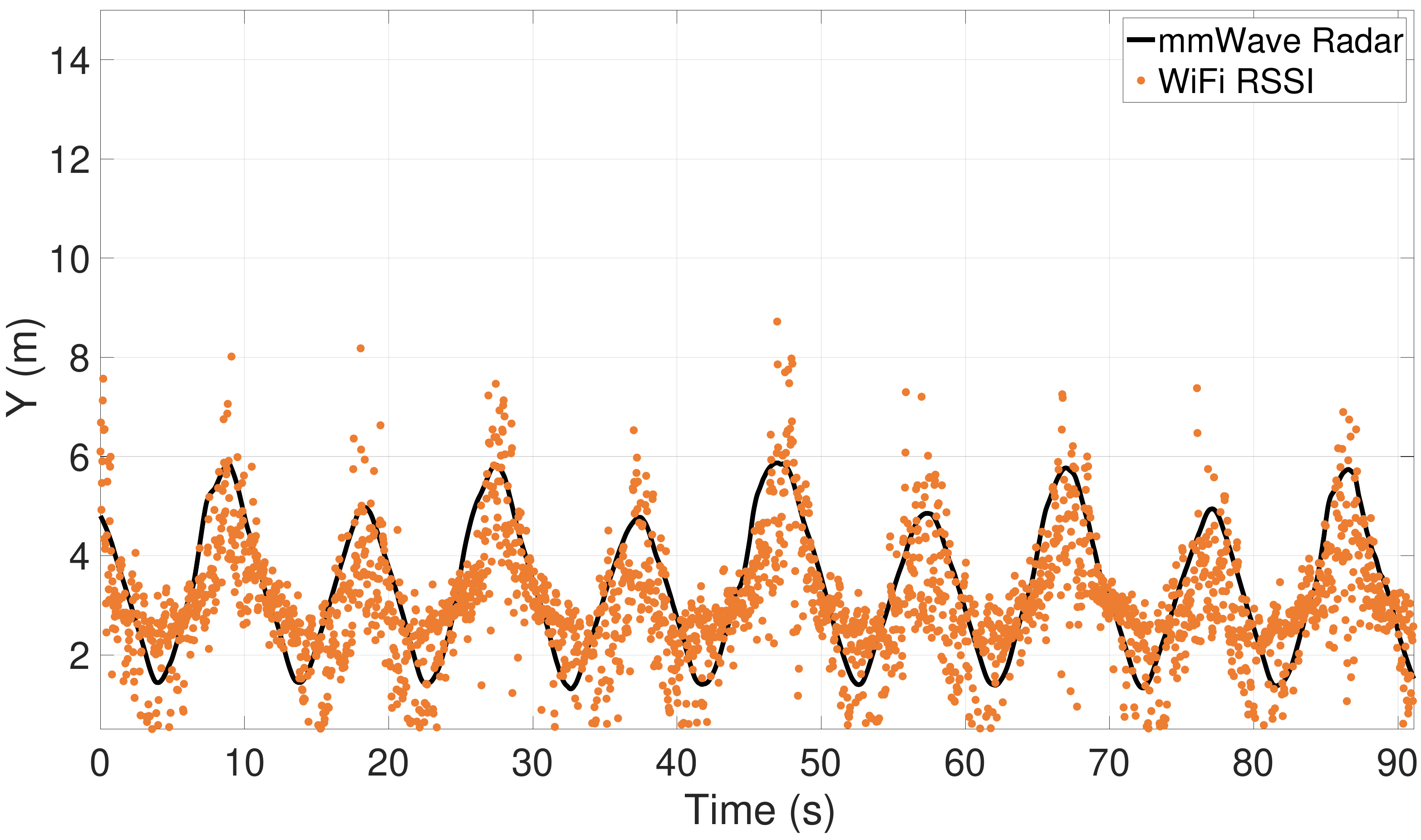}
            \vspace{-.5em}
            \subcaption{Raw y-axis coordinate (RSSI)}
        \end{subfigure}\\
        \begin{subfigure}{\textwidth}
            \centering
            \includegraphics[width=0.95\textwidth]{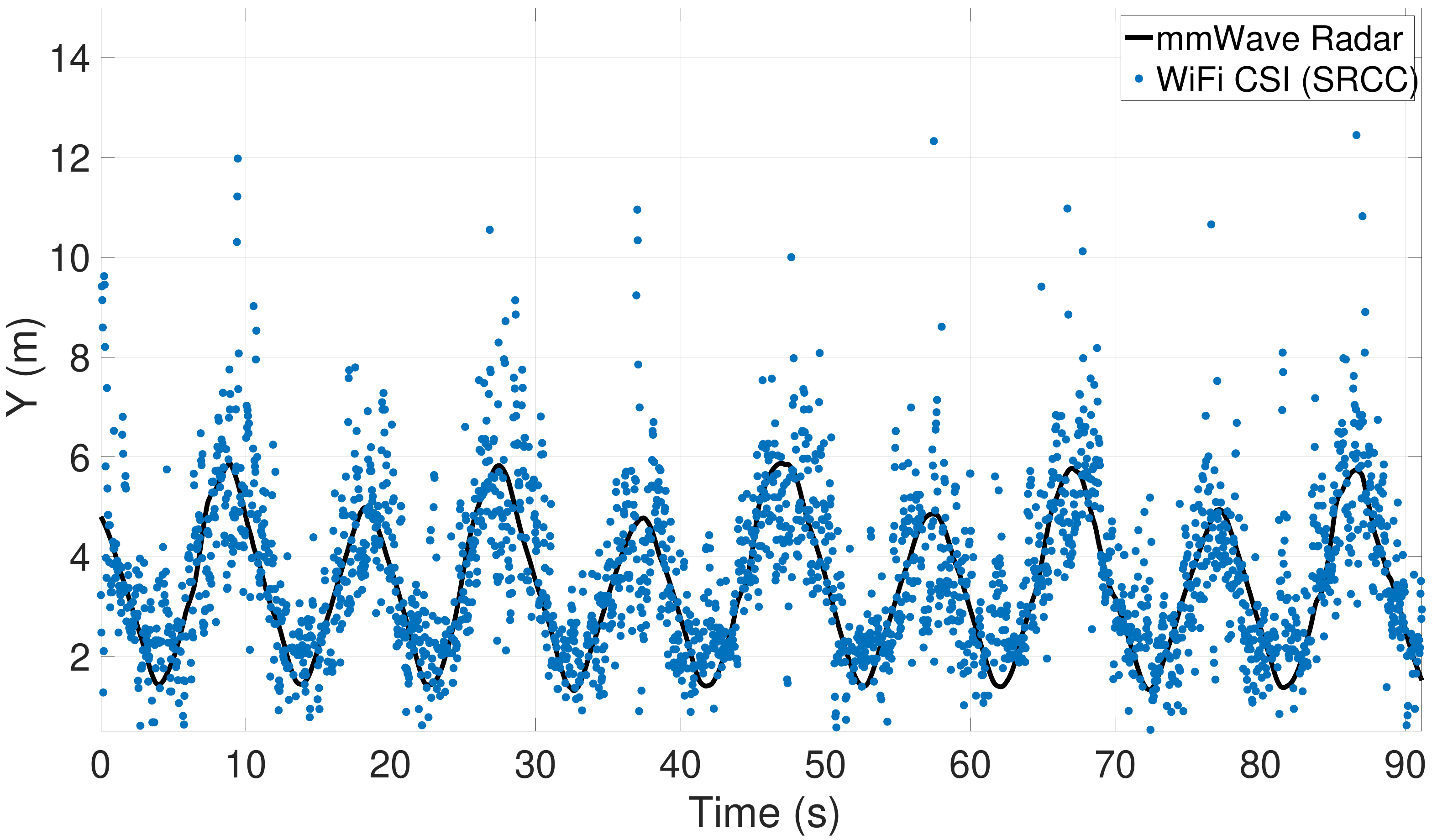}
            \vspace{-.5em}
            \subcaption{Raw y-axis coordinate (CSI)}
        \end{subfigure}\\
        \begin{subfigure}{\textwidth}
            \centering
            \includegraphics[width=0.95\textwidth]{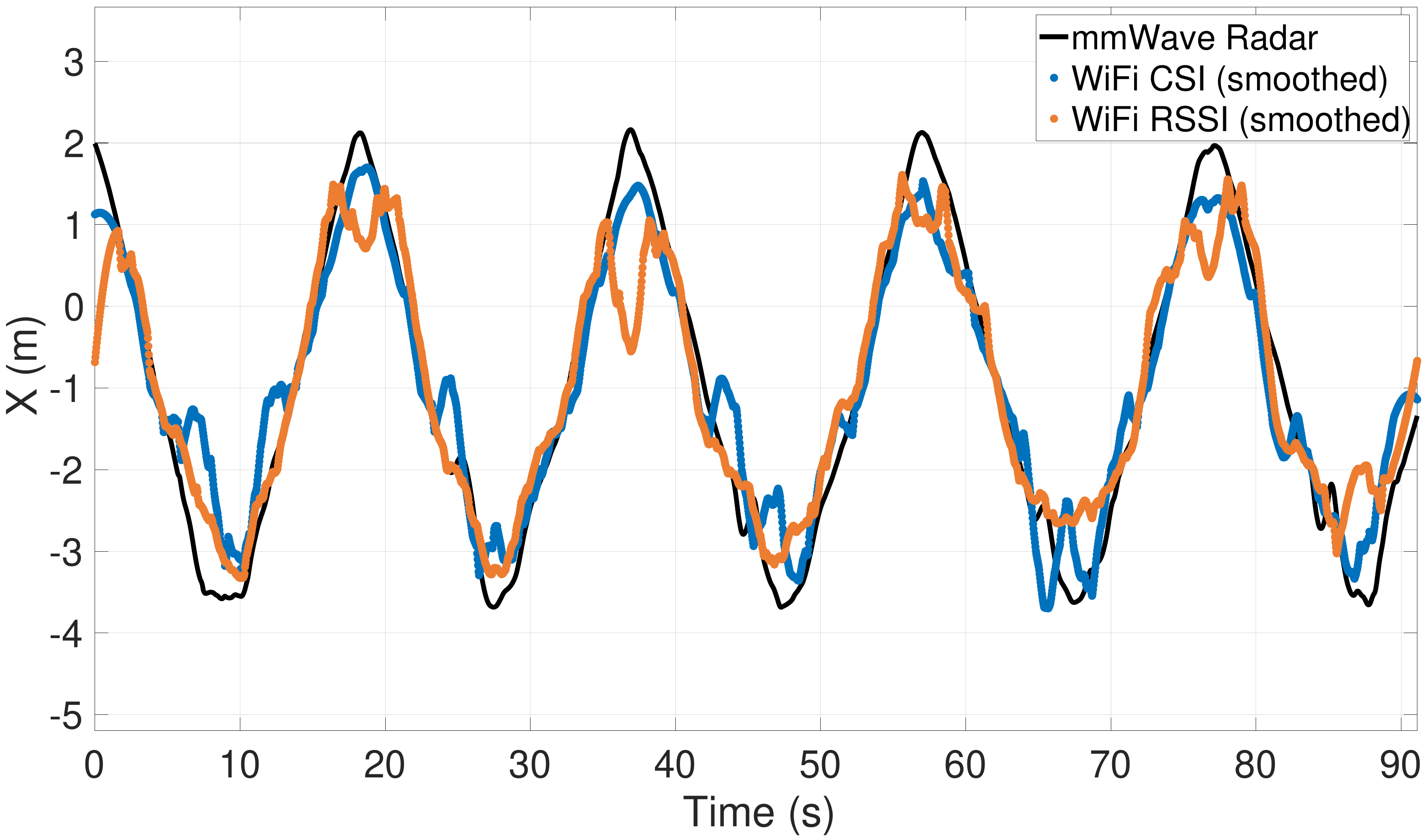}
            \vspace{-.5em}
            \subcaption{Smoothed x-axis coordinate}
        \end{subfigure}\\
        \begin{subfigure}{\textwidth}
            \centering
            \includegraphics[width=0.95\textwidth]{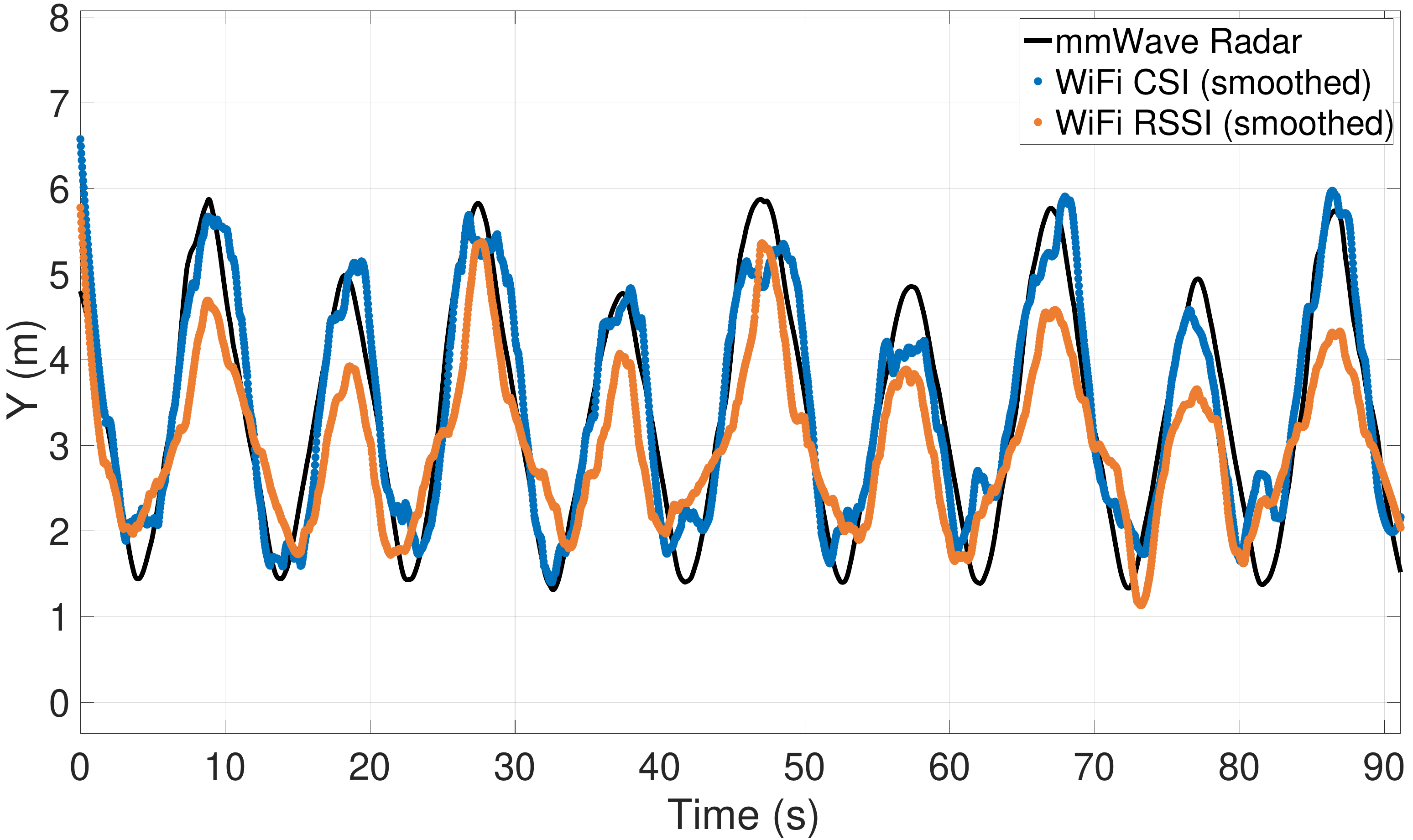}
            \vspace{-.5em}
            \subcaption{Smoothed y-axis coordinate}
        \end{subfigure}
        \caption{Linear trajectory.}
        \label{fig:raw_linear}
    \end{minipage}
    \begin{minipage}[t]{0.329\linewidth}
    \vspace{0pt} 
        \centering
        \begin{subfigure}{\textwidth}
            \centering
            \includegraphics[width=0.95\textwidth]{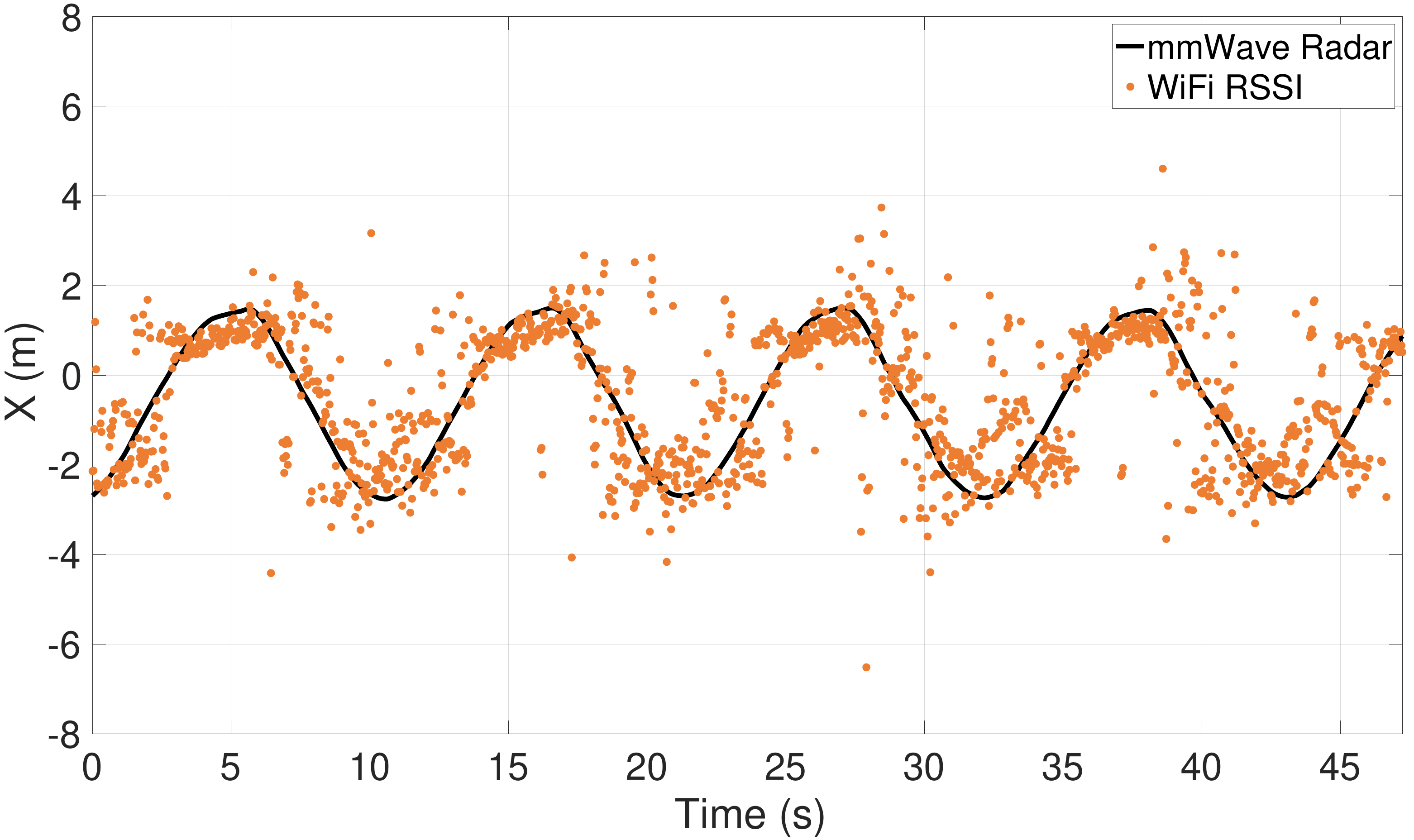}
            \vspace{-.5em}
            \subcaption{Raw x-axis coordinate (RSSI)}
        \end{subfigure}\\
        \begin{subfigure}{\textwidth}
            \centering
            \includegraphics[width=0.95\textwidth]{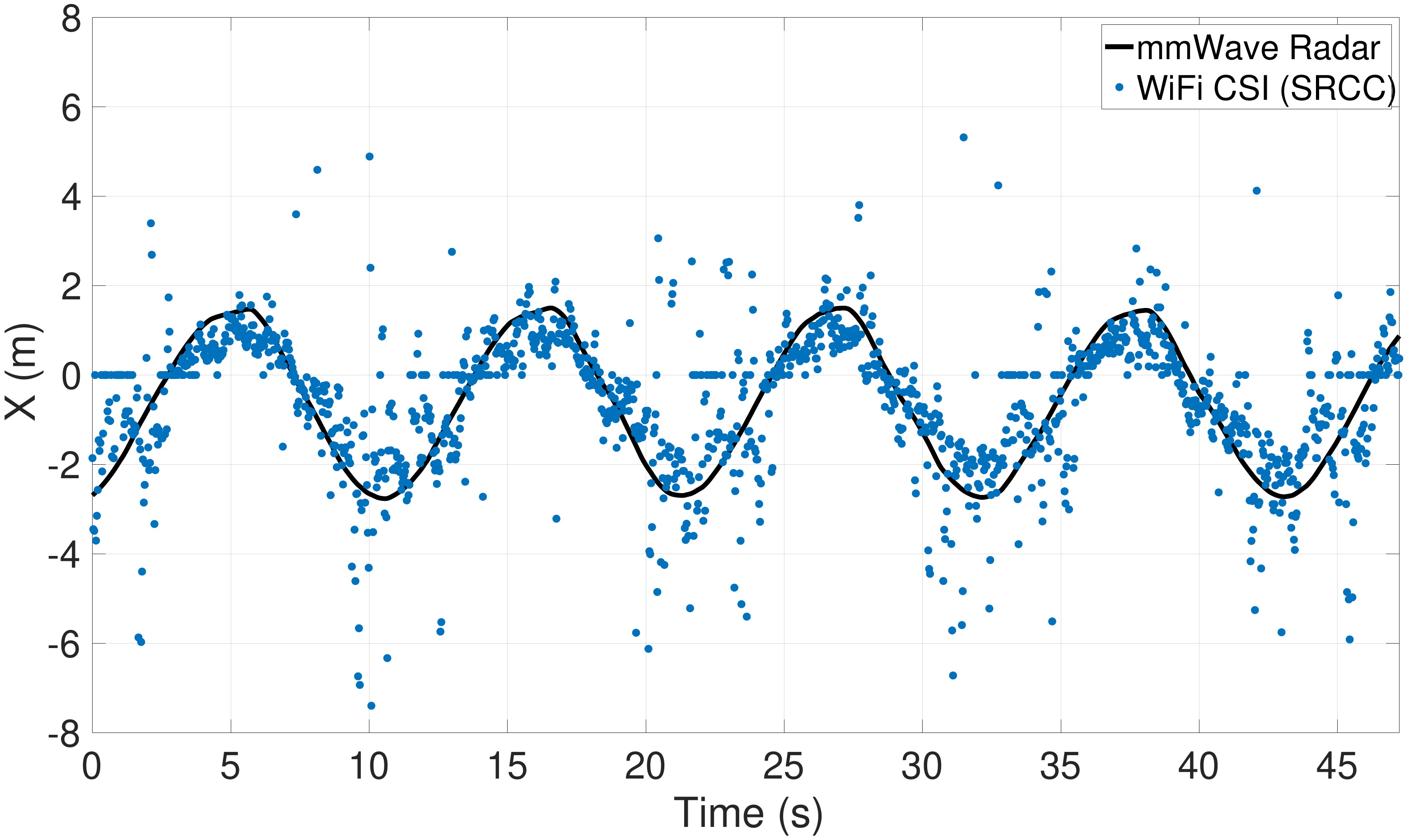}
            \vspace{-.5em}
            \subcaption{Raw x-axis coordinate (CSI)}
        \end{subfigure}\\
        \begin{subfigure}{\textwidth}
            \centering
            \includegraphics[width=0.95\textwidth]{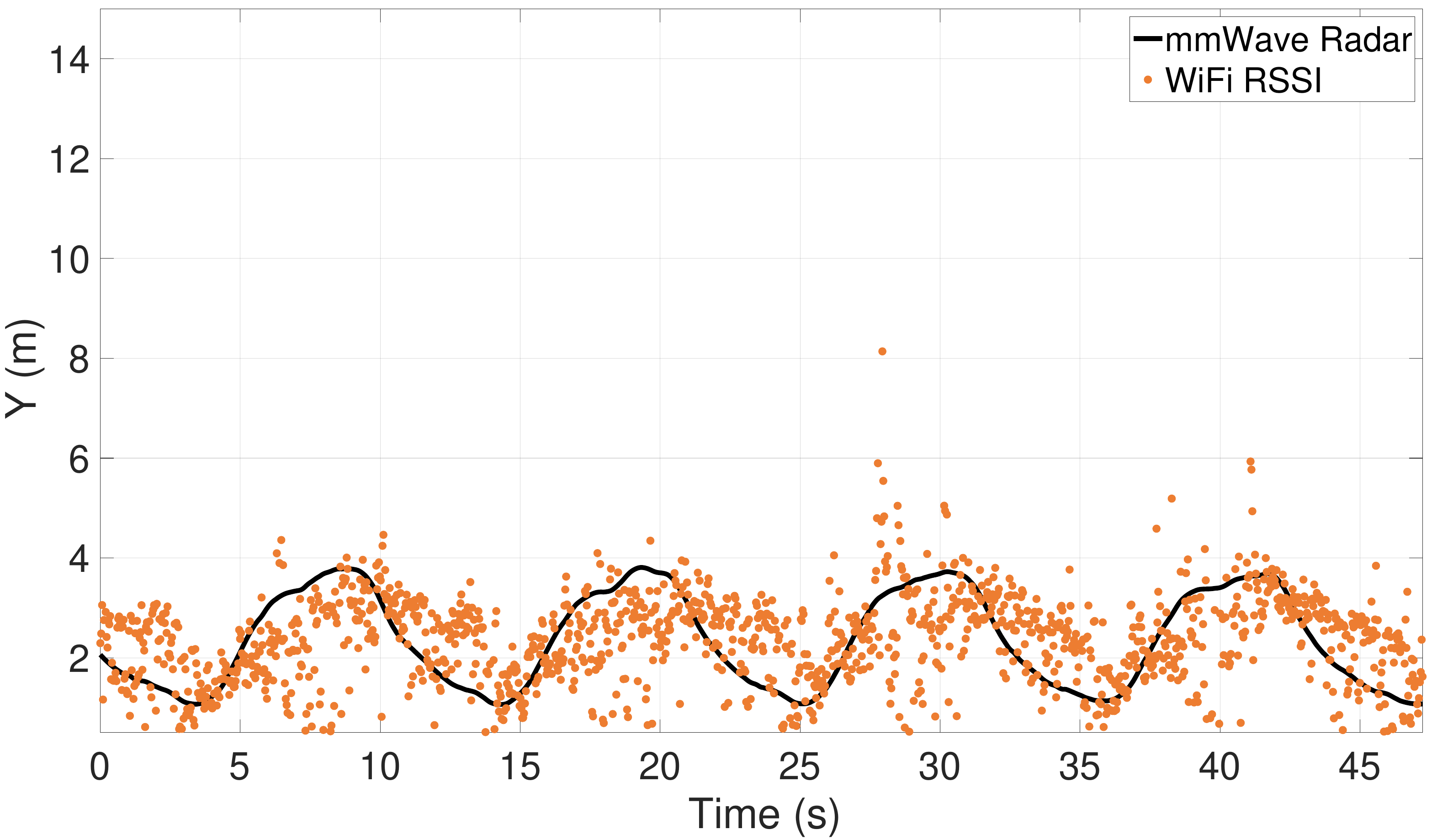}
            \vspace{-.5em}
            \subcaption{Raw y-axis coordinate (RSSI)}
        \end{subfigure}\\
        \begin{subfigure}{\textwidth}
            \centering
            \includegraphics[width=0.95\textwidth]{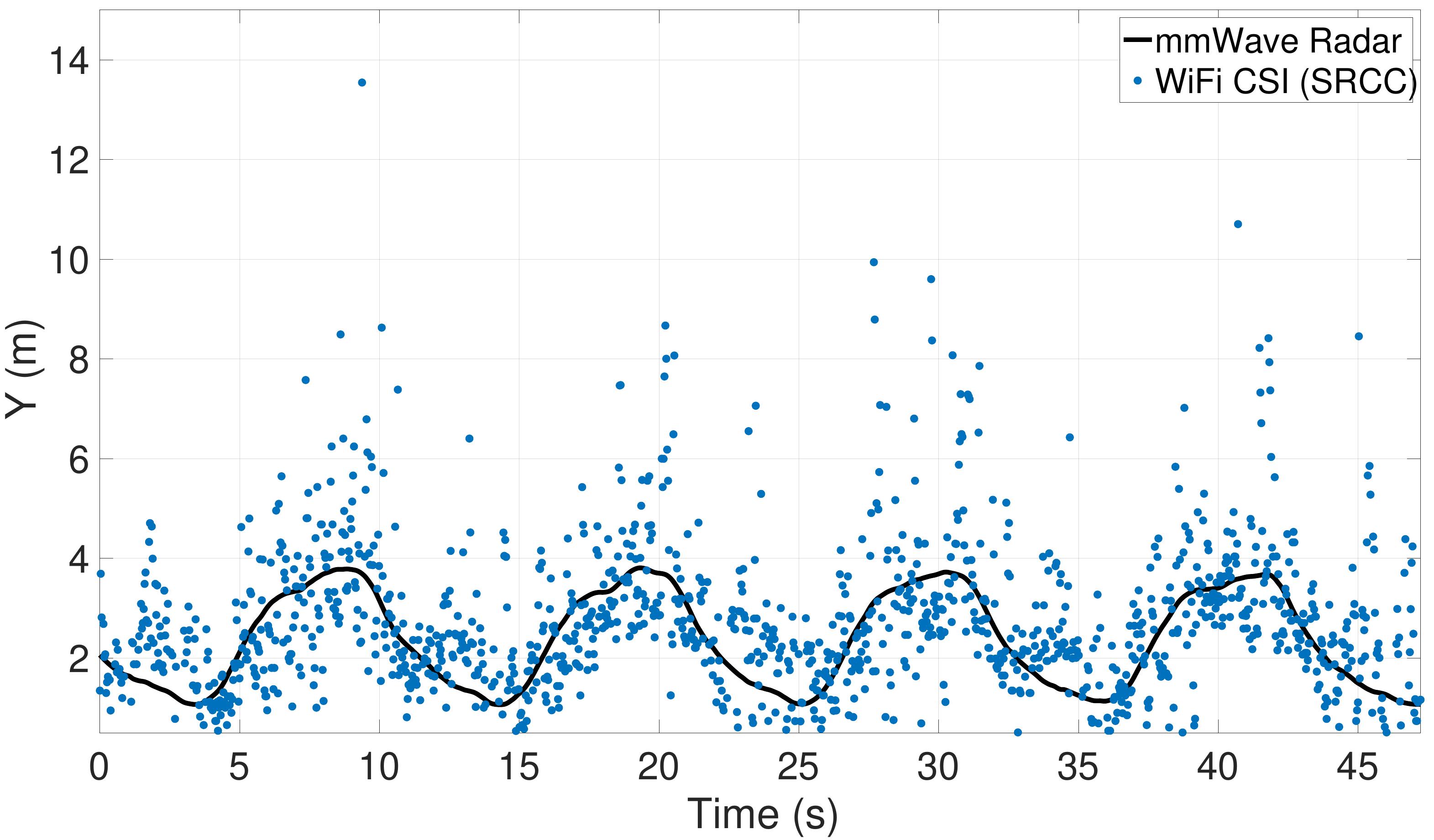}
            \vspace{-.5em}
            \subcaption{Raw y-axis coordinate (CSI)}
        \end{subfigure}\\
        \begin{subfigure}{\textwidth}
            \centering
            \includegraphics[width=0.95\textwidth]{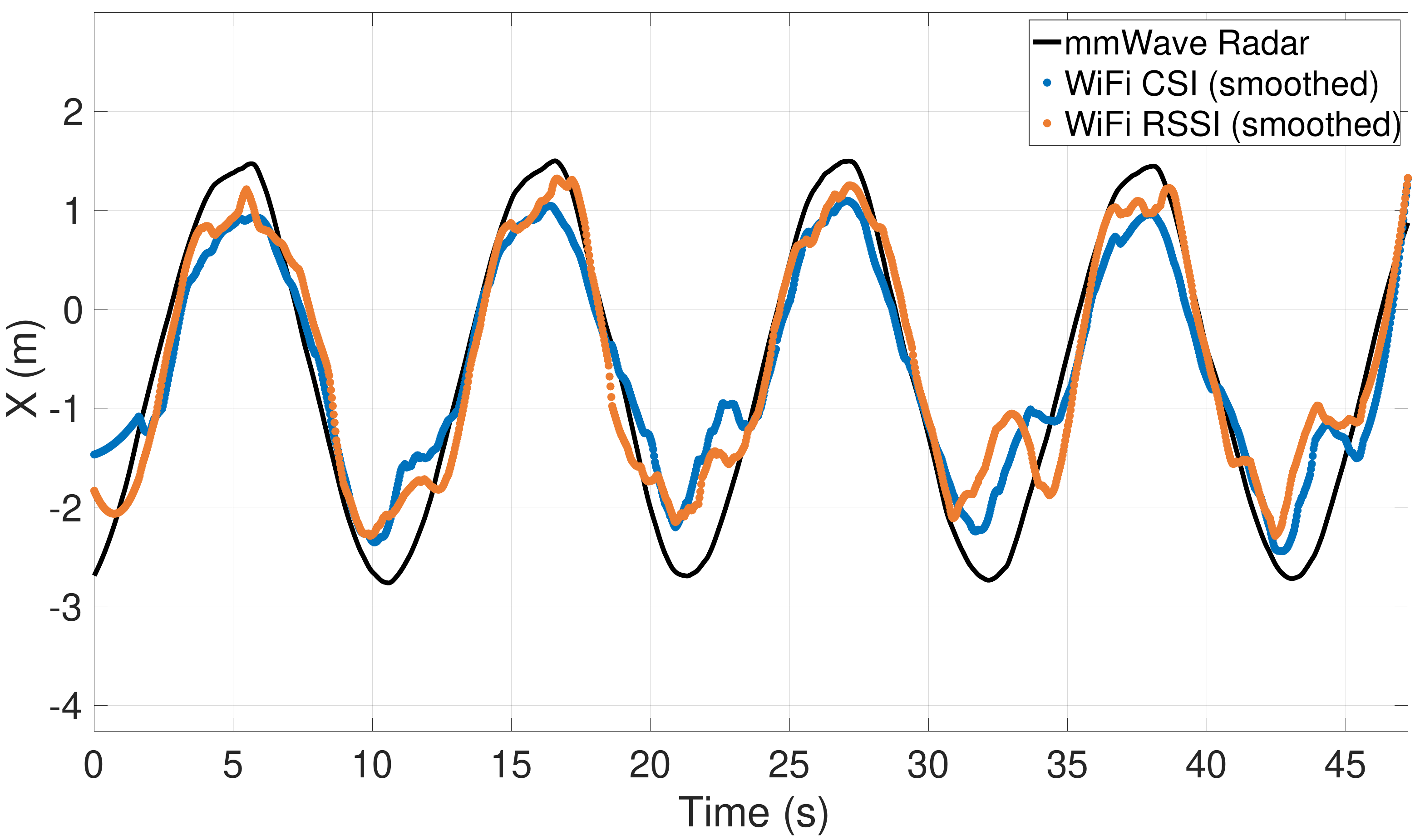}
            \vspace{-.5em}
            \subcaption{Smoothed x-axis coordinate}
        \end{subfigure}\\
        \begin{subfigure}{\textwidth}
            \centering
            \includegraphics[width=0.957\textwidth]{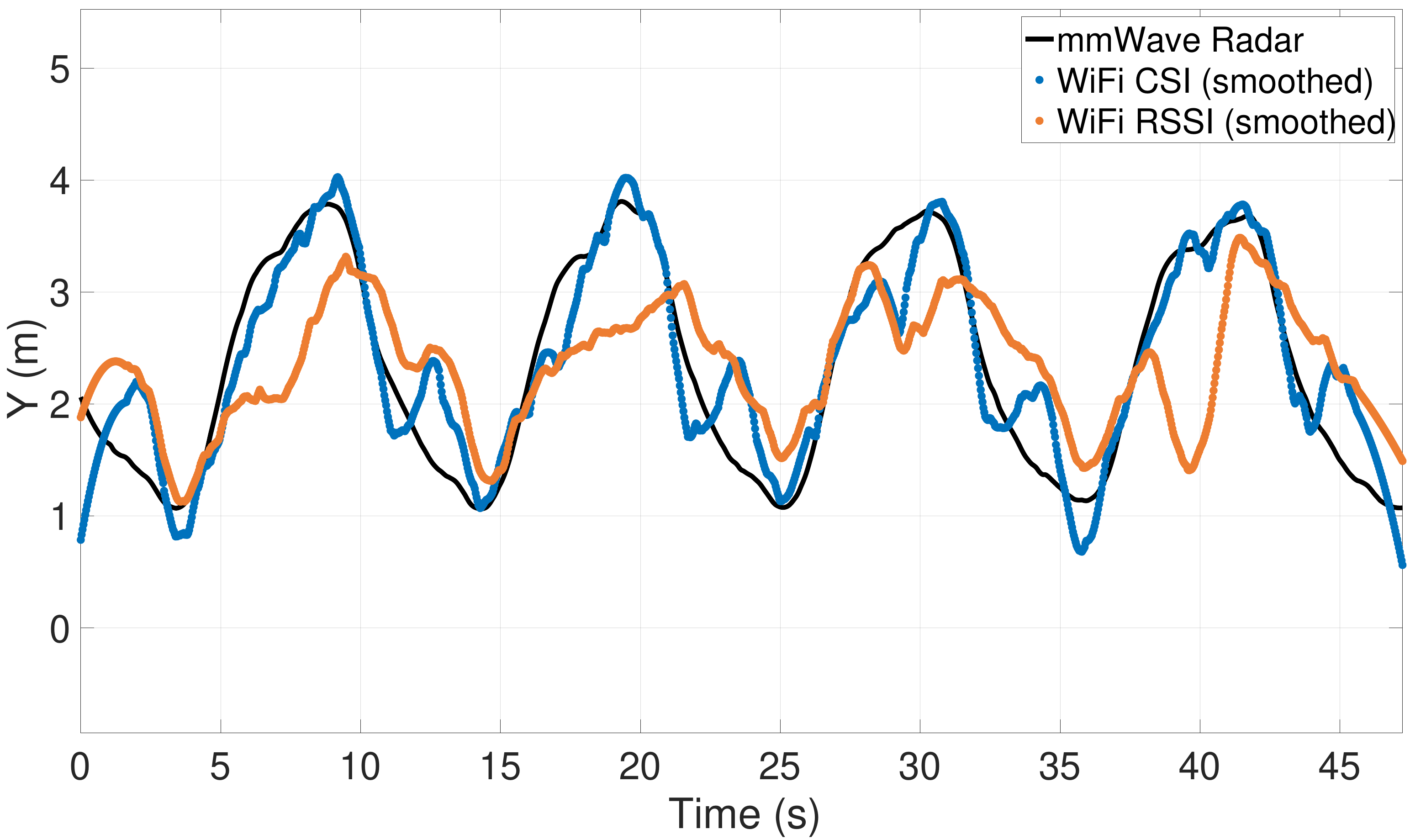}
            \vspace{-.5em}
            \subcaption{Smoothed y-axis coordinate}
        \end{subfigure}
        \caption{Rectangular trajectory.}
        \label{fig:raw_rectangle}
    \end{minipage}
    \label{fig:raw_xy_all}
\end{figure*}

\begin{figure*}
    \centering
    \begin{minipage}[t]{0.329\linewidth}
    \vspace{0pt}
        \centering
        \begin{subfigure}{\textwidth}
            \includegraphics[width=\textwidth]{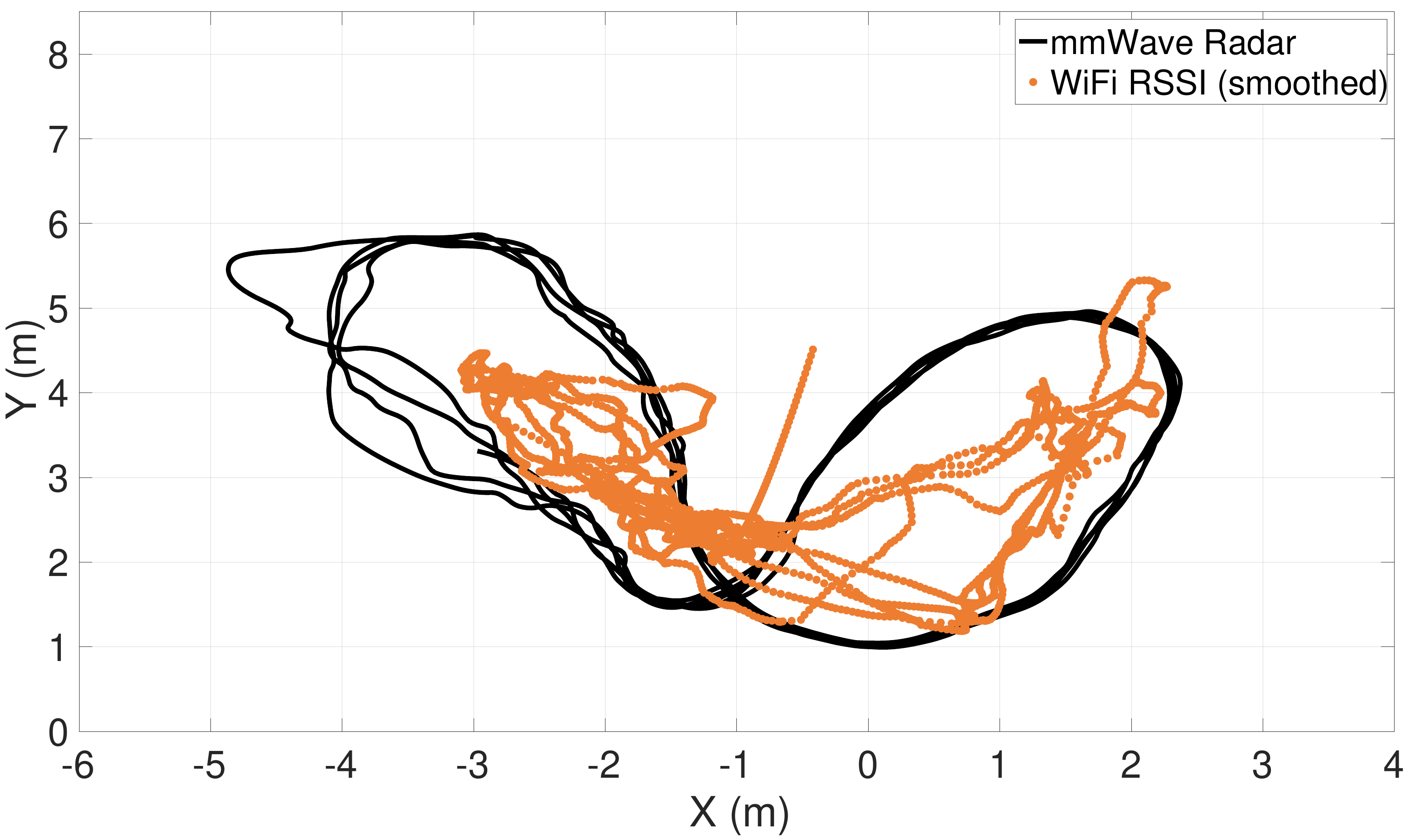}
            \subcaption{RSSI}
        \end{subfigure}\\
        \vspace{.5em}
        \begin{subfigure}{\textwidth}
            \includegraphics[width=\textwidth]{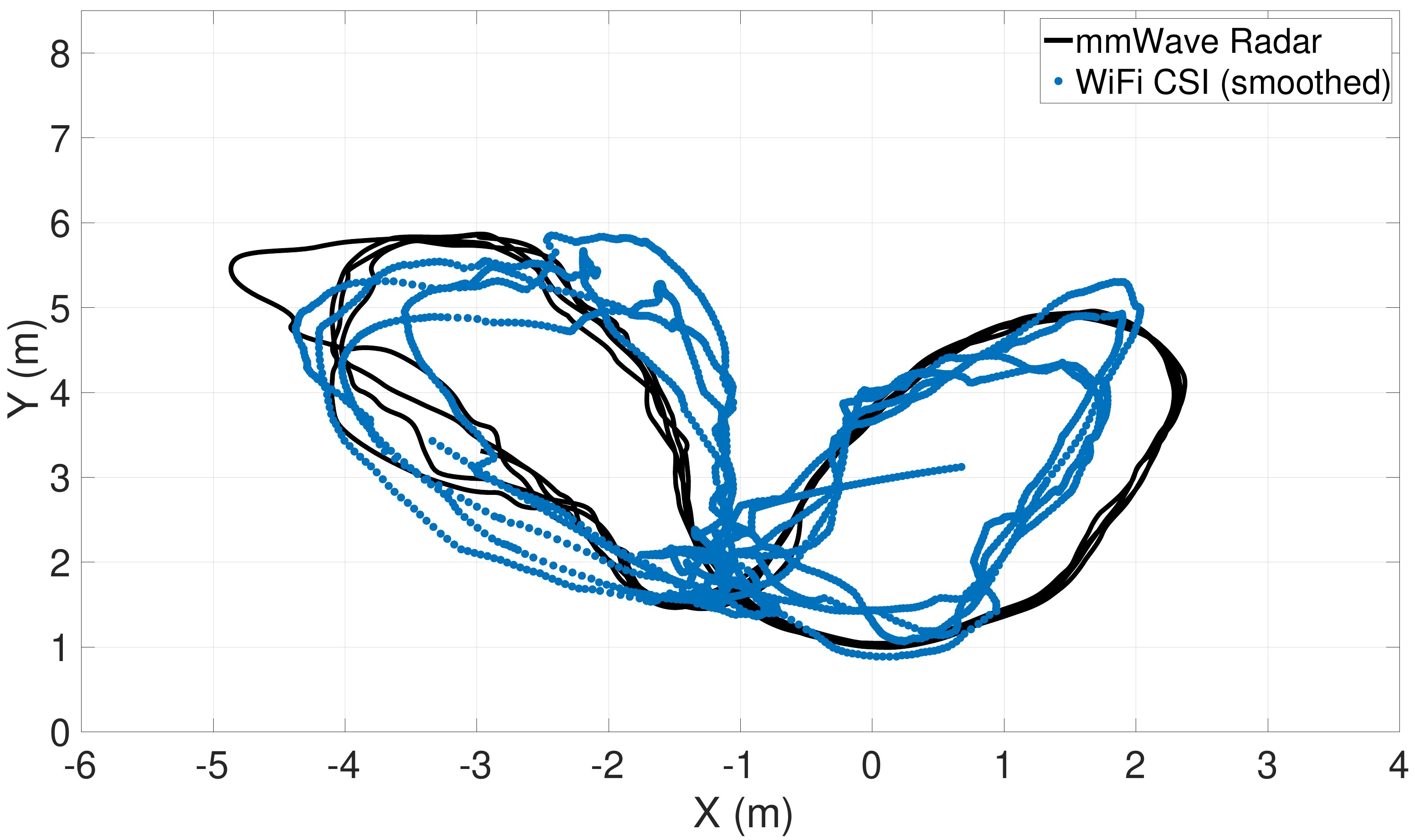}
            \subcaption{CSI}
        \end{subfigure}
        \caption{Elliptical trajectory.}
        \label{fig:traj_ellipse}
    \end{minipage}
    \begin{minipage}[t]{0.329\linewidth}
    \vspace{0pt}
        \centering
        \begin{subfigure}{\textwidth}
            \includegraphics[width=\textwidth]{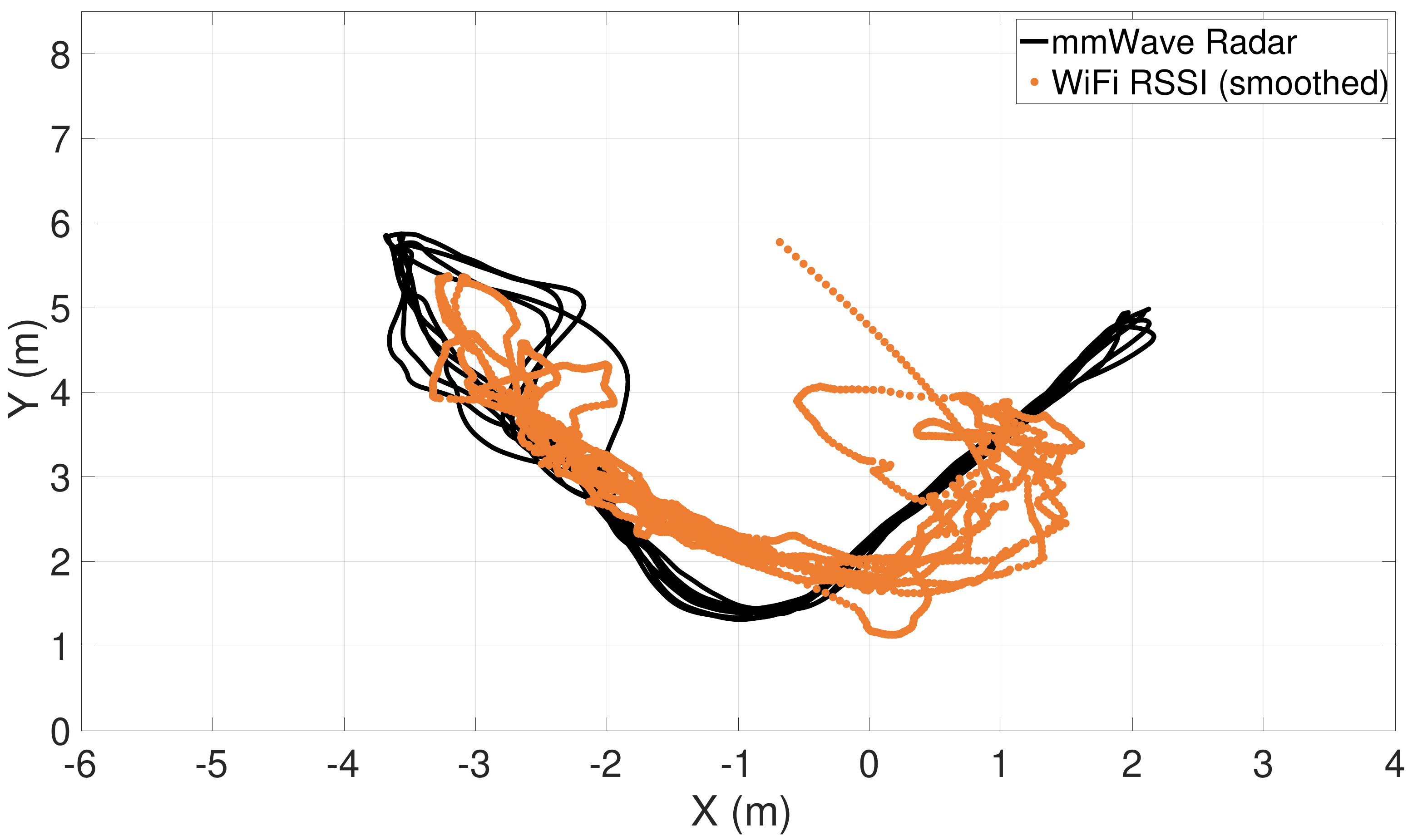}
            \subcaption{RSSI}
        \end{subfigure}\\
        \vspace{.5em}
        \begin{subfigure}{\textwidth}
            \includegraphics[width=\textwidth]{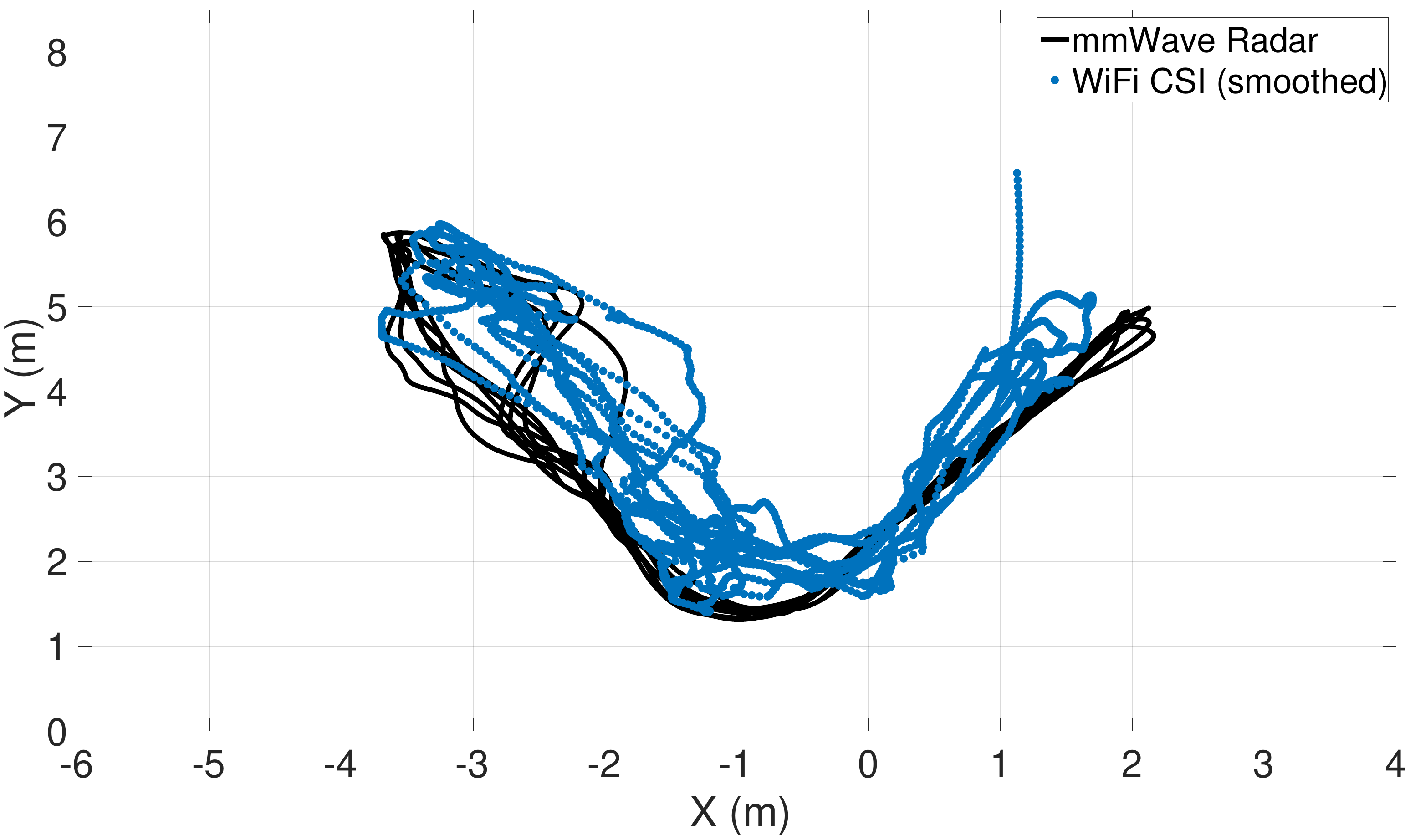}
            \subcaption{CSI}
        \end{subfigure}
        \caption{Linear trajectory.}
        \label{fig:traj_linear}
    \end{minipage}
    \begin{minipage}[t]{0.329\linewidth}
    \vspace{0pt}
        \centering
        \begin{subfigure}{\textwidth}
            \includegraphics[width=\textwidth]{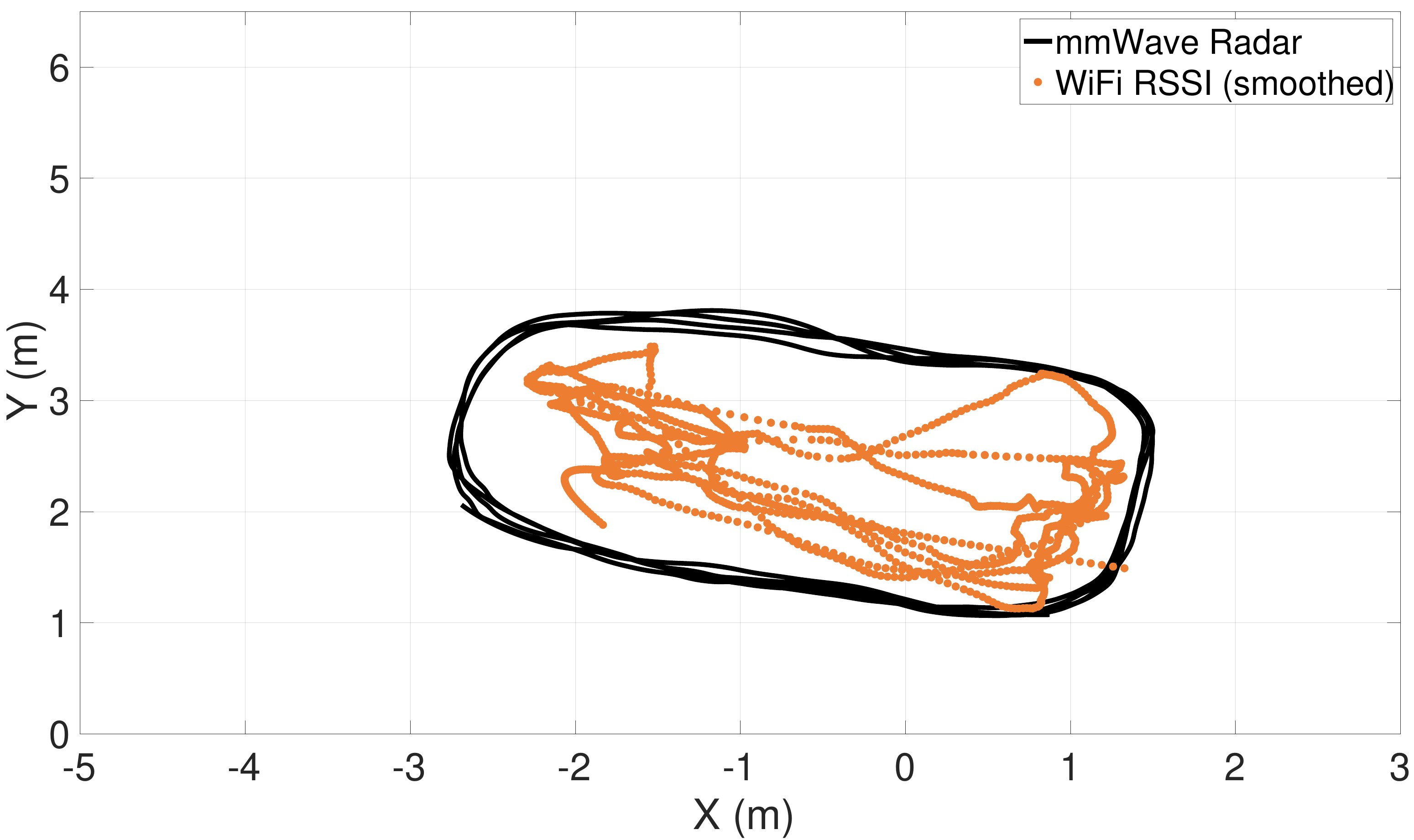}
            \subcaption{RSSI}
        \end{subfigure}\\
        \vspace{.5em}
        \begin{subfigure}{\textwidth}
            \includegraphics[width=\textwidth]{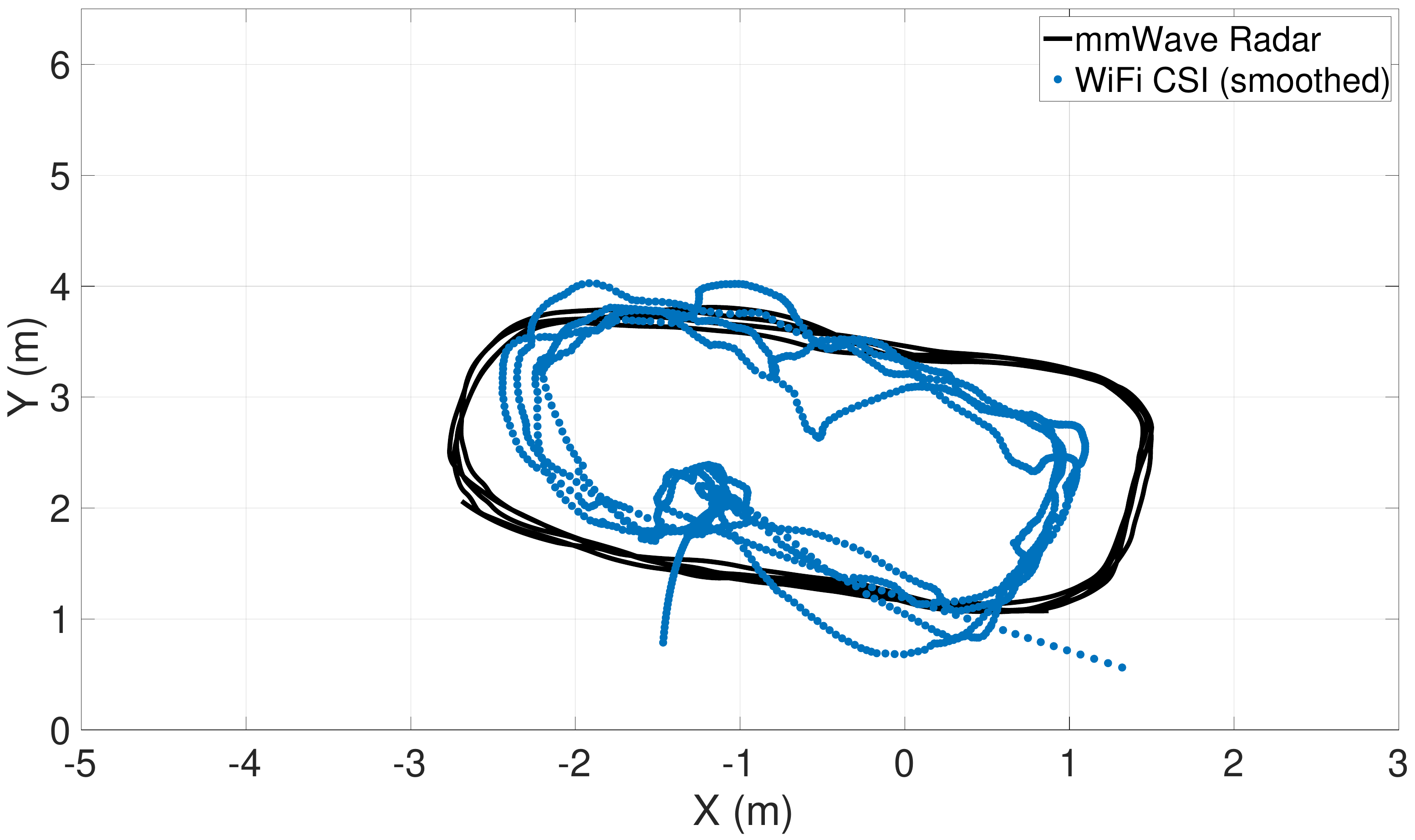}
            \subcaption{CSI}
        \end{subfigure}
        \caption{Rectangular trajectory.}
        \label{fig:traj_rectangle}
    \end{minipage}
    \label{fig:traj_all}
        \vspace{-1.5em}
\end{figure*}

\begin{figure*}
\centering
\begin{minipage}{0.329\textwidth}
    \centering
    \includegraphics[width=0.93\linewidth]{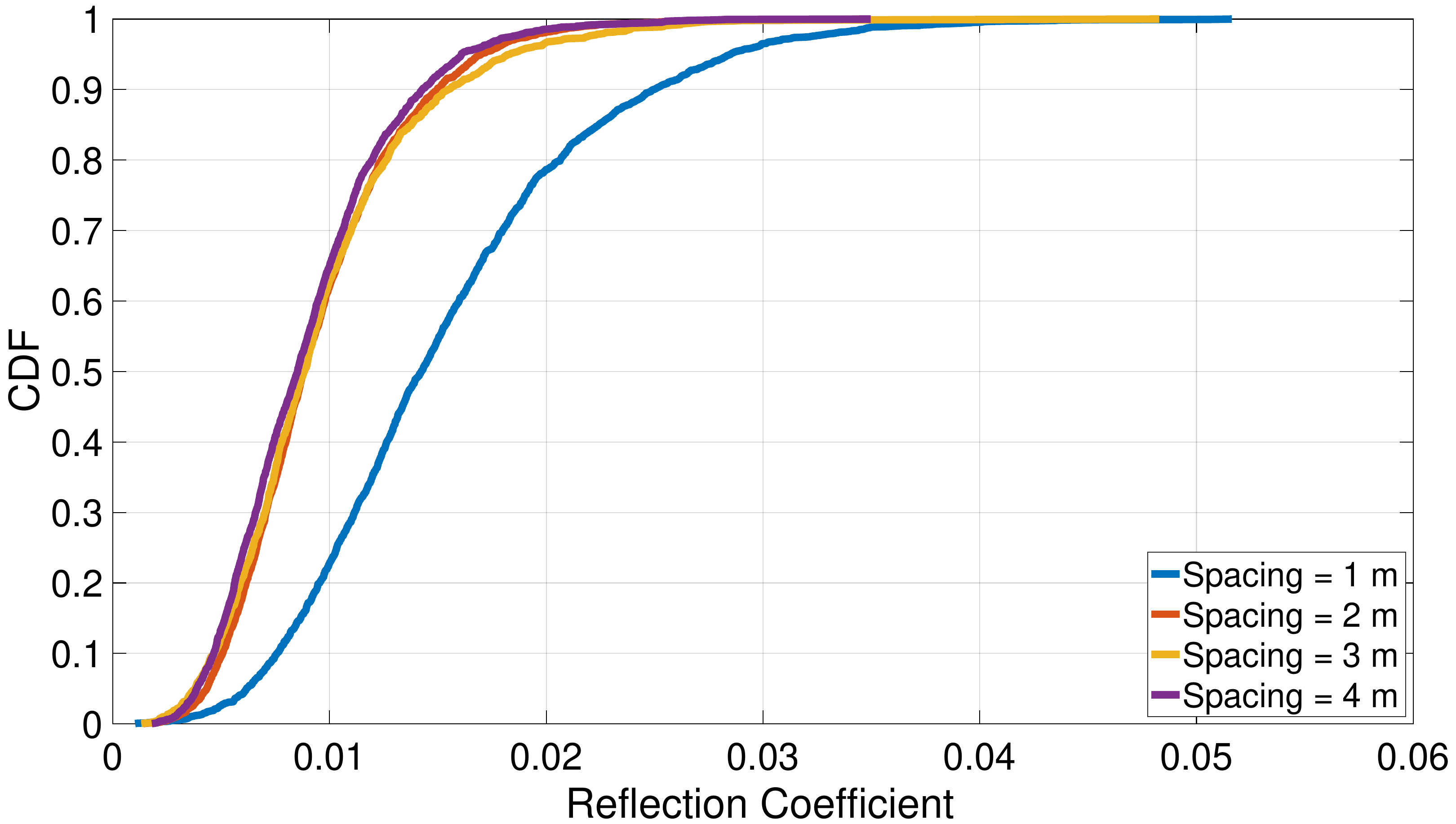}
    \caption{Overall human reflection coefficient.}
    \label{fig:cdf_range_ratio}
\end{minipage}
\begin{minipage}{0.329\textwidth}
    \centering
    \centering
        \includegraphics[width=0.92\textwidth]{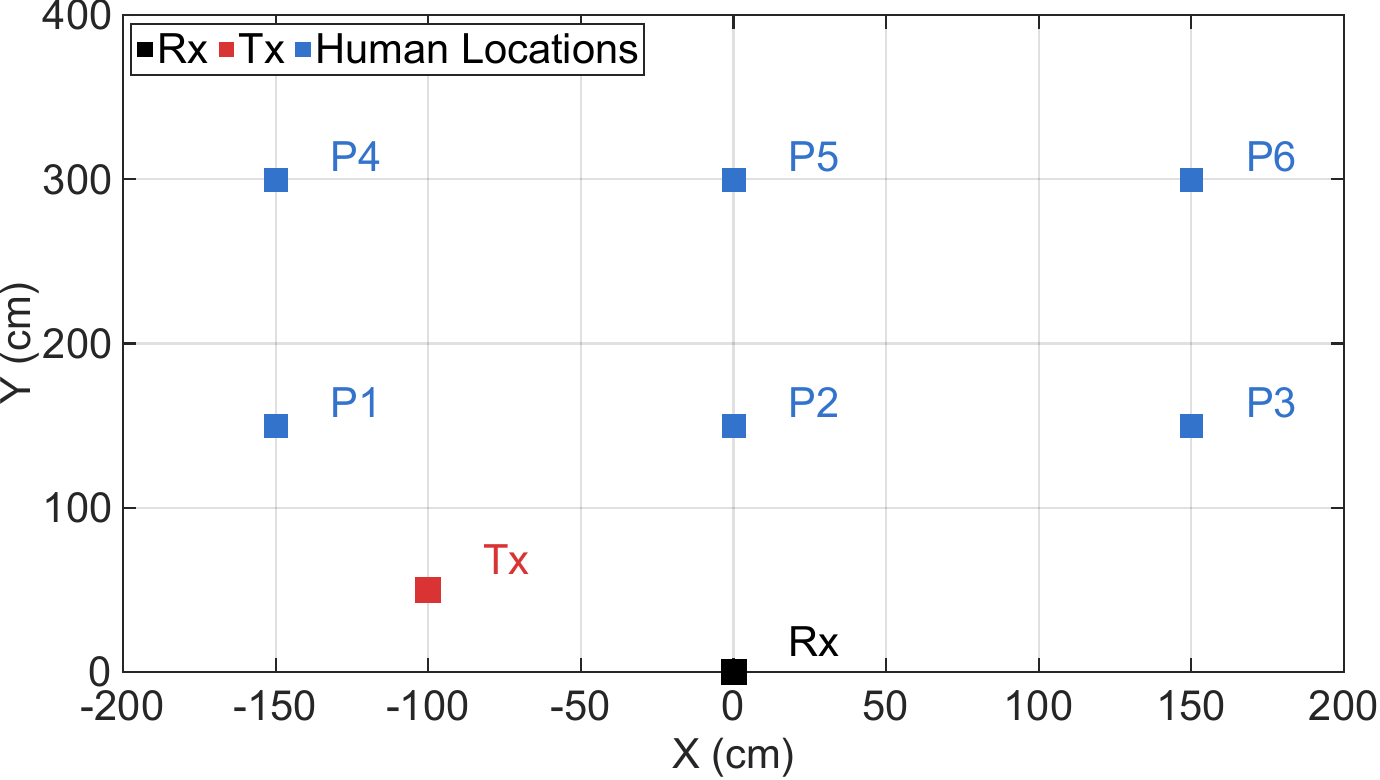}
        \caption{{Experimental geometry of Tx, Rx, and human positions.}}
        \label{fig:geom_map}
\end{minipage}
\begin{minipage}{0.329\textwidth}
    \centering
    \includegraphics[width=\textwidth]{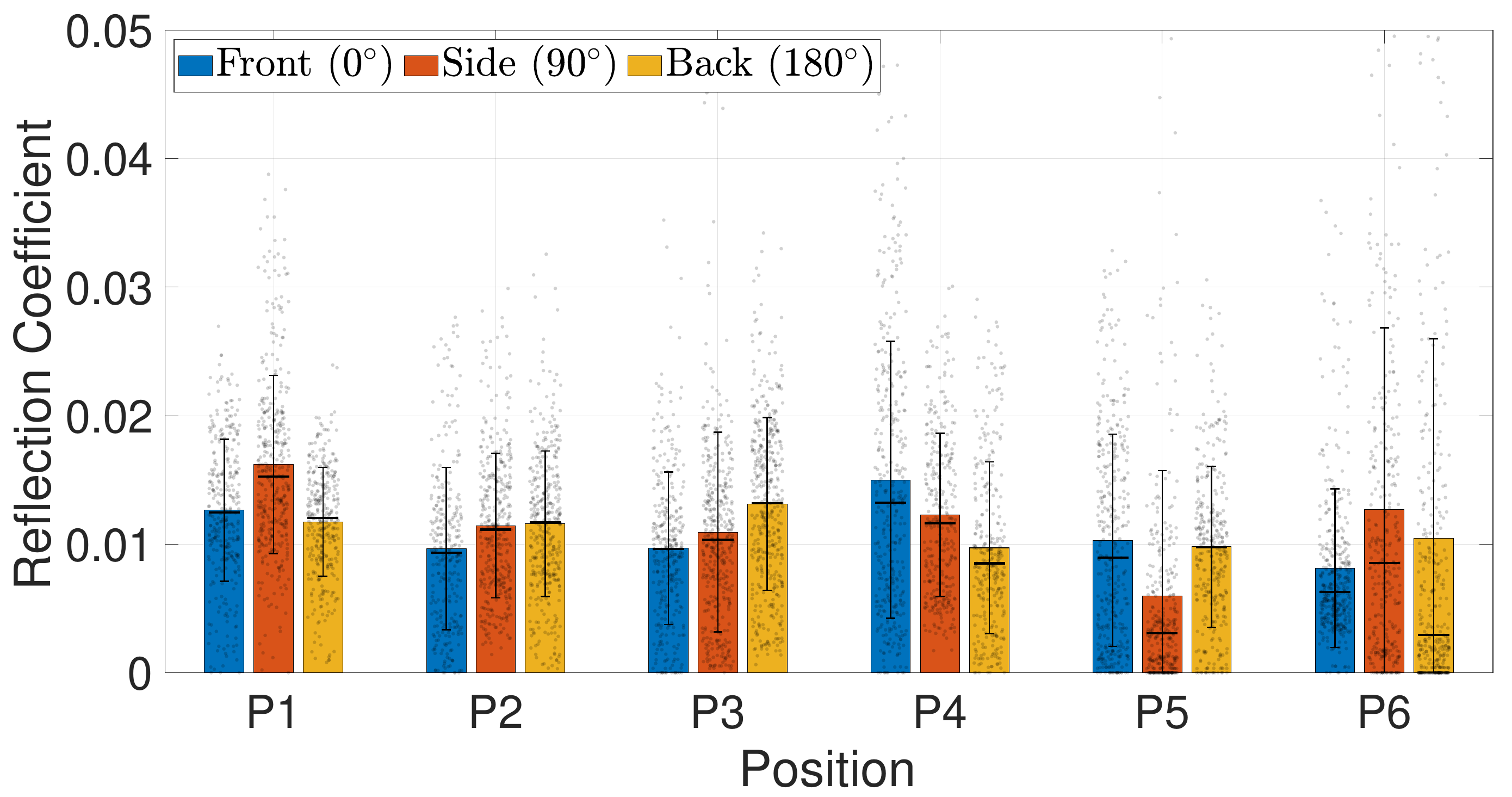}
        \caption{{Human reflection coefficient across positions and orientations.}}
        \label{fig:rcs_pd}
\end{minipage}
 \vspace{-1.5em}
\end{figure*}

\subsection{Overall Tracking Accuracy}
Fig.~\ref{fig:CDF_traj_combined} and Table~\ref{tab:CDF_errors} summarize the overall tracking accuracy of WiRSSI under three trajectories. The CDF curves indicate that RSSI-based tracking achieves sub-meter median accuracy across all motion patterns. The median XY errors (CDF = 0.5) are 0.905~m, 0.784~m, and 0.785~m, respectively. These results show that, despite the coarse resolution of RSSI, the proposed processing pipeline can reliably recover human motion at meter-level precision. {To quantify tail performance, we also report the 90th-percentile errors (CDF = 0.9), which are 1.678~m, 1.527~m, and 1.321~m for the three trajectories.} As expected, CSI-based tracking provides higher accuracy due to its finer resolution. The corresponding median XY errors are 0.574~m, 0.599~m, and 0.514~m. {The 90th-percentile XY errors are 1.139~m, 1.062~m, and 0.940~m, respectively.} Overall, CSI achieves higher tracking accuracy than RSSI, reducing the median XY error by about 0.26~m on average across the three trajectories.

To provide a clearer illustration of the estimated trajectories, Fig.~\ref{fig:raw_ellipse}, Fig.~\ref{fig:raw_linear}, and Fig.~\ref{fig:raw_rectangle} present the X- and Y-coordinate estimates before and after smoothing. As shown, the raw RSSI-based coordinates exhibit substantial noise and wide scattering, whereas the raw CSI-based coordinates are more concentrated. After applying the Hampel outlier removal and SG smoothing, both RSSI and CSI trajectories align closely with the mmWave radar ground truth. The RSSI traces, in particular, show a dramatic reduction in jitter and reveal a clear motion trend. These observations confirm the effectiveness of the proposed RSSI preprocessing and smoothing strategy in improving trajectory stability and consistency. In addition, Fig.~\ref{fig:traj_ellipse}, Fig.~\ref{fig:traj_linear}, and Fig.~\ref{fig:traj_rectangle} visualize the reconstructed 2D trajectories, respectively. The RSSI-based trajectories exhibit larger deviations and local distortions, but the overall motion shape is still preserved. In contrast, CSI-based trajectories follow the mmWave ground truth more closely. Overall, these results demonstrate that WiRSSI enables meaningful trajectory tracking using only RSSI measurements.

\subsection{Human Reflection Coefficient Estimation}
Fig.~\ref{fig:cdf_range_ratio} presents the estimated reflection coefficient ratio $\gamma$ for different Tx-Rx spacings (1-4 m) and across multiple subjects. For Tx-Rx spacings of 2-4 m, the distribution of $\gamma$ is highly concentrated, with median values around 0.01 and a variance of approximately 0.004. This indicates that, in typical mid-range Tx-Rx deployments, the human reflection coefficient remains relatively stable across different subjects and motions. In contrast, when the Tx-Rx spacing is reduced to 1~m, the estimated $\gamma$ increases noticeably, with a median value around 0.015 and a visibly larger spread. This result is consistent with the analysis in Appendix~\ref{appendix:NLOS}. Specifically, when the transmitter and receiver are close, the bistatic delay ratio $(\tau^{T \rightarrow X} + \tau^{X \rightarrow R}) / (\tau^{T \rightarrow X} \cdot \tau^{X \rightarrow R})$ becomes more sensitive to target position changes. As a result, the geometry-dependent term absorbed into $\gamma$ exhibits larger variations, leading to increased uncertainty in the estimated reflection coefficient. Overall, these results indicate that a pre-calibrated reflection coefficient ratio is feasible for amplitude-based delay estimation when the Tx-Rx spacing is not excessively small. Moderate Tx-Rx separations (e.g., 2-4 m) provide a favorable trade-off between maintaining a dominant propagation path and ensuring the stability of $\gamma$.

{
In addition to Tx-Rx spacing, we further examine the sensitivity of the reflection coefficient to target orientation and measurement position. Fig.~\ref{fig:geom_map} shows the experimental geometry with six locations, and Fig.~\ref{fig:rcs_pd} summarizes the estimated human reflection coefficient across these positions under three representative orientations (front, side, and back) in a different experimental environment from the previous experiments. We observe a median value close to 0.01, consistent with the results reported earlier, and the distributions remain relatively concentrated across both orientations and positions. While orientation changes can introduce local fluctuations because different body parts may dominate scattering at different times, the overall variation is moderate and does not change the dominant trend. These results support the quasi-static approximation adopted in Section \ref{subsec:delay_estimation} for typical indoor tracking, where a pre-calibrated human reflection coefficient can be treated as a prior for amplitude-based range estimation.
}

\begin{table*}[t]
\centering
\small
\caption{{Average processing time per CPI under the same CPI length and step size.}}
\label{tab:runtime}
\begin{tabular}{lcc}
\toprule
Method & WiRSSI (RSSI) & SRCC (CSI)\cite{wang2025towards}  \\
\midrule
Time per CPI (ms) & 0.2 & 12  \\
\bottomrule
\end{tabular}
\vspace{-1em}
\end{table*}

{
\subsection{Computational Complexity and Runtime}
WiRSSI mainly consists of a temporal FFT for Doppler extraction and an angle-spectrum computation for AoA estimation on three RSSI streams within each CPI, followed by peak selection and lightweight smoothing. With a CPI length of $M$ and an AoA grid size of $N$, the dominant cost is a Doppler FFT of length $M$ applied to three streams and an AoA evaluation over $N$ angular bins on a 3-element array, giving an overall complexity on the order of $\mathcal{O}(3M\log M + MN)$, where $M=128$ and $N=64$ in our default setting. To substantiate the computational overhead, we profile the average processing time per CPI for WiRSSI and a representative CSI-based baseline (SRCC) on the same CPU platform using identical CPI length and step size. The measured average per-CPI processing time is 0.2~ms for WiRSSI and 12~ms for SRCC. Both methods can run in real time under our default configuration, while WiRSSI is substantially more efficient. The higher runtime of SRCC mainly comes from its delay-domain processing, which performs MVDR-based beamforming over a discretized delay grid in addition to the Doppler FFT operations. This additional grid search and matrix operations introduce significantly higher computational overhead than the our FFT-based processing in WiRSSI.
}

\begin{figure*}
\centering
\begin{minipage}{0.38\textwidth}
    \centering
    \includegraphics[width=\linewidth]{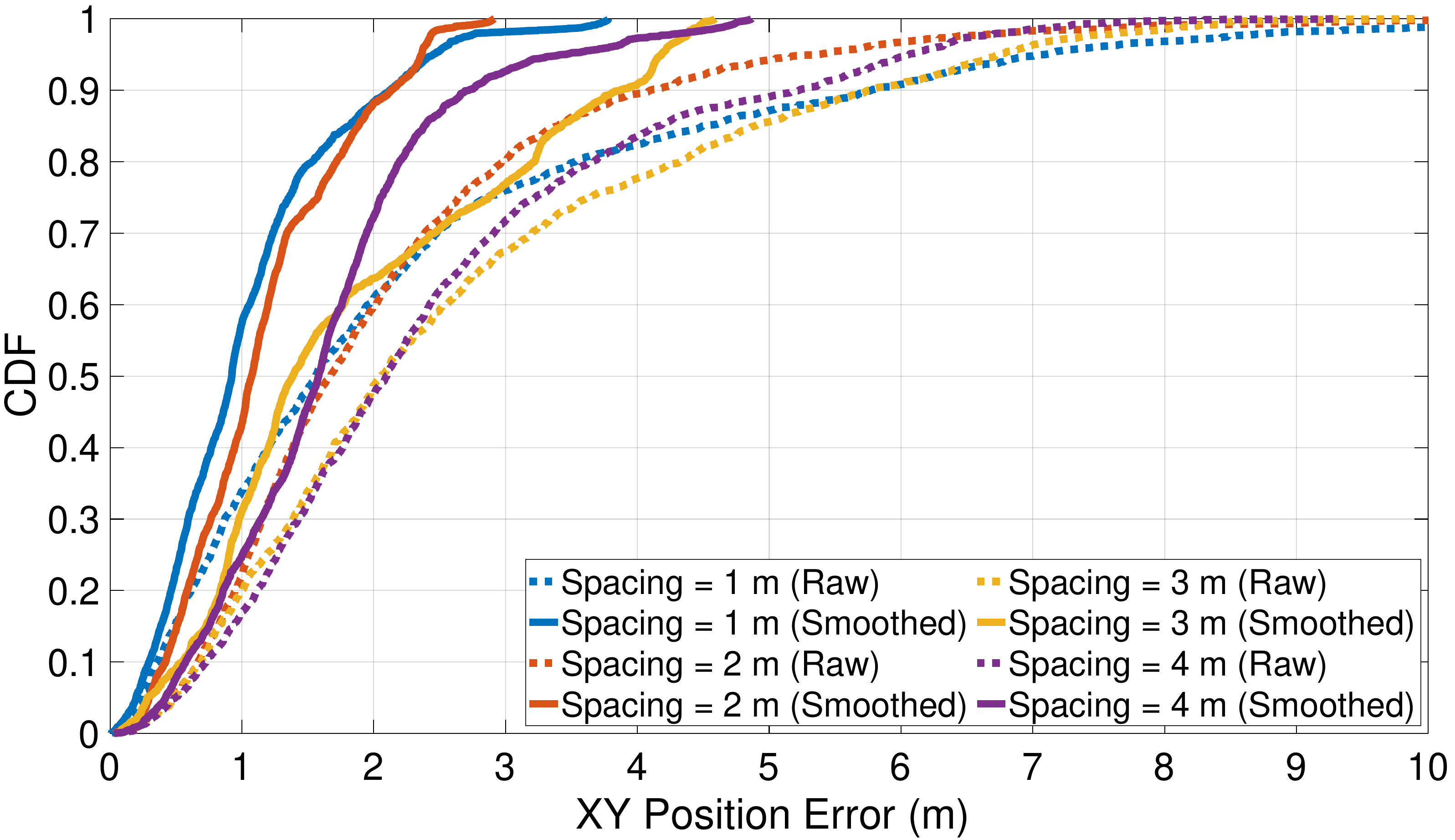}
    \caption{Impact of Tx-Rx distance.}
    \label{fig:cdf_rssi_xy}
\end{minipage}
~
\begin{minipage}{0.38\textwidth}
    \centering
    \includegraphics[width=\linewidth]{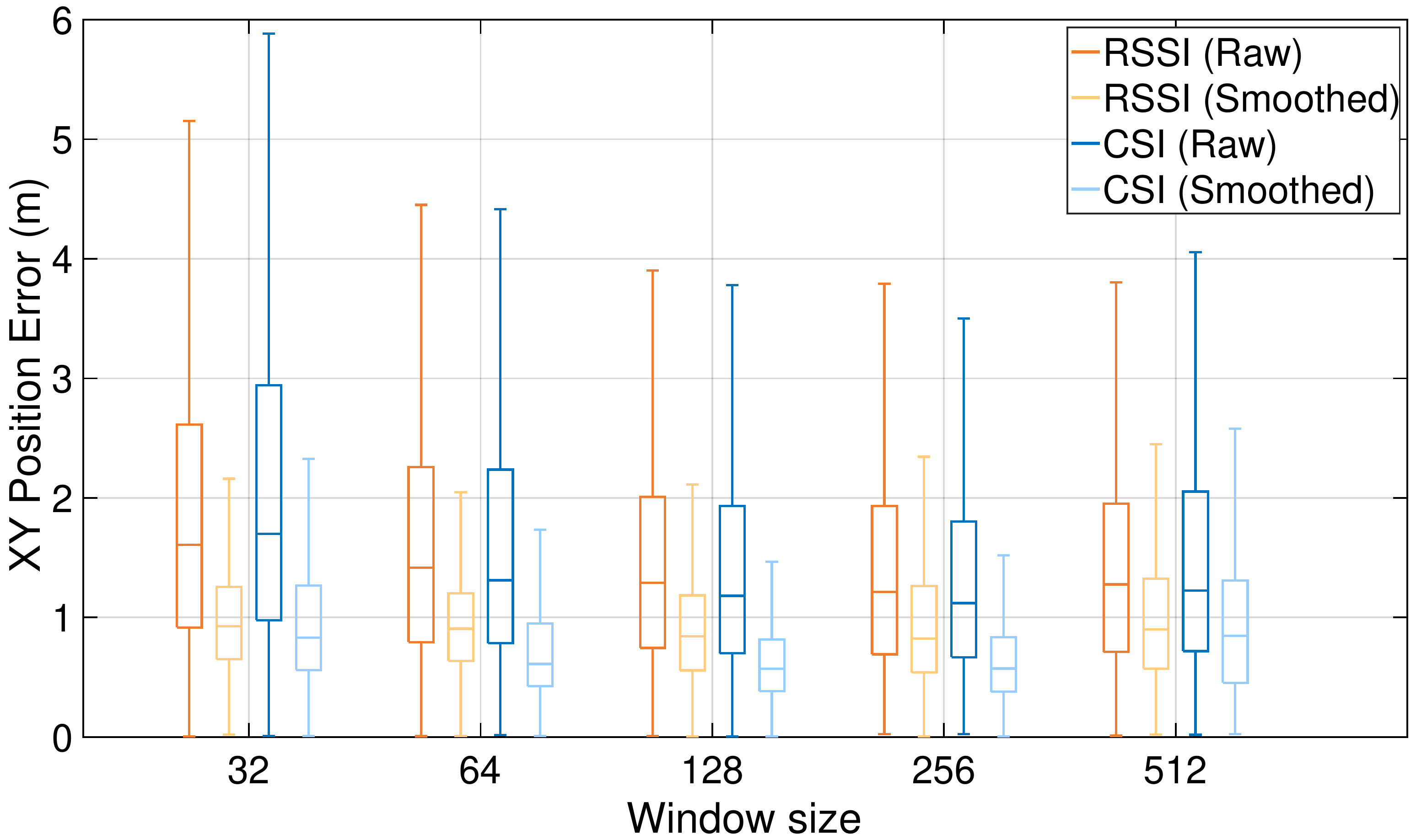}
    \caption{Impact of CPI size.}
    \label{fig:box_xy_window}
\end{minipage}
 \vspace{-1.5em}
\end{figure*}

\subsection{Impact of Tx-Rx Distance}
Fig.~\ref{fig:cdf_rssi_xy} illustrates the raw and smoothed XY tracking errors for Tx-Rx spacings ranging from 1 to 4 m. As the spacing increases, the tracking error gradually rises, primarily due to the increased impact of environmental multipath propagation. With larger separations, reflections from walls, floors, and surrounding objects become more prominent relative to the direct path, weakening the dominance of the static Tx-Rx component. In addition, the increased propagation distance leads to higher attenuation and noise, further degrading the reliability of AoA and delay estimation. Nevertheless, the proposed RSSI-based sensing approach remains robust under increased multipath interference.

\subsection{Impact of CPI Size}
Fig.~\ref{fig:box_xy_window} shows the XY tracking errors for CPI sizes ranging from 32 to 512 samples. Overall, both RSSI- and CSI-based tracking benefit from increasing the CPI size, as larger windows improve Doppler resolution and suppress short-term fluctuations. For RSSI, the smoothed median error decreases from 0.925 m at 32 samples to 0.824 m at 256 samples, after which the improvement saturates and slightly degrades at 512 samples (0.899 m). This suggests that while longer windows enhance Doppler-AoA stability, excessively large CPIs blur fast motion dynamics. A similar trend is observed for CSI. Consequently, a moderate CPI size (128-256 samples) provides a favorable trade-off for accurate sensing.

\begin{figure*}
    \centering
    \begin{minipage}[t]{0.329\linewidth}
    \vspace{0pt}
        \centering
        \begin{subfigure}{\textwidth}
            \includegraphics[width=\textwidth]{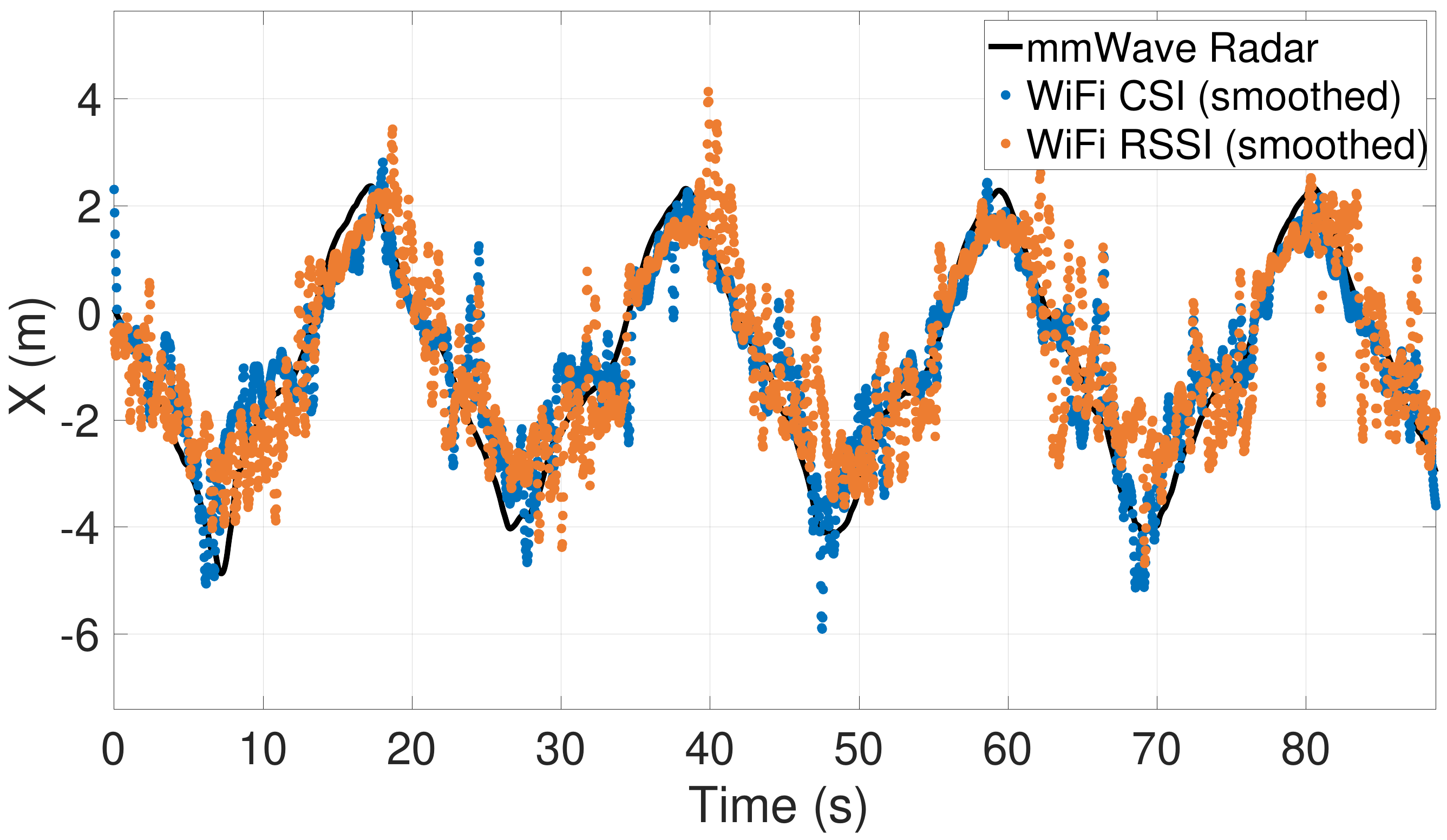}
            \subcaption{Smoothed x-axis coordinate}
            \label{fig:sg11_x}
        \end{subfigure}\\
        \vspace{.5em}
        \begin{subfigure}{\textwidth}
            \includegraphics[width=\textwidth]{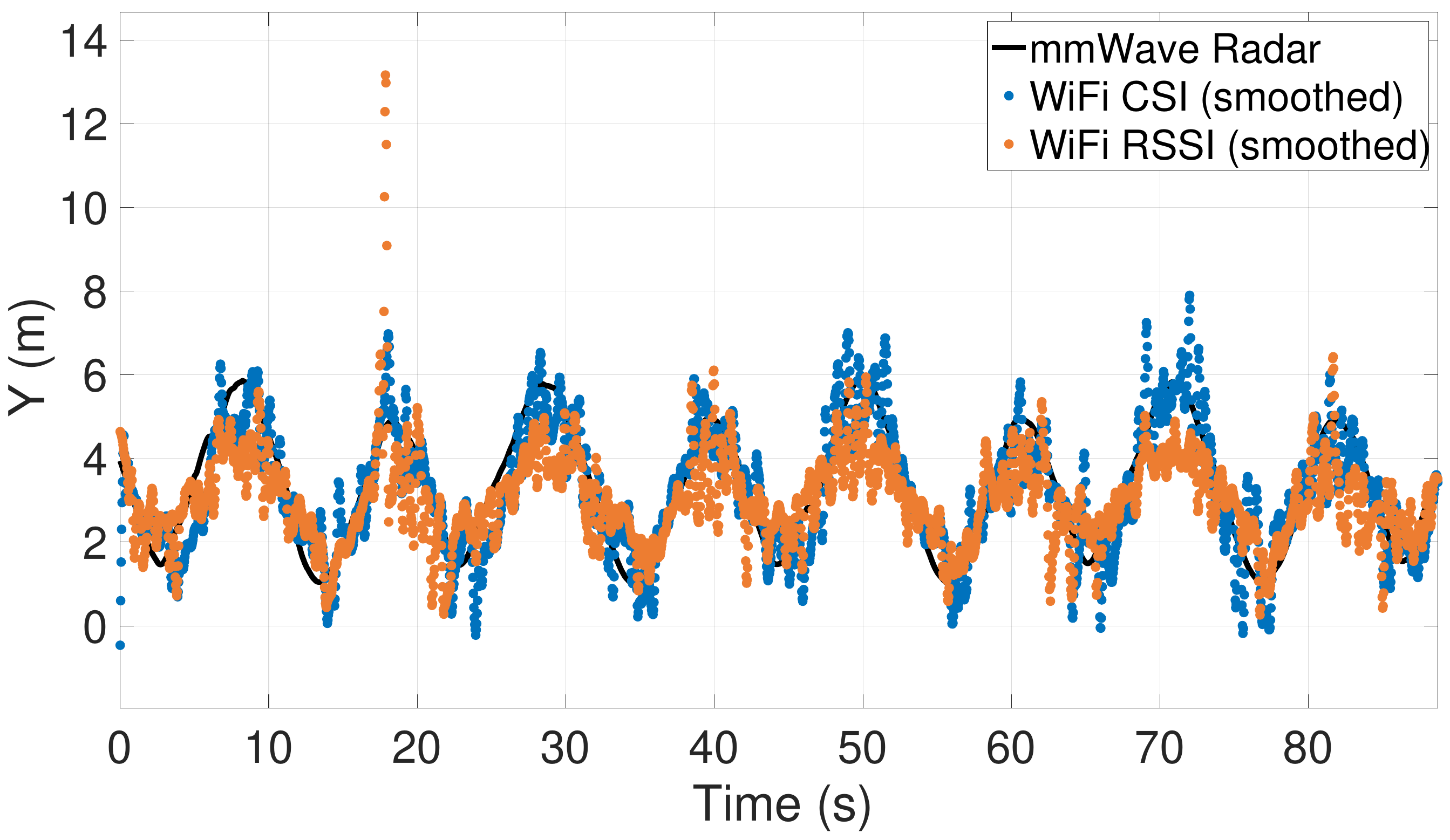}
            \subcaption{Smoothed y-axis coordinate}
            \label{fig:sg11_y}
        \end{subfigure}
        \caption{{SG window = 11 CPIs.}}
        \label{fig:sg11}
    \end{minipage}
    \begin{minipage}[t]{0.329\linewidth}
    \vspace{0pt}
        \centering
        \begin{subfigure}{\textwidth}
            \includegraphics[width=\textwidth]{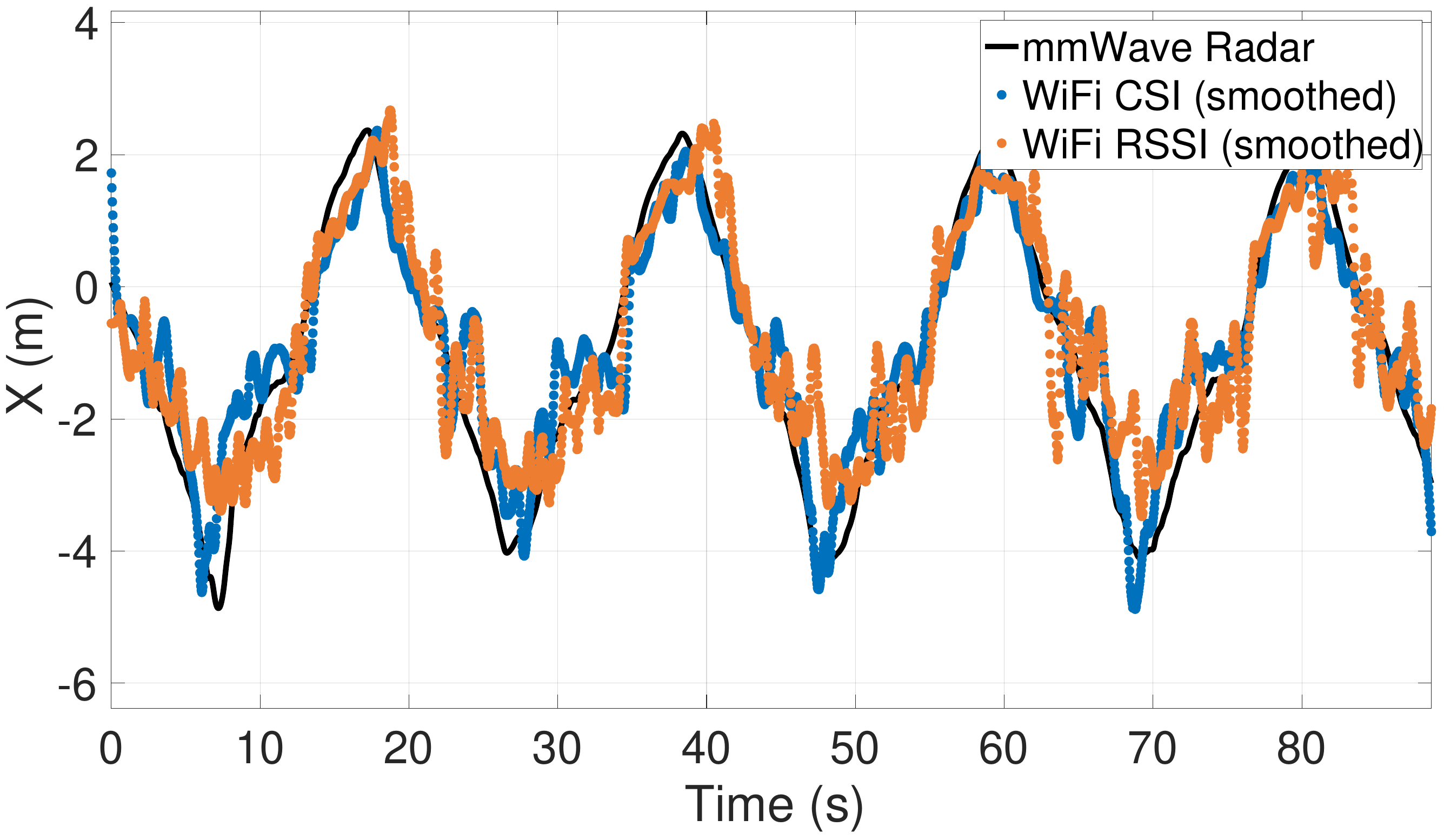}
            \subcaption{Smoothed x-axis coordinate}
            \label{fig:sg31_x}
        \end{subfigure}\\
        \vspace{.5em}
        \begin{subfigure}{\textwidth}
            \includegraphics[width=\textwidth]{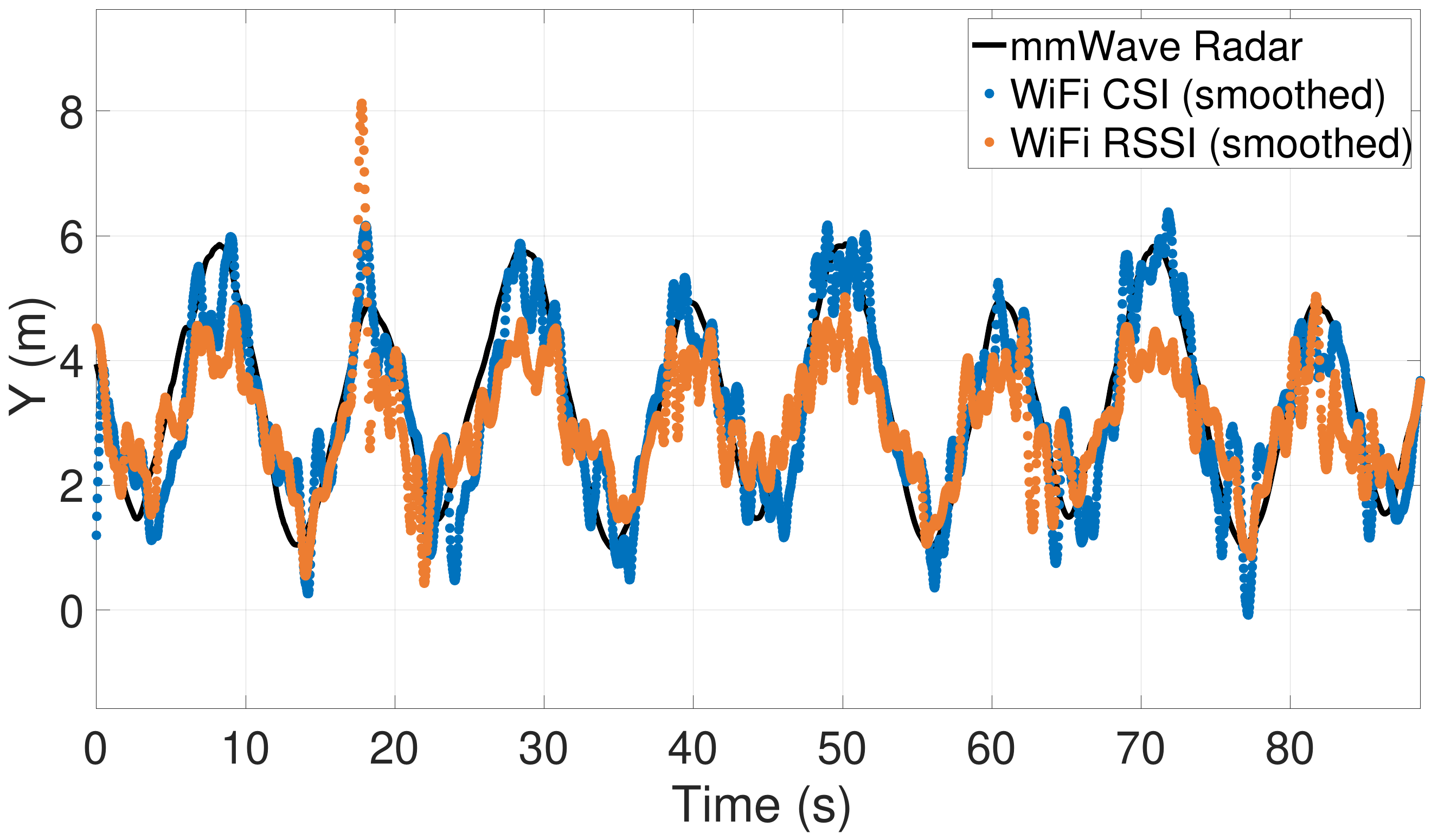}
            \subcaption{Smoothed y-axis coordinate}
            \label{fig:sg31_y}
        \end{subfigure}
        \caption{{SG window = 31 CPIs.}}
        \label{fig:sg31}
    \end{minipage}
    \begin{minipage}[t]{0.329\linewidth}
    \vspace{0pt}
        \centering
        \begin{subfigure}{\textwidth}
            \includegraphics[width=\textwidth]{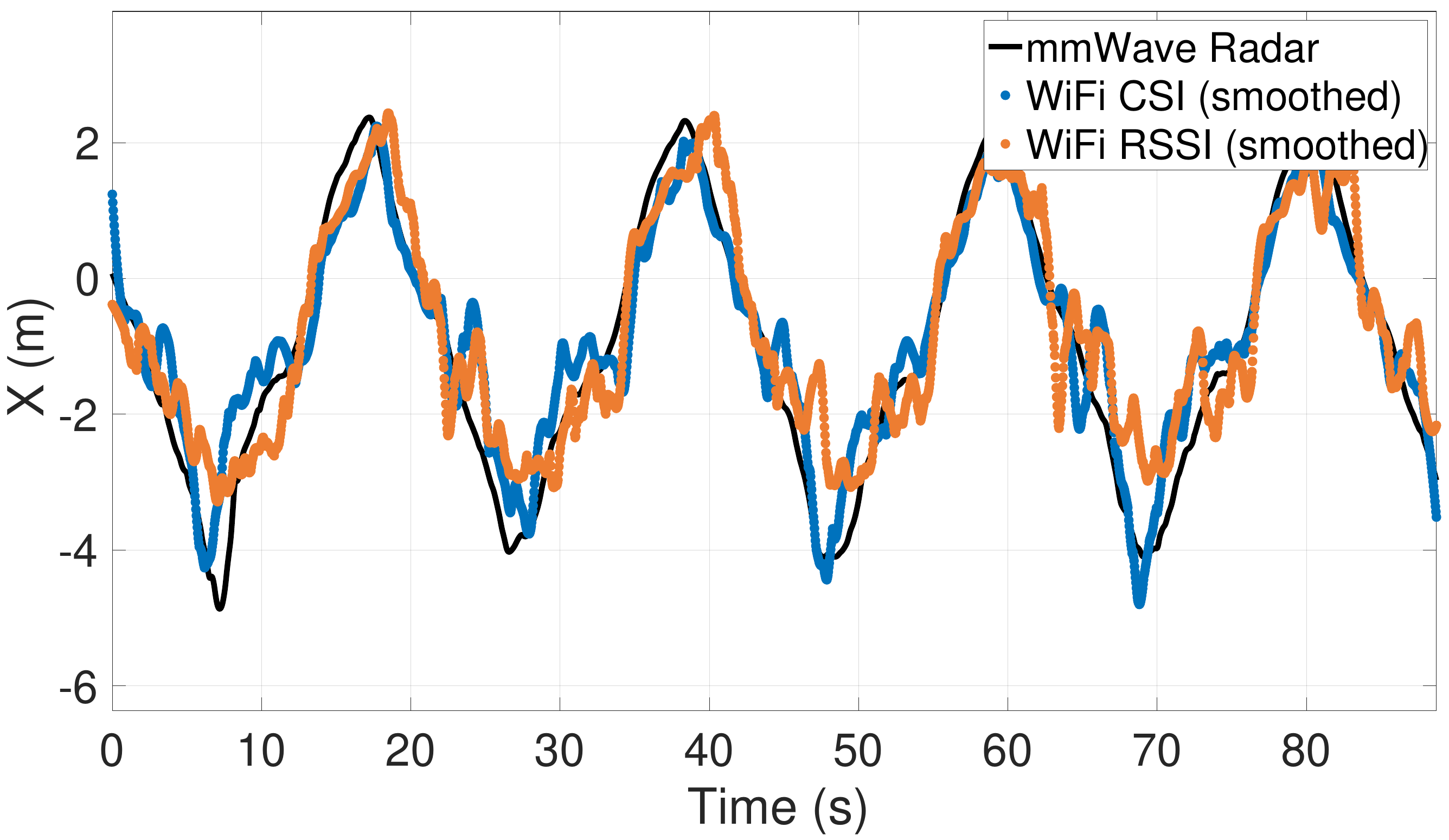}
            \subcaption{Smoothed x-axis coordinate}
            \label{fig:sg51_x}
        \end{subfigure}\\
        \vspace{.5em}
        \begin{subfigure}{\textwidth}
            \includegraphics[width=\textwidth]{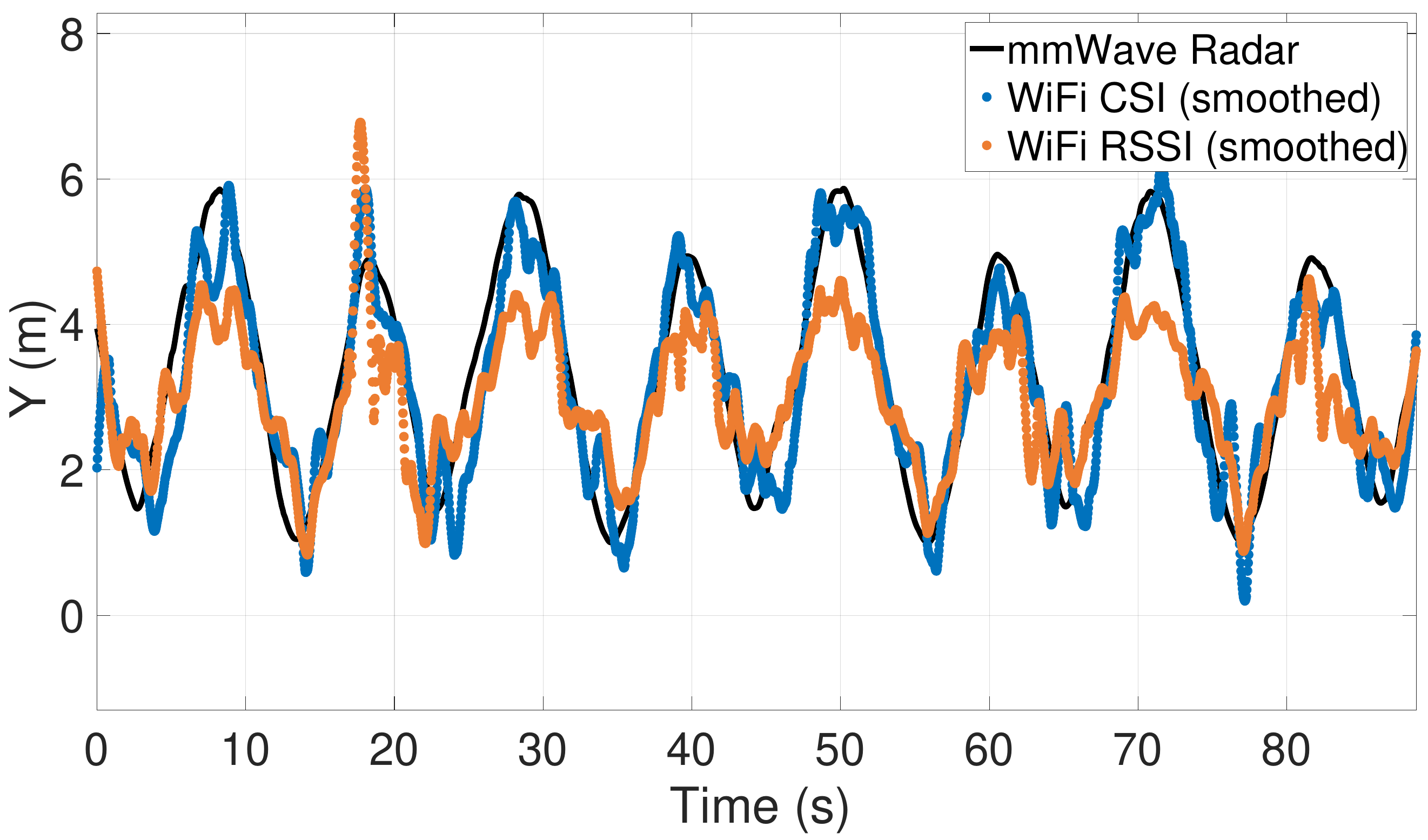}
            \subcaption{Smoothed y-axis coordinate}
            \label{fig:sg51_y}
        \end{subfigure}
        \caption{{SG window = 51 CPIs.}}
        \label{fig:sg51}
    \end{minipage}
    \vspace{-1.5em}
\end{figure*}

{
\subsection{Impact of Smoothing Window Length}
We evaluate the latency-stability trade-off of trajectory smoothing under the elliptical trajectory by varying the SG window length while keeping all other settings unchanged. With a CPI step size of 32~ms, window lengths of 11, 31, and 51 correspond to smoothing over approximately 0.35~s, 0.99~s, and 1.63~s of estimated data, respectively. Fig.~\ref{fig:sg11}, Fig.~\ref{fig:sg31}, and Fig.~\ref{fig:sg51} show the resulting $x$- and $y$-coordinate trajectories. As expected, larger windows suppress jitter and intermittent outliers more effectively but introduce higher latency. A short window yields lower delay with noisier trajectories, whereas moderate windows provide smoother and more stable tracking. These results indicate that WiRSSI can operate with substantially shorter smoothing windows when low latency is required.}

\begin{figure*}[t]
    \centering
    \begin{subfigure}[t]{0.245\textwidth}
        \centering
        \includegraphics[width=\textwidth]{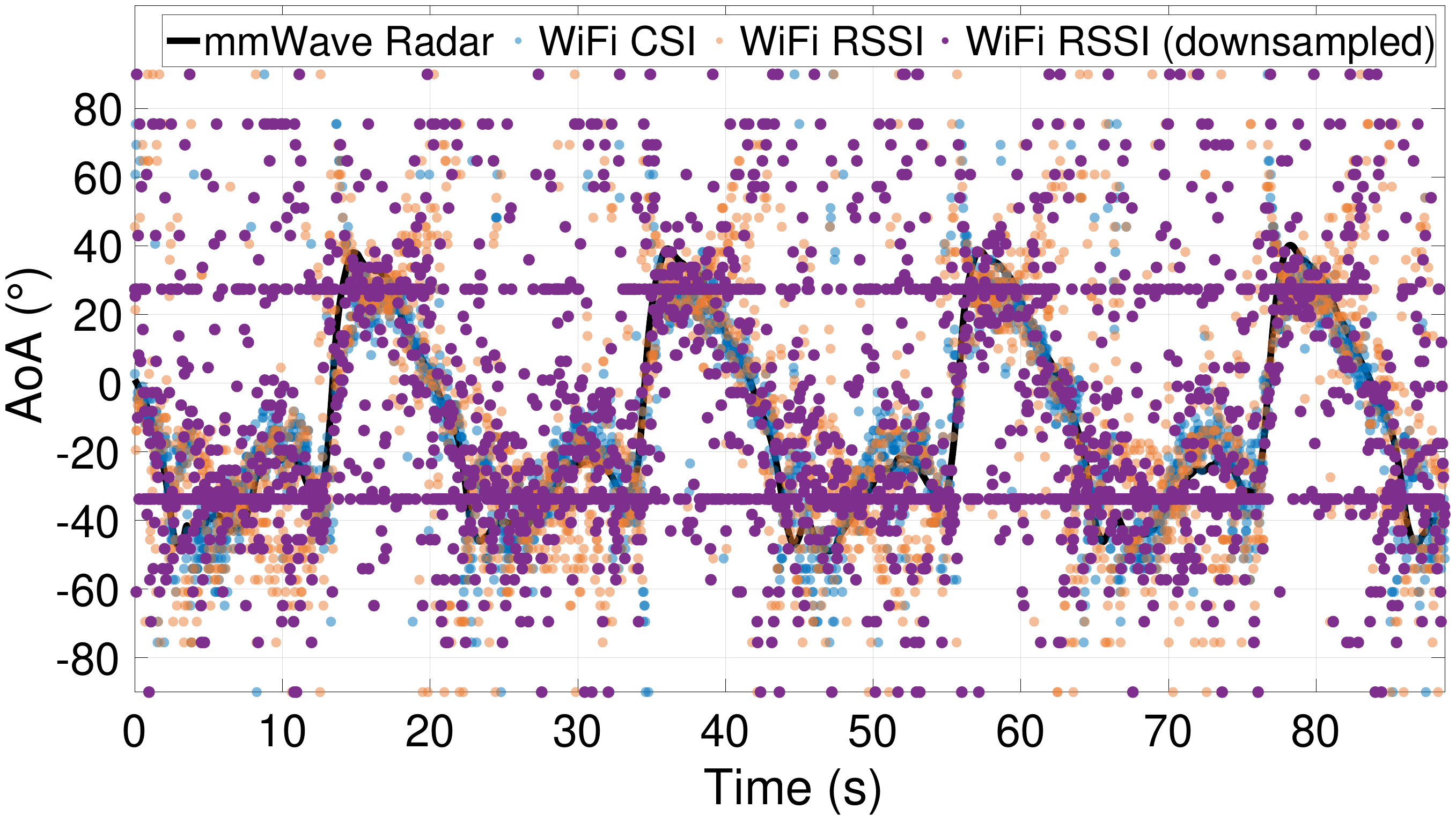}
        \subcaption{Raw AoA (3 samples)}
        \label{fig:sampling_aoa_incpi3}
    \end{subfigure}
    \begin{subfigure}[t]{0.245\textwidth}
        \centering
        \includegraphics[width=0.99\textwidth]{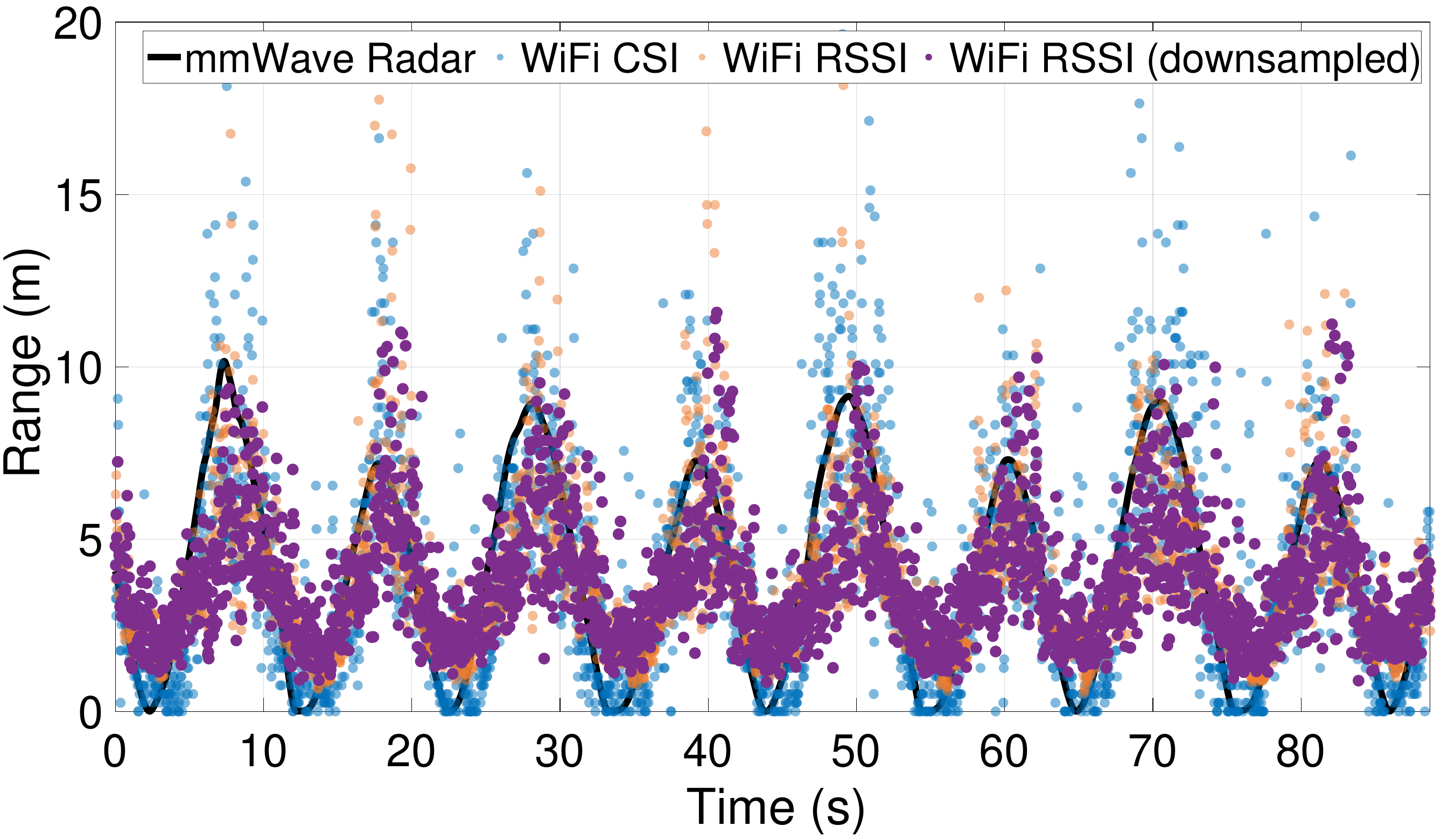}
        \subcaption{Raw Range (3 samples)}
        \label{fig:sampling_range_incpi3}
    \end{subfigure}
    \begin{subfigure}[t]{0.245\textwidth}
        \centering
        \includegraphics[width=\textwidth]{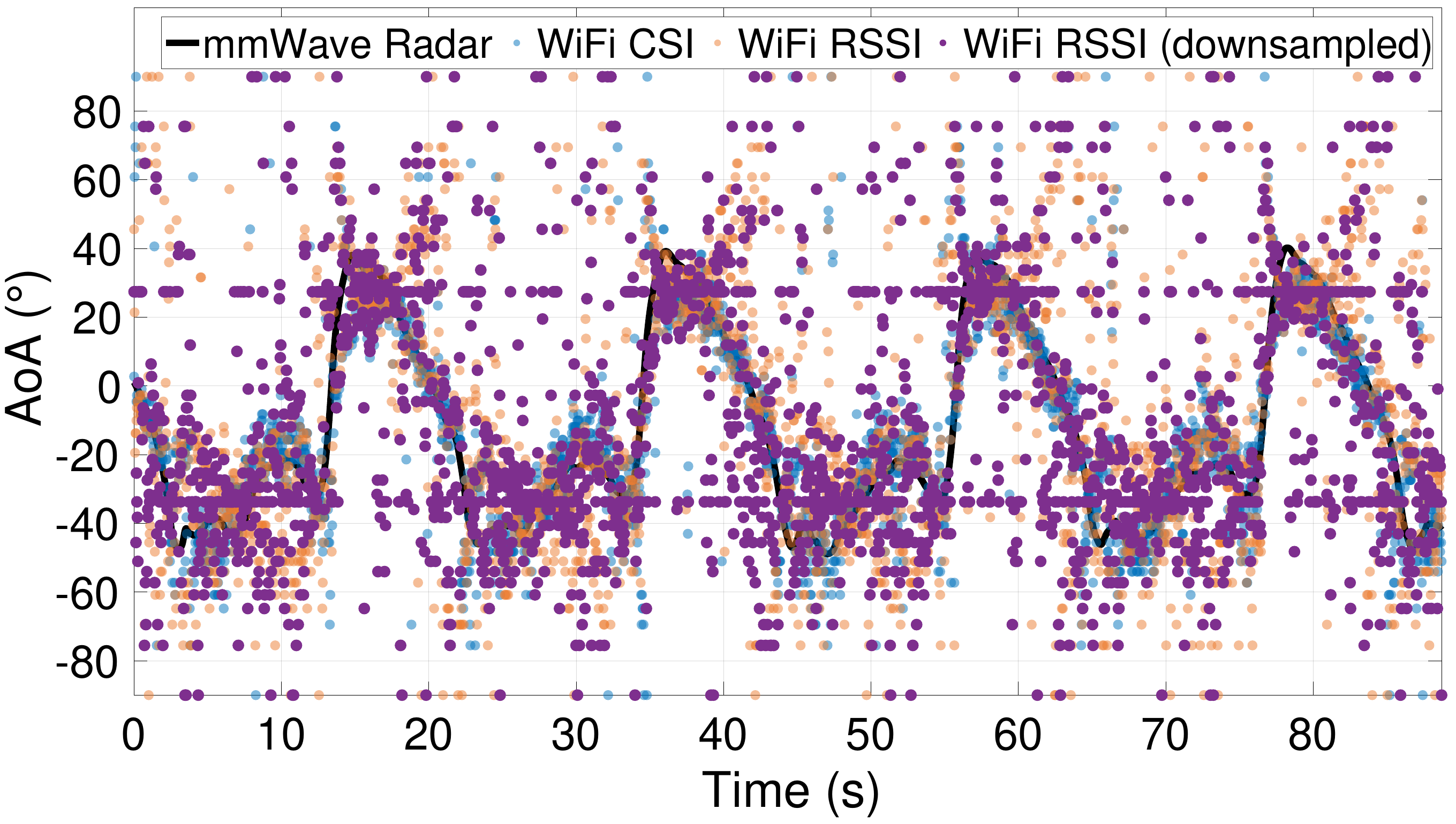}
        \subcaption{Raw AoA (16 samples)}
        \label{fig:sampling_aoa_incpi16}
    \end{subfigure}
    \begin{subfigure}[t]{0.245\textwidth}
        \centering
        \includegraphics[width=0.99\textwidth]{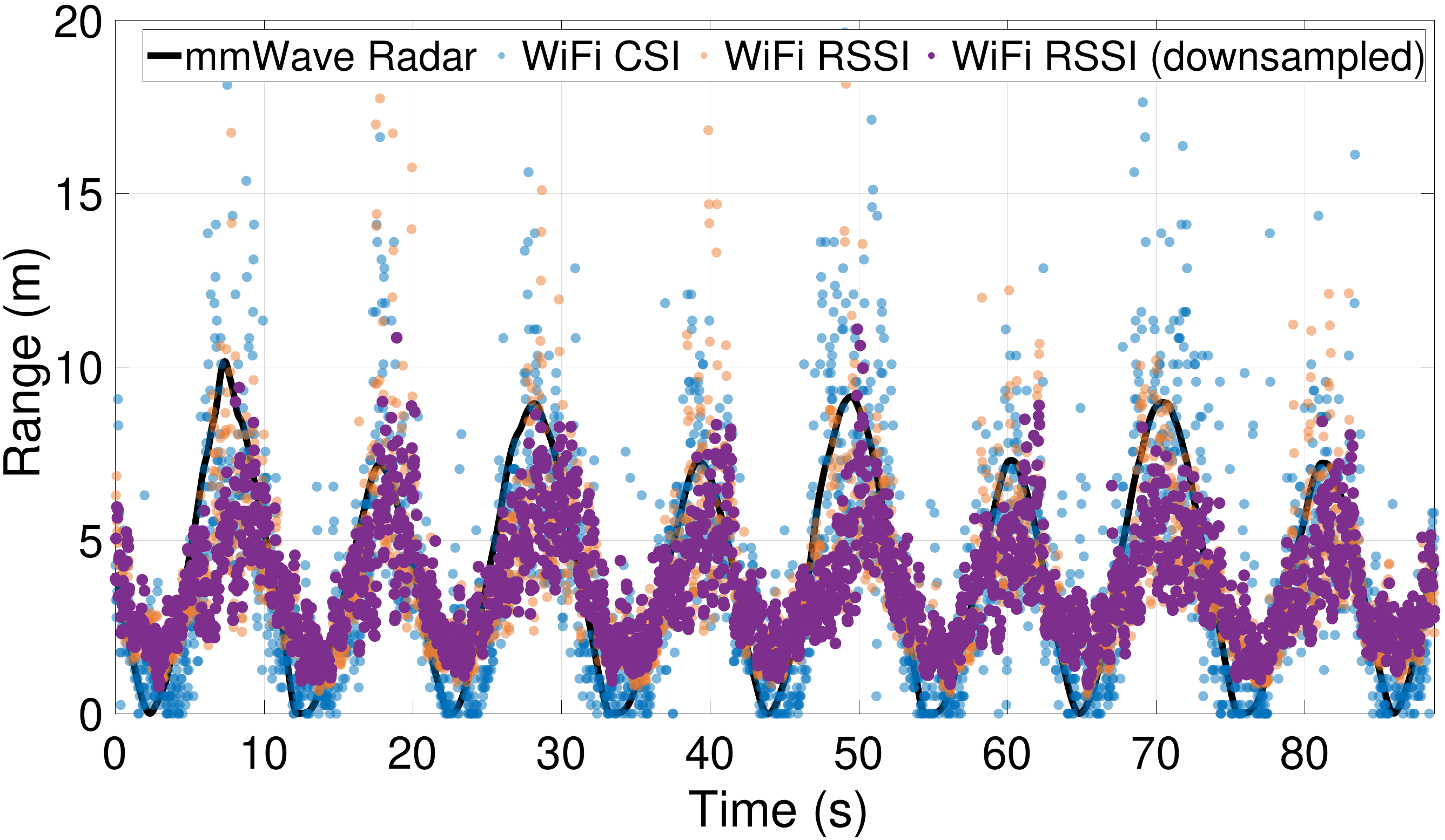}
        \subcaption{Raw Range (16 samples)}
        \label{fig:sampling_range_incpi16}
    \end{subfigure}
    \caption{{Impact of RSSI sampling rate on AoA and range estimation for the elliptical trajectory, comparing CPIs containing 3 and 16 RSSI samples, respectively.}}
    \label{fig:impact_sampling_rate_aoa_range}
    \vspace{-1.2em}
\end{figure*}

{
\subsection{Impact of RSSI Sampling Rate}
Our system records RSSI at 1~kHz in monitor mode. To isolate the impact of RSSI sampling rate, we keep the CPI length fixed to 128 samples and emulate lower-rate RSSI logging by down-sampling within each CPI. We consider two representative cases, where each CPI contains 3 or 16 RSSI samples, corresponding to much lower and moderate effective sampling rates, respectively, compared with the original setting that uses all 128 RSSI samples per CPI. The step size is kept at 32 samples, identical to the default setting, to generate a dense CPI sequence and enable a direct comparison of variation trends. This emulation is motivated by practical chipset constraints, as many commodity devices typically do not expose kHz-level RSSI reporting. In addition, under these lower sampling rates, we replace CPI-mean static suppression with an exponential moving average (EMA) to update the static component online and handle slow drifts. The EMA assigns a weight of 0.3 to the newly observed RSSI sample and 0.7 to the previous static estimate. All other processing steps are unchanged.

Fig.~\ref{fig:impact_sampling_rate_aoa_range} compare the AoA and range estimates for the elliptical motion under CPIs containing 3 and 16 RSSI samples. The original high-rate results using all 128 RSSI samples per CPI, together with the CSI results and the mmWave ground truth, are included as baselines. For range estimation, the human reflection coefficient is calibrated using the mmWave ground truth. As expected, both AoA and range become noisier with fewer samples per CPI due to reduced temporal resolution and weaker Doppler separation. Nevertheless, the dominant trends remain consistent with the mmWave ground truth, indicating that the proposed RSSI pipeline can still recover meaningful geometric cues with tens-of-Hz sampling. More frequent degradations are observed when the target radial velocity approaches zero, where Doppler separation from near-zero residual clutter is inherently weaker. Overall, these results highlight a practical performance trade-off with sampling rate while confirming that WiRSSI remains functional under realistic low-rate RSSI logging. 
}

\begin{table*}[t]
\centering
\small
\caption{{Performance comparison on Dataset 1 under a random split setting (70\% training, 30\% testing). \textit{Acc.}: Accuracy, \textit{Prec.}: Macro Precision, \textit{Rec.}: Macro Recall, \textit{F1}: Macro F1-score. All results use MobileViT-XXS.}}
\label{tab:interaction_single_performance}
\renewcommand{\arraystretch}{1.2}
\begin{tabular}{c|c|c|cccc}
\toprule
\textbf{Channel Metric} & \textbf{Feature} & \textbf{Minimum Setup} &
\textbf{Acc.} & \textbf{Prec.} & \textbf{Rec.} & \textbf{F1} \\
\midrule
\multirow{2}{*}{\textbf{CSI}}
& BVP (Doppler) \cite{zhang2021widar3} & Multi-Receiver & 0.850 & 0.849 & 0.849 & 0.849 \\
& \textbf{SRCC (Doppler-delay)}\cite{wang2025towards} & 1Tx-1Rx & \textbf{0.939} & \textbf{0.938} & \textbf{0.938} & \textbf{0.938} \\
\midrule
\textbf{RSSI}
& WiRSSI (Doppler-AoA) & 1Tx-3Rx & 0.753 & 0.751 & 0.751 & 0.751 \\
\bottomrule
\end{tabular}
\vspace{-1.5em}
\end{table*}

\begin{figure*}[t]
    \centering
    \begin{subfigure}[t]{0.325\textwidth}
        \centering
        \includegraphics[width=\textwidth]{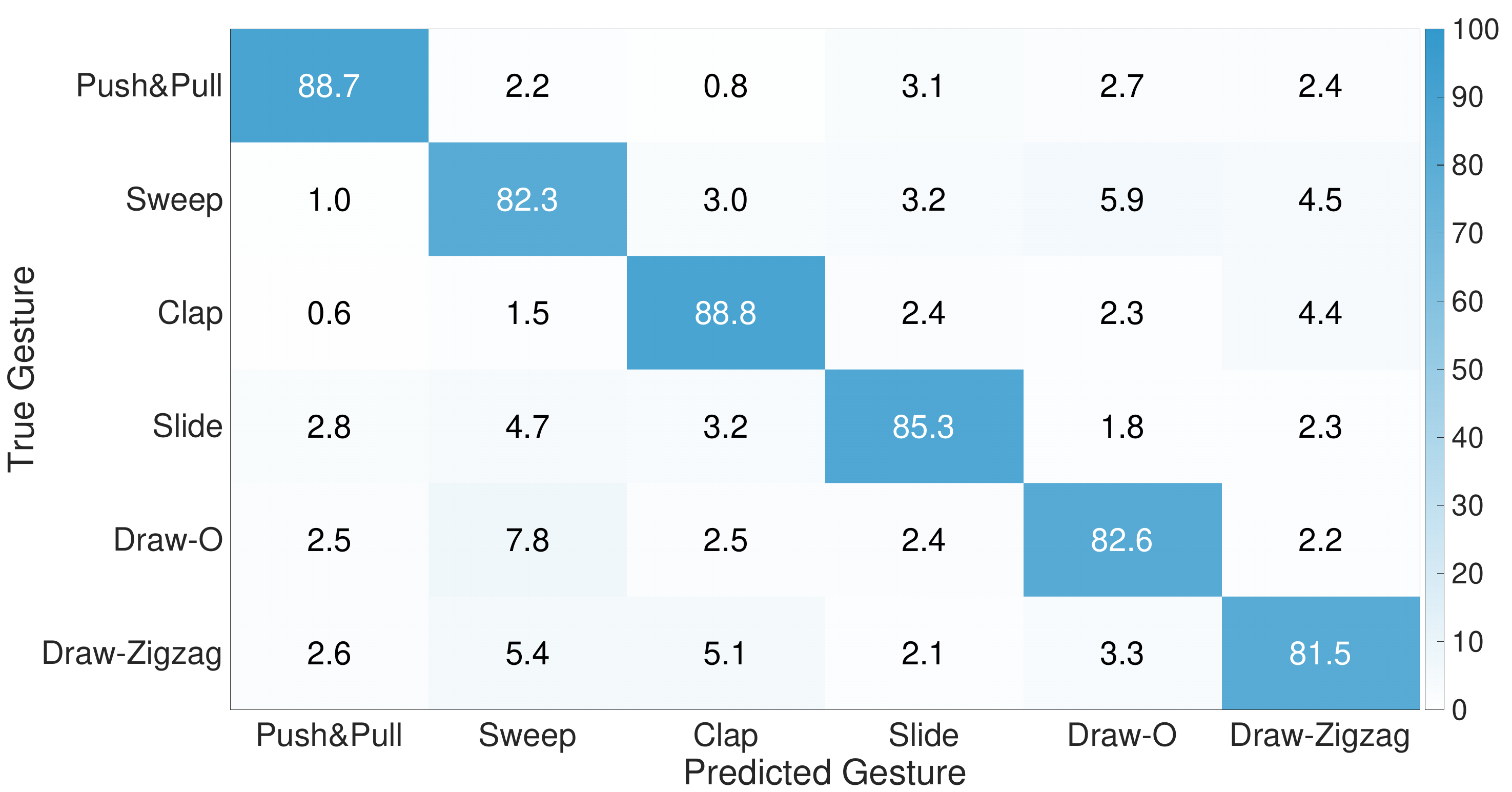}
        \subcaption{CSI-based BVP}
        \label{fig:cm_bvp}
    \end{subfigure}
    \begin{subfigure}[t]{0.325\textwidth}
        \centering
        \includegraphics[width=\textwidth]{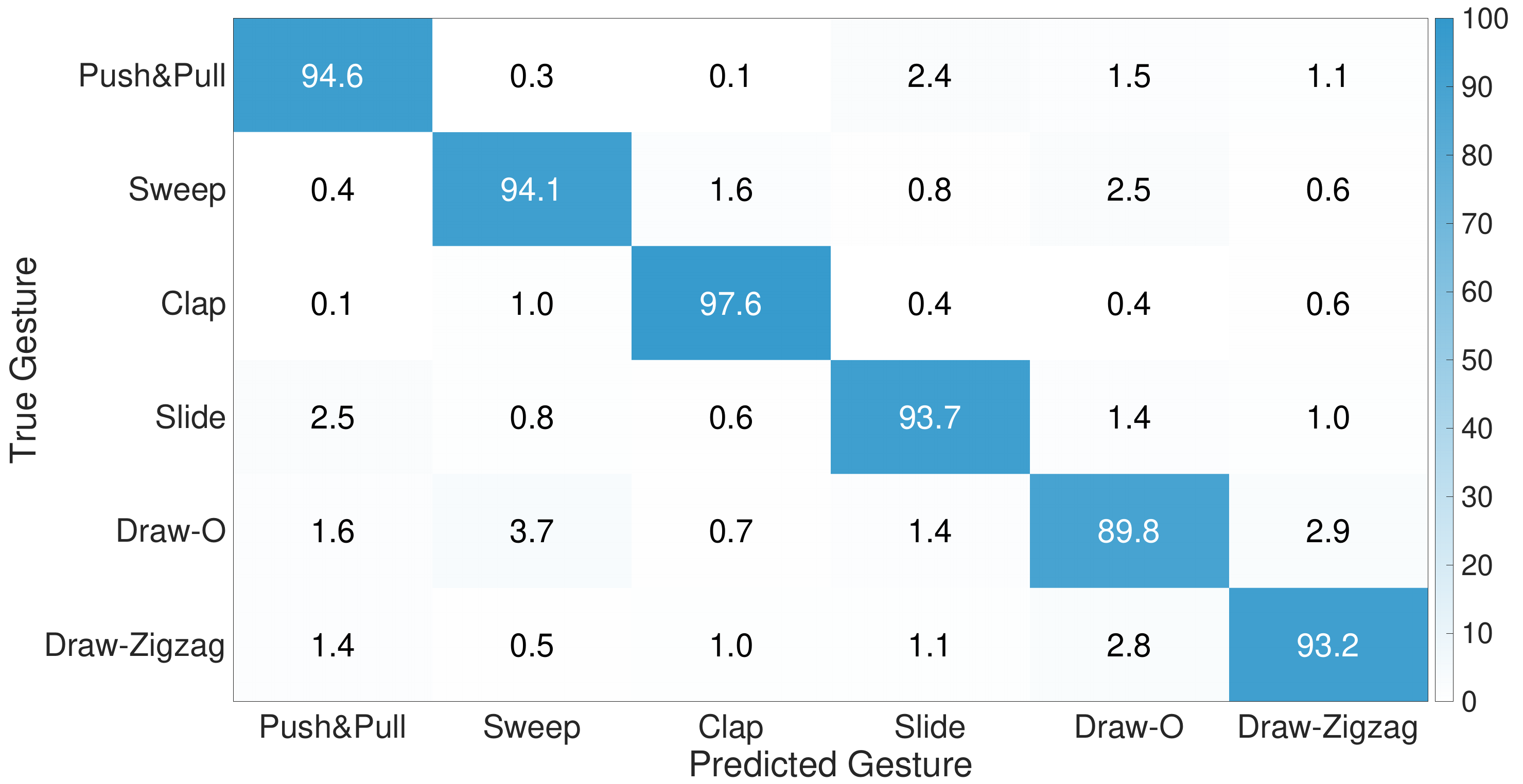}
        \subcaption{CSI-based SRCC}
        \label{fig:cm_srcc}
    \end{subfigure}
    \begin{subfigure}[t]{0.325\textwidth}
        \centering
        \includegraphics[width=\textwidth]{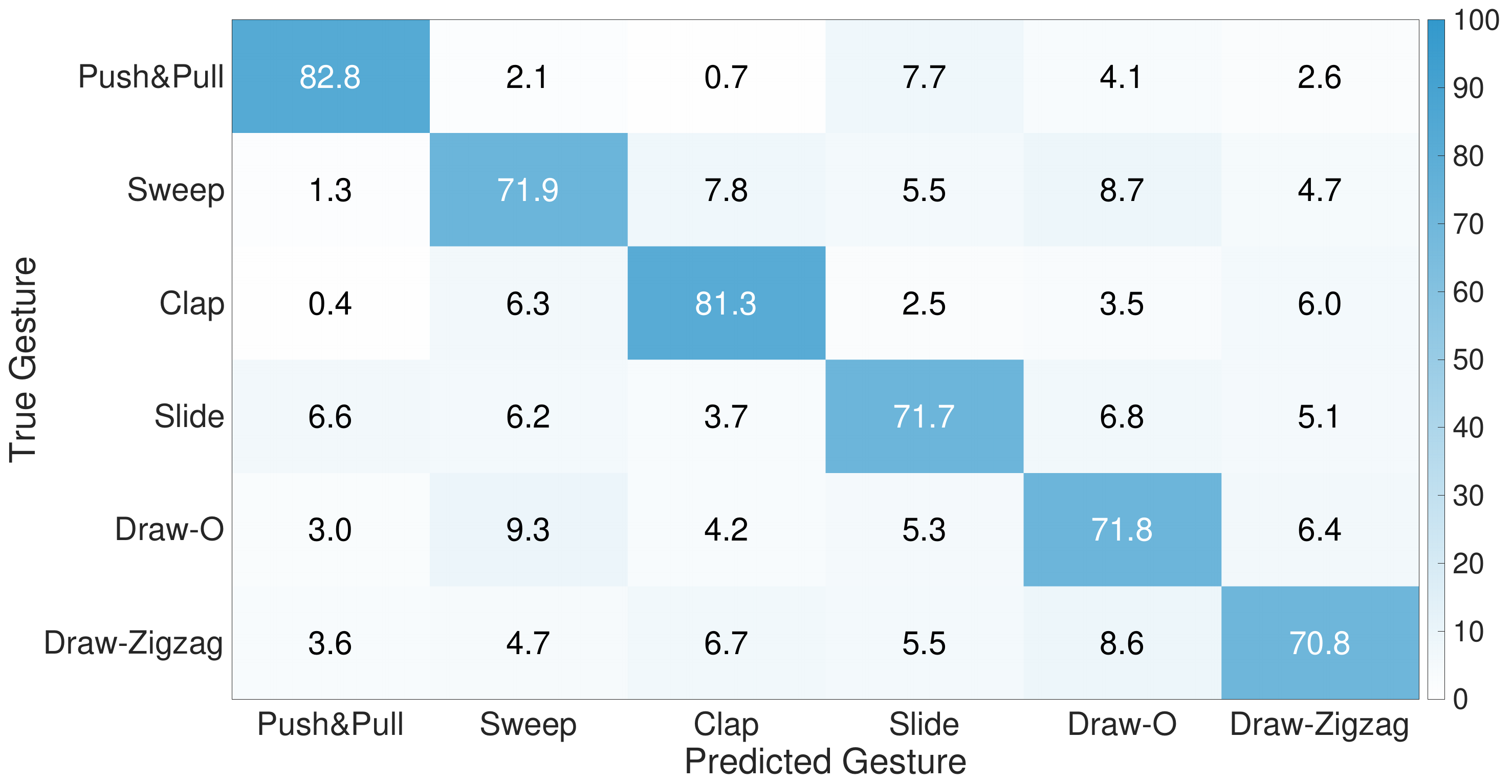}
        \subcaption{RSSI-based WiRSSI}
        \label{fig:cm_wirssi}
    \end{subfigure}
    \caption{{Confusion matrices on Dataset~1 under a random split setting (70\% training, 30\% testing) using MobileViT-XXS, comparing CSI-based features (BVP and SRCC) and the RSSI-based feature (WiRSSI).}}
    \label{fig:cm_dataset1_randomsplit}
    \vspace{-1.5em}
\end{figure*}

{
\subsection{Gesture Recognition}
\label{subsec:gesture_recognition}
We conduct a gesture recognition study on the Widar3.0 dataset (Dataset~1) \cite{zhang2021widar3} and compare WiRSSI features with representative CSI-based features.

\subsubsection{Overall Performance Comparison}
We first consider a conventional random split setting, where the samples of each gesture class are randomly divided such that 70\% are used for training and the remaining 30\% are used for testing. Table~\ref{tab:interaction_single_performance} summarizes the overall performance. CSI-based features achieve the best results, with SRCC (Doppler-delay) reaching an accuracy of 0.939 and a macro F1-score of 0.938, while the Doppler-based BVP feature attains an accuracy of 0.850 and a macro F1-score of 0.849. Despite its lower measurement resolution, the proposed RSSI-based WiRSSI (Doppler-AoA) remains effective, achieving an accuracy of 0.753 and a macro F1-score of 0.751. Fig.~\ref{fig:cm_dataset1_randomsplit} further visualizes the confusion matrices. Compared with CSI, WiRSSI shows more confusions between gestures with similar motion dynamics, which is expected given the reduced feature granularity of RSSI. Nevertheless, the matrices remain strongly diagonal-dominant, confirming that RSSI-only sensing can still provide discriminative information for gesture recognition.

\subsubsection{Generalization Evaluation}
We further evaluate cross-domain generalization under a strict setting that better reflects deployment challenges. Specifically, the model is trained only on samples collected at Location~\#1 with Receiver~\#1 (5,224 samples), and tested on samples from all other locations and receivers (160,500 samples). Table~\ref{tab:input_feature_summary} reports the results for multiple CSI-based inputs and our RSSI-based WiRSSI feature. Overall performance drops substantially for all methods due to the strong domain shift induced by changes in location, receiver, and environment. Among CSI-based approaches, SRCC (Doppler-delay) achieves the best generalization performance with an accuracy of 0.767 and a macro F1-score of 0.763, outperforming CSI amplitude and other correlation-based baselines. Under this strict setting, WiRSSI (Doppler-AoA) achieves an accuracy of 0.507 and a macro F1-score of 0.509, which is still competitive with several CSI-based features. While a performance gap to the best CSI method remains, the results indicate that RSSI-only features retain usable motion cues even under severe cross-location and cross-receiver shifts. These findings reinforce WiRSSI as a low-cost complementary sensing modality when CSI access is unavailable or limited, and also highlight that improving generalization, particularly for RSSI, remains an important direction for future work.}

\begin{table*}
\centering
\caption{{Performance comparison of different input features  based on MobileViT-XXS  under a strict generalization setting: the model is trained on samples from Location \#1 and Receiver \#1 (5,224 samples), and tested on samples from all other locations and receivers (160,500 samples).}}
\label{tab:input_feature_summary}
\renewcommand{\arraystretch}{1.2}
\resizebox{\textwidth}{!}{
\begin{tabular}{c|c|c|cccc}
\toprule
\textbf{Channel Metric} & \textbf{Input Feature} & \textbf{Minimum Setup} &
\textbf{Acc.} & \textbf{Macro Prec.} & \textbf{Macro Rec.} & \textbf{Macro F1} \\
\midrule
\multirow{6}{*}{\textbf{CSI}}
& CSI Amplitude \cite{yang2023sensefi}        & 1Tx-1Rx        & 0.366 & 0.362 & 0.357 & 0.355 \\
& BVP (Doppler) \cite{zhang2021widar3}                  & Multi-Receiver  & 0.422 & 0.440 & 0.417 & 0.418 \\
& CACC (Doppler) \cite{qian2018widar2}        & 1Tx-2Rx        & 0.601 & 0.601 & 0.600 & 0.594 \\
& CACC (Doppler-delay)                        & 1Tx-2Rx        & 0.614 & 0.618 & 0.612 & 0.613 \\
& CFCC (Doppler-delay) \cite{11079818}        & 1Tx-1Rx        & 0.565 & 0.565 & 0.561 & 0.558 \\
& \textbf{SRCC (Doppler-delay)} \cite{wang2025towards}               & \textbf{1Tx-1Rx} & \textbf{0.767} & \textbf{0.764} & \textbf{0.765} & \textbf{0.763} \\
\midrule
\multirow{1}{*}{\textbf{RSSI}}
& \textbf{WiRSSI (Doppler-AoA)}               & \textbf{1Tx-3Rx} & \textbf{0.507} & \textbf{0.510} & \textbf{0.510} & \textbf{0.509} \\
\bottomrule
\end{tabular}
}
\vspace{-1em}
\end{table*}

\section{Discussion}
This work challenges the prevailing perception that RSSI is inherently unsuitable for feasible WiFi sensing. Rather than treating RSSI as a coarse indicator for signal strength or distance, our results demonstrate that RSSI, when interpreted through an appropriate power-domain model, can implicitly preserve motion-related information that is traditionally accessed through CSI. By consolidating the experimental observations, several key insights and trade-offs emerge.

\subsection{Physical Basis of RSSI-based Sensing}
A central finding of this work is that the sensing capability of RSSI originates from the interaction between the dominant static path and the target-induced dynamic path in the power domain in practical multipath environments. Although RSSI discards explicit phase information by aggregating power across subcarriers, the resulting cross-term encodes Doppler-, AoA-, and delay-dependent variations. As a result, meaningful motion signatures remain observable even under coarse quantization. In contrast to conventional RSSI models that rely solely on large-scale path loss, the proposed framework exploits these fine-grained amplitude fluctuations to infer geometric information associated with target motion.

\subsection{Performance Gap Between RSSI and CSI}
Experimental results consistently show that CSI-based sensing outperforms RSSI-based sensing in terms of accuracy. This gap primarily stems from the loss of subcarrier-level phase diversity and the limited amplitude resolution of RSSI. In particular, Doppler estimation from RSSI is inherently noisier, which constrains target separability in the Doppler domain. Nevertheless, the RSSI-derived AoA and delay estimates exhibit sufficient temporal consistency to enable reliable trajectory reconstruction after lightweight smoothing. These observations suggest that RSSI does not aim to replace CSI, but rather complements it by offering a hardware-friendly alternative when CSI is unavailable or unreliable.

\subsection{Practical Deployment Considerations}
From a system perspective, the proposed WiRSSI is well suited for real-world deployment, as RSSI is ubiquitously available on commodity WiFi devices and the proposed FFT-based processing pipeline is computationally lightweight, enabling real-time operation without firmware or driver-level access. These properties make WiRSSI attractive for large-scale and low-cost deployments where CSI is unavailable. At the same time, performance degradation under strong multipath conditions and large Tx-Rx separations highlights the importance of environment-aware deployment and moderate spacing. The amplitude-based delay estimation relies on a reflection-coefficient ratio obtained through a brief pre-calibration step that must be repeated when the deployment configuration changes, due to its dependence on Tx-Rx geometry and the surrounding environment. In multi-target scenarios, the limited resolution and increased Doppler noise of RSSI constrain target separation primarily to the AoA domain; with a three-antenna array, at most two targets can be reliably resolved before extracting their corresponding delays for tracking. In contrast, CSI-based sensing enables joint discrimination across delay, Doppler, and AoA domains, offering greater scalability in complex scenes, while extending RSSI-based sensing to more challenging multi-target scenarios remains an important direction for future work.

{
\subsection{Hardware Availability and Generalization}
WiRSSI is RSSI-only and does not require CSI access. Its key requirement is the availability of multi-chain RSSI streams that provide spatial diversity for AoA estimation. Platforms exposing two receive chains can still apply the same processing pipeline, albeit with reduced angular resolution compared with the three-chain configuration used in this work on the Intel 5300 NIC. By contrast, devices that expose only a single RSSI stream (e.g., many smartphones and embedded modules) do not provide the spatial diversity needed for AoA processing, although they may still support coarse temporal sensing. In practice, per-chain RSSI access can be enabled on some engineered platforms, including certain modified smartphone stacks and a range of commercial routers/access points that expose receive-chain diagnostics (e.g., on OpenWrt-based systems).
}

\subsection{Broader Implications for ISAC}
Beyond RSSI-based sensing, the insights derived from this work have broader implications for ISAC. The results suggest that power-domain measurements can retain richer sensing information than commonly assumed when interpreted through appropriate signal models. This perspective may motivate new sensing paradigms that exploit aggregated or low-resolution measurements, thereby reducing hardware complexity and calibration requirements while maintaining acceptable sensing performance. Furthermore, the demonstrated relationship between RSSI and CSI power points to opportunities for hybrid sensing schemes that jointly leverage both measurements to enhance robustness and adaptability.


\section{Conclusion}
This work revisits the long-held assumption that RSSI is too coarse for high-resolution WiFi sensing. By deriving a power-domain relationship between CSI power and RSSI, we develop \textit{WiRSSI}, a bistatic RSSI-only framework that extracts Doppler and AoA via lightweight 2D FFT processing and infers bistatic delay from an amplitude--delay relationship. The delay inference requires only a one-time calibration of a reflection-coefficient ratio for absolute scaling. Extensive experiments show that WiRSSI achieves sub-meter median trajectory accuracy across multiple motion patterns, despite the limited resolution of RSSI. {Beyond tracking, we further demonstrate that WiRSSI features provide useful motion signatures for RSSI-only sensing tasks such as gesture recognition.} Overall, these results suggest that RSSI can support practical WiFi sensing as a complementary and hardware-friendly option, particularly when CSI is restricted, unreliable, or privacy-sensitive.

\backmatter
\backmatter

%
%
%
%
%
%

\section*{Declarations}
\subsection*{Funding}
This research did not receive any specific grant from funding agencies in the public, commercial, or not-for-profit sectors.

\subsection*{Conflict of Interest}
The authors declare no competing interests.

\subsection*{Data Availability}
The datasets generated and analysed during the current study are available from the corresponding author on reasonable request and will be made publicly available upon acceptance of the manuscript.

\subsection*{Code Availability}
The code used in this study is available from the corresponding author upon reasonable request.

\subsection*{Author Contributions}
Z.W. conceived the study, developed the signal model and algorithms, conducted the experiments, and drafted the manuscript. J.A.Z. supervised the research, contributed to the technical analysis, and revised the manuscript. K.W. assisted with data analysis and contributed to manuscript revision. Y.J.G. provided overall guidance and contributed to the interpretation of the results. All authors reviewed and approved the final manuscript.

\begin{appendices}
{
\section{Exploiting Transmitter AoA to Resolve Doppler--AoA Symmetry}
\label{appendix:asymmetry}
Because RSSI is real-valued, its Doppler transform exhibits conjugate symmetry. For the RSSI sequence $\mathcal{R}_{i,k}\in\mathbb{R}$ on the $i$-th receive chain, the Doppler FFT
\begin{equation}
X_i(f^{D})=\mathcal{F}_{D}\{\mathcal{R}_{i,k}\}
\end{equation}
satisfies
\begin{equation}
X_i(-f^{D})=\overline{X_i(f^{D})},
\label{eq:appendix_dopp_sym}
\end{equation}
which implies identical magnitudes at $\pm f^{D}$ and creates a mirror ambiguity if no additional spatial reference is used.

WiRSSI introduces a spatial reference through the known transmitter direction $\theta^{S}$. After Doppler FFT, we form an angle response by applying a steering phase across the array:
\begin{equation}
Y(f^{D},\theta^{X})
=\sum_{i=0}^{N-1} X_i(f^{D})\,e^{-\bm{J}\pi(i-1)(\sin\theta^{X}-\sin\theta^{S})}.
\label{eq:appendix_aoa_fft}
\end{equation}
Intuitively, the term $e^{\bm{J}\pi(i-1)\sin\theta^{S}}$ imposes a nonzero phase progression across antennas, which breaks the mirror symmetry between $(f^{D},\theta^{X})$ and $(-f^{D},-\theta^{X})$.

To see the degeneracy condition, evaluate the response at the mirrored pair $(-f^{D},-\theta^{X})$. Using \eqref{eq:appendix_dopp_sym} and $\sin(-\theta^{X})=-\sin\theta^{X}$,
\begin{equation}
Y(-f^{D},-\theta^{X})
=\sum_{i=0}^{N-1}\overline{X_i(f^{D})}\,e^{-\bm{J}\pi(i-1)(-\sin\theta^{X}-\sin\theta^{S})}.
\label{eq:Y_neg_new}
\end{equation}
Comparing Eq. \eqref{eq:Y_neg_new} with $\overline{Y(f^{D},\theta^{X})}$ shows that they can be identical up to a global phase factor only when the $\theta^{S}$-dependent steering terms coincide, which requires
\begin{equation}
\sin\theta^{S}=0.
\label{eq:blind_condition}
\end{equation}
Therefore, the mirror ambiguity degenerates only in the blind configuration where the transmitter provides no effective spatial phase reference (i.e., $\sin\theta^{S}=0$, under our AoA convention). Otherwise ($\sin\theta^{S}\neq 0$), the transmitter-induced phase progression breaks the conjugate symmetry in the joint Doppler-AoA spectrum, enabling unambiguous peak selection.
}

\section{Amplitude-Delay Relationship of the LOS Path}
\label{appendix:LOS}
We first consider a LOS propagation path between a transmitter and a receiver. Let the corresponding propagation delay be denoted by $\tau^{T \rightarrow R}$. Under free-space propagation, the complex channel coefficient of the LOS path can be expressed as
\begin{equation}
h^{\mathrm{LOS}}
= \Gamma^{S} e^{-j 2\pi f_c \tau^{T \rightarrow R}},
\end{equation}
where $f_c$ denotes the carrier frequency. The coefficient $\Gamma^{S}$ is a real-valued amplitude factor that absorbs transmit power, antenna gains, polarization mismatch, and other system- and hardware-dependent constants.

Taking the magnitude yields the LOS path amplitude
\begin{equation}
\rho^{S} \triangleq |h^{\mathrm{LOS}}|
= \frac{\Gamma^{S}}{\tau^{T \rightarrow R}},
\end{equation}
which reveals an inverse dependence between the received signal amplitude and the propagation delay.

\section{Amplitude-Delay Relationship of the NLOS Path}
\label{appendix:NLOS}
We next consider a NLOS propagation path induced by a human target.
Although the target may generate distributed scattering, the received signal is
typically dominated by a single effective reflection associated with the
strongest scattering center. Accordingly, the NLOS propagation is modelled as a
bistatic path consisting of two segments: from the transmitter to the target,
and from the target to the receiver. Let $\tau^{T \rightarrow X}$ and
$\tau^{X \rightarrow R}$ denote the propagation delays of the two segments,
respectively. Following a multiplicative Green's-function-based formulation~\cite{rappaport2002wireless}, the complex NLOS channel coefficient can be expressed as
\begin{equation}
h^{\mathrm{NLOS}}
= \Gamma^{X}
\frac{e^{-\bm{J} 2\pi f_c (\tau^{T \rightarrow X} + \tau^{X \rightarrow R})}}
{\tau^{T \rightarrow X} \, \tau^{X \rightarrow R}},
\end{equation}
where $\Gamma^{X}$ is an effective reflection-related amplitude coefficient that
captures target-dependent scattering characteristics, antenna gains,
polarization effects, and other system- and hardware-related factors.

Taking the magnitude yields the NLOS path amplitude
\begin{equation}
\rho^{X} \triangleq |h^{\mathrm{NLOS}}|
= \frac{\Gamma^{X}}{\tau^{T \rightarrow X} \, \tau^{X \rightarrow R}}.
\label{eq:rho_nlos_raw}
\end{equation}

To relate the amplitude to the total bistatic delay, the inverse delay product can
be rewritten as
\begin{equation}
\frac{1}{\tau^{T \rightarrow X} \, \tau^{X \rightarrow R}}
=
\frac{1}{\tau^{T \rightarrow X} + \tau^{X \rightarrow R}}
\cdot
\zeta,
\label{eq:tau_product_to_sum}
\end{equation}
where
\begin{equation}
\zeta \triangleq
\frac{\tau^{T \rightarrow X} + \tau^{X \rightarrow R}}
{\tau^{T \rightarrow X} \, \tau^{X \rightarrow R}}
=
\frac{1}{\tau^{T \rightarrow X}}
+
\frac{1}{\tau^{X \rightarrow R}}
\label{eq:zeta_def}
\end{equation}
is a geometry-dependent factor determined by the bistatic propagation geometry.

We next examine how the geometry-dependent factor $\zeta$ varies under target
motion. When the target undergoes small displacements, the two propagation
segments experience small delay variations $\Delta \tau_1$ and
$\Delta \tau_2$. In this case, $\zeta$ defined in Eq. \eqref{eq:zeta_def} can be
approximated by a first-order Taylor expansion as
\begin{equation}
\begin{aligned}
\Delta \zeta
&\approx
\frac{\partial \zeta}{\partial \tau^{T \rightarrow X}} \, \Delta \tau_1
+
\frac{\partial \zeta}{\partial \tau^{X \rightarrow R}} \, \Delta \tau_2 \\
&=
-\frac{\Delta \tau_1}{(\tau^{T \rightarrow X})^2}
-
\frac{\Delta \tau_2}{(\tau^{X \rightarrow R})^2}.
\end{aligned}
\label{eq:zeta_first_order}
\end{equation}
This characterizes the local sensitivity of $\zeta$ to
motion-induced changes in the bistatic geometry, revealing an inverse-square
dependence on the segment delays.

For a fixed Tx-Rx deployment, the target is confined to a bounded sensing
region and cannot approach the transmitter or receiver arbitrarily closely.
Therefore, there exists a strictly positive lower bound $\tau_{\min} > 0$ such
that
\begin{equation}
\tau_{\min}
\le
\tau^{T \rightarrow X}, \;
\tau^{X \rightarrow R}.
\label{eq:tau_min}
\end{equation}
Using this bound together with Eq.\eqref{eq:zeta_first_order}, the variation of
$\zeta$ can be upper-bounded as
\begin{equation}
\begin{aligned}
|\Delta \zeta|
&\le
\frac{|\Delta \tau_1|}{(\tau^{T \rightarrow X})^2}
+
\frac{|\Delta \tau_2|}{(\tau^{X \rightarrow R})^2}\le
\frac{|\Delta \tau_1| + |\Delta \tau_2|}{\tau_{\min}^2}.
\end{aligned}
\label{eq:zeta_bound}
\end{equation}
This bound shows that, as long as the target remains within the sensing region,
the geometry-dependent factor $\zeta$ evolves slowly over time.

In contrast, the total bistatic propagation delay is given by the sum of the two segment delays, and its variation directly reflects the target-induced
path-length changes:
\begin{equation}
\Delta\!\left(
\tau^{T \rightarrow X} + \tau^{X \rightarrow R}
\right)
\approx
\Delta \tau^{T \rightarrow X} + \Delta \tau^{X \rightarrow R}.
\label{eq:tau_sum_variation}
\end{equation}
This indicates that the total bistatic delay varies on the same time scale as
the target motion, whereas the geometry-dependent factor $\zeta$ is much less
sensitive to such motion-induced variations.

Combining the observations from Eq. \eqref{eq:zeta_bound} and Eq. \eqref{eq:tau_sum_variation},
the product
\begin{equation}
\tilde{\Gamma}^{X} \triangleq \zeta \, \Gamma^{X}
\end{equation}
can be treated as a constant and obtained via a one-time calibration under the same operating conditions. Under this approximation, the NLOS path amplitude can be simplified as
\begin{equation}
\rho^{X}
=
\frac{\tilde{\Gamma}^{X}}{\tau^{T \rightarrow X} + \tau^{X \rightarrow R}},
\end{equation}
indicating that the dynamic-path amplitude is primarily governed by the total
bistatic propagation delay.

\end{appendices}


\bibliography{sn-bibliography}

\end{document}